\documentclass[superscriptaddress,twocolumn,pre,nofootinbib,showpacs]{revtex4}

\usepackage{amsmath}
\usepackage{graphicx,color}
\usepackage{amssymb}
\usepackage{amsfonts}
\usepackage{bm}

\newcommand{\be}{\begin{equation}}
\newcommand{\ee}{\end{equation}}
\newcommand{\bea}{\begin{eqnarray}}
\newcommand{\eea}{\end{eqnarray}}

\begin{document}
\sloppy


\title{Maximum mass of relativistic self-gravitating Bose-Einstein condensates with repulsive
or attractive $|\varphi|^4$ self-interaction}

\author{Pierre-Henri Chavanis}
\affiliation{Laboratoire de Physique Th\'eorique, Universit\'e de Toulouse,
CNRS, UPS, France}

\begin{abstract}

We derive an approximate analytical expression of the maximum mass of 
relativistic self-gravitating Bose-Einstein condensates with repulsive or
attractive $|\varphi|^4$ self-interaction.
This expression interpolates between the general relativistic
maximum mass of noninteracting bosons stars, the general relativistic
maximum mass of bosons stars with a repulsive self-interaction in the
Thomas-Fermi limit, and the Newtonian maximum mass of
dilute axion stars with an attractive self-interaction [P.H. Chavanis, Phys.
Rev. D
{\bf 84}, 043531 (2011)]. We
obtain the general structure  of our formula from simple considerations
and determine the numerical
coefficients in order to recover the exact asymptotic expressions of the maximum
mass in particular limits. As a result, our formula should provide a relevant
approximation of the maximum mass of relativistic boson stars for any value
(positive and negative) of
the self-interaction parameter. We discuss the evolution of the system above
the maximum mass and consider application of our results to dark matter halos
and inflaton clusters. We also make a short review of
boson stars and
Bose-Einstein condensate dark matter halos, and point out analogies with models
of
extended elementary particles.

\end{abstract}


\maketitle


\section{Introduction}

The concept of boson stars was introduced by Kaup \cite{kaup} and Ruffini and
Bonazzola \cite{rb} (see also Refs. \cite{das,bp,fmk}) in the 1960s in a
rather academic manner without
explicit connection to any astrophysical object. They were just hypothetical
stars governed by the laws of general relativity and quantum
mechanics. In a
sense, boson stars are the 
descendents of the so-called {\it geons} of Wheeler \cite{geons} except that
they
are built from scalar particles of spin $0$ instead of electromagnetic field,
i.e., spin-$1$ bosons.\footnote{The {\em geon}, which is a {\it g}ravitational
{\it e}lectromagnetic entity, was originally introduced
by Wheeler \cite{geons} as a localized nonsingular 
solution of the Einstein-Maxwell equations. A geon
consists of a
spherical shell of electromagnetic radiation held together by its own
gravitational attraction.
It realizes to some extent  the proposal of Einstein \cite{einsteinelec} and
Einstein and Rosen \cite{eiro}:
``Is an atomistic theory of matter and electricity conceivable which, while
excluding singularities in the field, makes use of no other fields than those
of the gravitational field and  those of the 
electromagnetic field in the
sense of Maxwell?", and some of  the goals of the unitary field theory
\cite{rosenelec1,rosenelec2,finkelstein,flr,rosenrosenstock,rosenstock,ffk}.
These objects, however, were found to be unstable
\cite{geons,power} leading to
the belief that gravitational collapse is inevitable. In the following
years, transferring the ideas 
of Mach and Einstein to the microcosmos, many researchers   tried to find a 
model  which describes an elementary particle in terms of a semiclassical field
coupled to the Einstein equations. 
These extended particles  resemble  geons \cite{geons} or
wormholes \cite{regge,misner,wormhole}. Kaup \cite{kaup} presented the
notion of Klein-Gordon geon that was later called boson star.} 
Kaup \cite{kaup} and Ruffini and Bonazzola \cite{rb} considered the $T=0$ limit
in which bosons form Bose-Einstein condensates (BECs). In
that case, all the
bosons are in the same quantum state described by a unique complex wave function
$\varphi(x^{\mu})$ satisfying the Klein-Gordon-Einstein  (KGE)
equations.\footnote{See Ref.
\cite{chavmatos} for a short account of the early history of wave mechanics
(Schr\"odinger, 
Klein-Gordon, Dirac, Gross-Pitaevskii equations) and an exhaustive list of
references.} This
semi-classical equation is valid in the Hartree-Fock approximation for the
second quantized two-body problem \cite{rb}.  Boson stars can be regarded as
macroscopic quantum states that are only prevented from collapsing
gravitationally by the Heisenberg uncertainty principle. No Schwarzschild type 
event horizon occurs in these objects since their density profile extends to
infinity. Kaup \cite{kaup} and Ruffini and
Bonazzola \cite{rb} showed that equilibrium states can exist only below a
maximum mass $M_{\rm max}^{\rm GR}=0.633 M_P^2/m$ set by general relativity,
where $M_P=(\hbar
c/G)^{1/2}$ is the Planck
mass. Above this maximum mass
the star is expected to collapse and form a
black hole (or evaporate). Their results were 
re-derived and confirmed in Refs.  \cite{bgr,bbh,bgz,tayo,tdlee,flpd,lpstab}.
The maximum mass of boson stars is the counterpart of the maximum mass
of fermion stars such as white dwarf stars and neutron stars. The maximum mass
of white dwarf stars  $M_{\rm max}^{\rm WD}=3.10\, M_P^3/(\mu H)^2$,
where $H$ is the proton mass and $\mu$ the molecular weight, was found by
Chandrasekhar \cite{chandra31} and the maximum mass of 
neutron stars $M_{\rm max}^{\rm NS}=0.384\, M_P^3/m^2$, where
$m$ is the neutron mass, was found by Oppenheimer and Volkoff \cite{ov}.

Boson stars can  be viewed as complex scalar fields (SFs) or anisotropic fluids
\cite{kaup,rb} while fermion stars are isotropic fluids
\cite{htww}. Despite
this distinguished
feature, there exist remarkable 
similarities between boson stars  and fermion stars. The mass-central density
relation $M(\rho_0)$ and the
radius-central density relation $R(\rho_0)$ of boson stars exhibit damped
oscillations
\cite{kaup,rb,bgz,flpd,vdbg,gleiser,jetzer,gw,ss90,kms1,kms2,ksa} and the
mass-radius relation
$M(R)$ has a snail-like (spiral)
structure \cite{bgz,ss90,ksa}.
The series of equilibria becomes unstable  after the first mass peak and a new
mode of stability is lost at each subsequent turning point of mass like in the
case of neutron stars \cite{htww}. These stability  results can be established
from the study of the 
pulsation equation \cite{gleiser,jetzer,gleiserE,gw,jetzerex}\footnote{The
pulsation equation for boson stars was derived by Gleiser \cite{gleiser}
and Jetzer \cite{jetzer} who generalized  the approach developed by
Chandrasekhar \cite{chandra64} for an isotropic fluid in general relativity.
Since the SF is complex,
they
obtained a system of two coupled eigenvalue equations. Using the method of test
functions they could only prove that equilibrium states with a central density
much larger than the critical density (corresponding to the maximum mass) are
unstable. They first suggested that the instability point may not coincide with
the
maximum mass because of anisotropic effects. But soon after,  Gleiser
\cite{gleiserE}, Gleiser and Watkins \cite{gw}  and Jetzer \cite{jetzerex},
following the work of Lee and Pang \cite{lpstab}, showed that the instability
actually occurs at the maximum mass and that a new mode becomes unstable at
each successive critical points. This is due to the fact that the
functions $M(\rho_0)$
and $N(\rho_0)$ are extremal at the same points (see below), implying that the
pulsation vanishes at these points \cite{lpstab}.} or
from the energy
principle  \cite{flpd} stating that a stable
equilibrium state is a minimum of mass-energy $Mc^2$ at fixed particle number
$N$ (the charge or the number of bosons minus anti-bosons is conserved for a
complex SF). The variational principle for the first variations $\delta
M-\alpha\delta N=0$ implies that
$M(\rho_0)$ and $N(\rho_0)$ are extremal at the same points, precisely where a
mode of pulsation vanishes \cite{flpd,lpstab}. The mass-particle number relation
$M(N)$ is plotted in Refs.
\cite{flpd,lpstab,kms1,kms2,ksa}. The
$M(N)$ curve presents cusps at the critical points where $M(\rho_0)$
and $N(\rho_0)$ are
extremal. It forms a zig-zag course of smaller and smaller extension
associated with the spiral nature of the mass-radius relation. Kusmartsev {\it
et al.} \cite{kms1,kms2,ksa} used the $M(N)$ curve interpreted as a bifurcation
diagram to investigate the dynamical stability of boson stars by invoking
Arnold's classification of singularities in catastrophe theory
\cite{post,arnold}
and the Whitney theorem \cite{whitney} which were
previously applied in Ref.
\cite{kusmartsev} to nongravitational solitons. If the
mass becomes larger after a cusp, a new mode of instability appears. Inversely,
if the mass becomes smaller after a cusp, a mode of instability
disappears.\footnote{One can also
deduce the stability of boson stars from the Poincar\'e theorem
\cite{poincare,katzpoincare} (see, e.g.,
Appendix C of \cite{acf} for a brief exposition of this theorem). This approach
was previously applied to the series of equilibria of isothermal
self-gravitating systems which present features similar to those of fermion and
boson stars (damped oscillations,
spirals, zigzags, cusps...) \cite{ijmpb,acf}. Using the Poincar\'e
turning point argument, the stability of boson
stars
can be directly infered from Fig. 8 of \cite{flpd}, returning the results of
\cite{kms1,kms2,ksa}. One can also use the Poincar\'e
turning point argument to  determine the stability of fermion
stars (white dwarfs and neutron stars) \cite{katzrevue}, returning the results
of \cite{htww} based on the Wheeler $M(R)$ theorem.} Seidel and Suen
\cite{ss90},
Kusmartsev {\it et al.} \cite{kms1,kms2,ksa} and Guzm\'an \cite{g1,g2}
showed that a stable boson star which is slightly
perturbed will oscillate with a fundamental frequency, emit SF
radiation, and settle down into a new equilibrium configuration with less mass
while an unstable boson
star will in general collapse to a black hole or migrate to the stable branch.
They also mentioned 
that the
binding
energy $E_b=(M-Nm)c^2$ can become positive signaling the existence of
configurations with excess energy.  Such configurations are always unstable
against a collective transformation in which they are dispersed into free
particles at infinity \cite{kaup,vdbg}.  These results
are remarkably similar to those obtained for neutron stars \cite{htww}.
However,
there also exist crucial differences between boson and fermion stars. In
particular, boson stars are stabilized
by the Heisenberg
uncertainty principle while fermion stars are stabilized by the Pauli exclusion
principle. This difference is reflected in the scaling
with $m$ of the maximum mass of stable configurations. The maximum mass of boson
stars scales as $M_{\rm max}^{\rm GR}\sim  M_P^2/m$ \cite{kaup,rb} instead of 
$M_{\rm max}^{\rm GR}\sim  M_P^3/m^2$ \cite{chandra31,ov}
for fermion stars. As a result, for the same particle mass $m$, the maximum mass
of
boson stars is smaller than the mass of fermion stars by a factor $m/M_P\ll 1$.
For example, for $m\sim 1\, {\rm GeV/c^2}$ (of the order of the
neutron mass)
for which $m/M_P\sim 10^{-19}$, the maximum mass of boson stars is $M\sim
10^{-19}\, M_{\odot}$ while the maximum mass of neutron stars is of the order
of
the solar mass. This leads to the concept of mini boson stars
\cite{bgz,tayo,lpstab} or
mini soliton stars \cite{tdlee,flpd,lpstab} with small mass, small
radius and extremely high densities. This
property may facilitate
the formation of small black holes made of cold invisible axions (bosonic
black holes \cite{bbh} or axion black
holes \cite{tayo}) in the axion dominated universe. The
maximum
mass of boson stars becomes of the order of the solar mass if the mass of the
bosons is very small, typically $m\sim 10^{-10}\, {\rm eV/c^2}$.

Friedberg {\it et al.} \cite{flpd} (see also
\cite{misc,fmk,demi,mielke543}) computed the radial
solutions of the static KGE equations  with nodes, which correspond to excited
states. They found that the  critical mass grows
approximately linearly with the number of nodes (see also \cite{vdbg}).  Jetzer
\cite{jetzerex} showed that the excited Bose star configurations are stable for
central densities up to the critical density.\footnote{Lee
and Pang \cite{lpstab} previously  argued that excited modes are
always unstable but their claim is incorrect because
they did not require the
particle number to be constant in their stability analysis.}
As a result, as noted by Lee \cite{leecomment,tdlee}, by increasing the node number $n$ there exists an equilibrium state for any value of
the mass (at least classically). Since a boson
star is a stationary solution of
the KG equation  in its own 
gravitational field, a boson star is also 
 called a {\it gravitational atom} \cite{fegl}. The energy levels of these
gravitational atoms present an interesting fine structure for high values
of the  principal quantum number. This
fine splitting of the
energy levels may indicate a rich particlelike structure of the quantized
geons \cite{misc}.   In the last stages of boson star  formation,
one 
expects that a highly excited configuration first forms with several nodes and that it eventually decays
into the ground state (with no node)  by a combined emission of scalar 
radiation and gravitational radiation. This mechanism has been analyzed
by Ferrell and Gleiser \cite{fegl} in a Newtonian approximation.

Colpi {\it et al.} \cite{colpi}   considered 
the case of bosons stars with a repulsive $\frac{\lambda}{4\hbar c}|\varphi|^4$
($\lambda>0$) self-interaction and found that the resulting configurations
differ markedly from the noninteracting
case.\footnote{The first investigations of the KGE
equations with a
self-interaction potential were carried out by Mielke and Scherzer \cite{misc}
(see also \cite{mielke530,demi,mielke4525,mielke543,mielke189}). 
They used a $-|\varphi|^4+|\varphi|^6$  potential 
obtained from a nonlinear Heisenberg-Pauli-Weyl \cite{heisenberg,weyl1,weyl2}
spinor equation in a curved spacetime produced by the energy-momentum tensor of
the spinor fields via the Einstein equations. This work
was done in the context of particle physics
(independently from the context of boson stars) in
order
to describe classically extended particles consisting of confined quarks. In
that case, the
Einstein-type field equations account for strong interactions on a curved space
time of hadronic dimensions characterized by a modified Planck length of strong
gravity
\cite{isham1,tennakone,sivaram,salam1,salam2,sivaram77,salam3,salam4,
caldirola,sivaramsinha}. Algebraic complications resulting
from the spinor  structure were avoided  by considering SFs coupled to gravity. 
In order to maintain a  similar dynamics, a scalar  self-interaction
$V(\varphi)$ was formally obtained by ``squaring'' the fundamental nonlinear
spinor equation.  This provides a model for a unitary field theory of extended
particles resembling the geons of Wheeler which are 
held together by self-generated gravitational forces and are composed of 
localized fundamental classical fields described by the Einstein-Maxwell
equations. The coupling of gravity to neutrino fields
had been  previously  considered by Brill and Wheeler \cite{brill}. Their work 
provided the appropriate groundwork for an extension 
to nonlinear {\it spinor  geons} satisfying the 
Dirac-Einstein equations considered in \cite{misc}. On the
other hand, Lee and co-workers \cite{tdlee,sss,fss}
investigated a Higgs-type potential \cite{higgs} with symmetry breaking
of the degenerate vacuum form.  They called the solutions of their KGE equations
non-topological soliton
stars and found that the maximum mass  of these scalar \cite{sss} or fermion \cite{fss} soliton stars has
units of $M_P^4/m^3$ which is huge -- of the order of the mass of the
universe $\sim 10^{20}\, M_{\odot}$ -- in
comparison with a boson or a neutron star (for the case of comparable boson and
fermion masses $m\sim 1\, {\rm GeV/c^2}$).
Nishimura and Yamaguchi \cite{niya} constructed
a neutron
star
using an equation of state of an isotropic fluid built  from Higgs bosons. Their
work
showed that the
limit for such boson stars can possibly exceed the limiting mass of $3.2\,
M_\odot$ for neutron stars \cite{rhoades}. }  In the Thomas-Fermi (TF) limit,
the maximum
mass set by
general relativity is  $M_{\rm max}^{\rm GR}=0.0612 \sqrt{\lambda}
M_P^3/m^2$.\footnote{Tkachev \cite{tkachev} independently considered the case of
bosons stars with a repulsive $|\varphi|^4$ self-interaction and obtained the
scaling of the maximum mass  $M_{\rm max}^{\rm GR}\sim \sqrt{\lambda}
M_P^3/m^2$ from qualitative arguments.} 
For $\lambda\sim 1$ it
exhibits the same scaling
as the maximum mass of fermion stars (Chandrasekhar's mass).  This leads to
much bigger
structures
than in the noninteracting case, making them much more astrophysically
interesting. They are called ``massive boson stars''.
In a sense, the repulsive self-interaction for bosons plays a role
similar to the quantum pressure arising from the Pauli exclusion principle for
fermions.  Like for fermion stars and noninteracting boson stars, the
$M(\rho_0)$, $R(\rho_0)$ and $N(\rho_0)$
curves exhibit damped
oscillations 
\cite{colpi,gleiser,jetzer,kms1,kms2,ksa,niya,schunckliddle,gb,bss,
schuncktorres,chavharko}, the
$M(R)$ curve has a snail-like
(spiral) structure \cite{ksa,chavharko}, and the $M(N)$ curve displays cusps and
forms a
zig-zag course of smaller and smaller
extension
\cite{kms1,kms2,ksa,schuncktorres}. A detailed discussion of the  analogy
between boson stars and neutron stars is given in Ref. \cite{ksa}. Colpi {\it et
al.} \cite{colpi} (see also \cite{tkachev}) showed that, in the TF limit, a 
boson star with a repulsive
quartic self-interaction is equivalent to an isotropic barotropic fluid with a
well-defined equation of state $P(\epsilon)$.\footnote{See Ref. \cite{action}
and Appendix \ref{sec_ra} for a detailed derivation of this equation of state.
The
hydrodynamic decription of boson stars with a repulsive self-interaction in the
TF limit has been studied in detail by Chavanis and Harko \cite{chavharko} in
the frawework of the Oppenheimer-Volkoff equations.} This
equation of state reduces to that of an $n=1$ polytrope at low densities
(nonrelativistic limit) and to the linear law $P\sim \epsilon/3$ similar to the
equation of state of radiation at high
densities (ultrarelativistic limit).  This strengthens the analogy between boson
stars and neutron stars which display the same asymptotics at high densities.
In these objects the speed of sound is always less than the speed of light.
However, ultrarelativistic configurations are dynamically
unstable according to the Poincar\'e criterion because they are located
after the first turning point of mass \cite{chavharko}. Cold mixed
boson-fermion stars have been studied by 
Henriques {\it et al.} \cite{hlm1,hlm2,hlm3} and 
Jetzer \cite{jetzerbf}.

Jetzer and Scialom \cite{js,jsarx} showed that the static solutions of the KGE
equation for a real SF have a  naked singularity
at the origin and that they are always dynamically unstable. This result is in
agreement
with the cosmic
censorship conjecture which 
excludes spacetimes with naked singularities \cite{penrose}. In particular,
general
relativistic real massless SFs are
unstable.\footnote{Real massless SFs coupled to Einstein gravity were
first
considered by Bergmann and Leipnik \cite{berleip} and Yilmaz \cite{yilmaz} (see
also \cite{treder}). They are known to
admit exact static solutions which were discovered by Buchdahl \cite{buch} for a
special case. In the framework of the
Jordan--Brans--Dicke--Thiry theory, these solutions  already
appeared   in Ref. \cite{ehlers} and correspond to those found by 
Majumdar \cite{majumdar} for the Einstein--Maxwell system. Later, the solutions 
 discovered by Buchdahl \cite{buch} were 
rederived by Wyman \cite{wyman} (see also \cite{bmhh,scsc}) and generalized to 
spacetimes of arbitrary dimensions  by
Xanthopoulos and Zannias \cite{xz}.} This conclusion is in agreement with the
result of 
Christodoulou \cite{christodoulou1,christodoulou2} (see also
\cite{piran,choptuik})
who showed that all time-dependent spherically symmetric solutions of the KGE equations with  $m=\lambda=0$ 
must either disperse to infinity or form a black hole.  Real
SFs (like axions) do not admit regular static solutions to the KGE equations
because there is no conserved Noether current leading to particle number (or
charge) conservation.\footnote{Note that particle number conservation is
approximately
restored in the nonrelativistic limit so that Newtonian boson stars made of a
real SF,
like dilute axion stars, 
become stable in that limit \cite{phi6,tunnel}.} It is possible to
construct regular solutions which are periodic in time 
when  $M<M_{\rm max}^{\rm GR}=0.606\, M_P^2/m$ \cite{ssreal,alcubierre}.
However, on
a
long timescale (which can nevertheless exceed the age of the universe), these
``oscillatons'' are
unstable and disperse to infinity or form a
black hole.

Nontopological solitons
\cite{flr,rosenrosenstock,rosenstock,ivanenko,petiau,skyrme1,skyrme2,enz,
derrick,rosen1269,rosen2066,
rosen2071,rosen573,rosen996,derrickkong,
rosen1186,scott,rubinstein,barone,
anderson,dashen1,gj,dashen2,jac,leewick,christlee,fls,soler,demi,
mielke4525,mielke543,mielke189} may
be
regarded as the nongravitational precursors
of boson stars.\footnote{The KG equation with an attactive
$|\varphi|^4$ self-interaction was introduced by Finkelstein {\it et al.}
\cite{flr} and by Rosen and Rosenstock \cite{rosenrosenstock,rosenstock}
(see Schiff \cite{schiff1,schiff2} and Malenka \cite{malenka} for the case of a
repulsive $\varphi^4$ 
self-interaction in the context of nonlinear meson theory for heavy nuclei; see
also
Goldstone \cite{goldstone}, Higgs \cite{higgs} and Nielsen and Olesen \cite{no}
for a Mexican hat potential
with $m^2<0$ and $\lambda>0$). The
sine-Gordon equation was introduced by Petiau
\cite{petiau}, Skyrme \cite{skyrme1,skyrme2} and Enz \cite{enz} in the context
of particle
physics (see also the analogy with Bloch walls in magnetic crystals
\cite{doring} and the motion of a slide dislocation in a
crystalline structure \cite{fk,ks1,ks2,ks3}).
The name
``sine-Gordon equation''
first appeared in Ref. \cite{rubinstein} and was coined by Kruskal (it is more
precise than the name ``nonlinear KG equation'' \cite{scott}). The
concept and the
name  ``soliton'' were introduced by Zabusky and Kruskal \cite{zk}
in the context of the Korteweg-deVries (KdV) equation \cite{kdv}. The analogy
between the sine-Gordon equation and the KdV equation was first pointed out by
Rubinstein \cite{rubinstein}.} For a specific
Higgs  type self-interaction potential
$V(\varphi)$, they are localized solutions of a 
nonlinear KG equation in flat spacetime. These solitons were
meant to
represent elementary particles extending in space (i.e. particles with a
structure),
pursuing the initial goal of Einstein and Rosen
\cite{einsteinelec,eiro,rosenelec1,rosenelec2}, Finkelstein
\cite{finkelstein,flr,ffk},
Dirac \cite{dirac1,dirac2,dirac2b,dirac3}, and de Broglie \cite{debroglie}. In
flat
spacetime, according to Derrick's
theorem \cite{derrick},
no stable time-independent solution of finite energy exists for a nonlinearly
coupled real SF in dimension larger than one.
Nontopological solitons may,
however, have a very
long lifetime if their rate 
of dissolution is small \cite{rosen1269}. Objects similar to nontopological
solitons are called Q-balls \cite{qball},   fermion
Q-balls
\cite{lynn,bls1,bls2,bls3},
neutrino balls \cite{holdon}, and quark
nuggets \cite{witten} in the case of
spinors. Q-balls \cite{qball} are stabilized by their conserved charge $Q$.
Bound
further by their self-gravity, Q-stars \cite{lynn,bls1,bls2,bls3}  may model 
neutron stars   with a mass larger than $\sim 3\, M_\odot$.
Other SFs arise
from axion \cite{rees,kt,kt2,kt3,kt2apj}, inflaton or dilaton
fields \cite{grk} with their corresponding compact objects axion stars
\cite{tkachev,tkachevrt,kt} and
dilaton stars \cite{grk}. The dilaton field
arises in the process of a Kaluza-Klein type dimensional reduction of
supergravity or superstring models. These real SFs couple to gravity similarly
to the Brans-Dicke field of scalar-tensor theories. 
The dilaton stars \cite{grk} are  stable because of a conserved  dilaton current
and charge in such models. In an extended model 
\cite{tx} with Higgs field and
dilaton coupling, there  occur, however, unstable branches with diverging mass
for high 
central values  of the Higgs field. Nonsingular, time-dependent,
spherically symmetric solutions of nonlinear SF theories were
constructed in \cite{pulson1,pulson2,oscillon,cgm} and called
``pulsons'' \cite{pulson1,pulson2} or
``oscillons'' \cite{oscillon,cgm}.
They characterize 3D nongravitational self-interacting real  SFs described by
the 
KG equation. Although
unstable, they are extremely
long-lived.  Spherically symmetric 
solutions (including excited states) of the KG equations for a real or complex
SF in a prescribed Schwarzschild, de Sitter
or Friedmann-Lema\^itre-Robertson-Walker metric were
constructed
in Refs.~\cite{okolowski,demi,mielke189,elizalde1,elizalde2,elizalde3,salam4} in
the context of  black holes,  particle physics, and cosmology. More
recently,
Rosen
\cite{rosenkg,rosenproca} revived his old
idea of an
elementary  particle built out of SFs  within the framework of the KG
or Proca equations coupled to the Einstein equations.\footnote{See also the
concept of Proca stars introduced in
\cite{radu1,radu2}.} He developed a model of
particles  (interpreted as gravitational
solitons) closely related to the
model of boson stars, being apparently unaware of the vast literature on this
subject.\footnote{It is fascinating to
realize that the KG
equation coupled to the Maxwell and Einstein equations have been introduced
independently by
different
communities to describe either boson stars in astrophysics (with
the usual gravitational constant $G$) or elementary
particles (with a modified gravitational constant $G_f$). This was
foreseen by Einstein \cite{einsteinelec} in 1919: ``There are reasons for
thinking that the
elementary formations which go to make up the atom are held together by
gravitational forces''. Models of extended elementary particles have been
constructed from the
Maxwell equations with Poincar\'e stress
\cite{abraham,lorentz,poincareE,dirac3,hk1,hk2},
from nonlinear
Maxwell equations
\cite{mie,born1933,borninfeld,schiffnm}, from the Maxwell-Einstein equations
\cite{einsteinelec,geons,geonbh},  from the KG-Maxwell equations
\cite{finkelstein,rosenelec1,rosenelec2}, from the (multifield) self-interacting
KG equations
\cite{flr,rosenrosenstock,rosenstock,ivanenko,petiau,skyrme1,skyrme2,enz,
derrick,rosen1269,rosen2066,rosen2071,rosen573,rosen996,derrickkong,
rosen1186,scott,rubinstein,barone,
anderson,dashen1,gj,dashen2,jac,leewick,christlee,fls,soler,demi,
mielke4525,mielke543,mielke189}, from the (massless) self-interacting Dirac
equation
\cite{weyl50,thirring,flr,ffk,heisenberg53,heisenberg54,gursey,
ivanenko,heisenberg57,thirring58,das63,soler,ranadasoler,ranada,mielke2034,
mielke530}, from the 
Dirac-Maxwell equations
\cite{kaempffer,wakano,lisi,bohun,biguaa}, from the KG-Maxwell-Einstein
equations
\cite{das,ovono,cr,ak}, from the (multifield) self-interacting
KG-Maxwell equations \cite{rosenkgmn,cr,cooperstock,rybakov}, from the
self-interacting
Dirac-Maxwell equations \cite{soler2,rs2,rrsv}, from the self-interacting
Dirac-KG equations \cite{leewick,bcdwy}, from the self-interacting
Dirac-KG-Maxwell equations \cite{ranadavazquez}, from the self-interacting
KG-Einstein
equations \cite{kodama1,kodama2,kodama3,misc}, from the KG-Einstein equations
\cite{salam4,rosenkg}, from the 
Einstein-Maxwell equations for a charged fluid with a vacuum
equation of state $P=-\epsilon$
\cite{tiwari,gautreau,gron,isro1,isro2}, from the Yang-Mills-Einstein
equations \cite{bartnik,sz}, from the Proca equations \cite{rosenproca},
 from the 
Weyl equations
\cite{isro4}, from the Dirac-Einstein equations 
\cite{finster}, from the Dirac-Maxwell-Einstein equations \cite{fsy,fsy2}, from
the  self-interacting (Higgs)
KG-Maxwell-Einstein equations 
\cite{burinskiiH}, and from the gravitational Dirac-Maxwell \cite{km} equations.
In Maxwell's theory of electromagnetism, charged particles appear as
singularities in the field. The field equations break down at
singular points,
and so, separate equations of motion have to be prescribed for the particles.
Einstein and Infeld \cite{eiin} emphasized that a
proper field theory knows only fields and
not particles and that particles should only emerge from the fields
themselves. They write: ``Could we not reject
the concept of matter and build a pure field
physics? We could regard matter as the region in space where the field is
extremely strong.''
In the unitary field theories listed above, particles appear not
as singularities but as small volumes in which the energy and the charge of the
field are concentrated. These unitary field models are necessarily nonlinear
in order to avoid singularities.
All the properties of
the particles such as
their equation of motion follow from the field equations.
Other authors
\cite{isham1,tennakone,sivaram,salam1,
salam2,sivaram77,salam3,salam4,geonbh,lopez84,lopez,isro3,dkn,burinskii}
developed analogies
between elementary particles and Schwarzschild
\cite{schwarzschild1,schwarzschild2},
Kerr-Newmann \cite{kerr,newman1,newman2},
and Reissner-Nordstr\"om \cite{reissner,nordstrom} black holes.
Carneiro \cite{carneiro} writes: ``The strong gravity
approach \cite{salam3,salam4,caldirola,sivaramsinha} tries to derive the
hadron
properties from a scaling down of gravitational theory, treating particles as
black hole type solutions. It is based on the scale invariance of general
relativity. With this philosophy, we can think the universe as a self-similar
structure, with the same physical laws appearing at different scales''.
Kodama \cite{kodama2,kodama3} found a stable static, singularity-free,
finite energy solution of the KGE equations with a $\phi^4$ potential that
extends the usual 1D kink solution \cite{dashen1,gj}. The spacetime geometry
implied resembles that of the Einstein-Rosen bridge \cite{eiro} of the
Schwarzschild geometry, i.e., two asymptotically flat spaces connected by a
bridge (a ``black soliton'' or a kind of wormhole in the sense of Wheeler
\cite{geons,regge,misner,wormhole}).} In
a
sense, Wheeler's geon concept has 
anticipated the nonintegrable soliton solutions of classical nonlinear field
theory. We refer to specific reviews on solitons
in particle physics \cite{rajaraman,makhankov,rajaramanbook}, soliton stars
\cite{leepang} and boson stars 
\cite{jetzerREVUE,liddlemadsen,sm1998,sm1999,sm2000,sm2003,liebling,krippendorf,
visinelliREVUE} for  more  information on the subject.

Boson stars associated with a complex SF can be formed from a dissipationless 
relaxation process called {\it gravitational cooling} \cite{seidel94}. This
process has a counterpart in stellar dynamics.
Collisionless stellar  systems are known to
experience 
a process of violent relaxation \cite{lb} during which they form a  centrally
dense core by sending some stars at large distances in an extended halo.
Similarly, a bosonic cloud will settle to a 
unique boson star by ejecting part of the scalar matter.\footnote{See
\cite{wignerPH} for a description of the analogy between gravitational cooling
and
violent relaxation.} Since there is no 
viscous term in the KG equation, the radiation of  the SF is the
only mechanism. The emission of gravitational waves also occurs in 
the absence of spherical symmetry.  Boson star solutions in their ground state
or in excited
configurations have been used in the context of dark matter (DM) to fit the 
observed rotation
curves of dwarf 
and spiral galaxies (see the introduction of Ref.  \cite{prd1} for a short
review of early works on
this topic). Indeed, if bosons are
ultralight, with a mass $m\sim 10^{-22}\, {\rm
eV/c^2}$, they
can form compact objects
of galactic size (see \cite{hui,jeansapp} and references therein).\footnote{More
massive
bosons with a  mass $ 10^{-22}\, {\rm eV/c^2}\le m\le 10^{-3}\, {\rm eV/c^2}$
can also form compact objects of galactic size provided that they are
self-interacting with a dimensionless self-interacting constant in the range
$0\le \lambda\le 10^{-15}$ \cite{jeansapp}.} Therefore, ultralight axions
(ULAs) have been invoked in
models of DM halos. In the  context of
DM, Newtonian
gravity is usually a good approximation so we can use the Schr\"odinger-Poisson 
or GPP equations instead of the KGE equations. A Newtonian boson star is a stationary
solution of these equations. However, models of DM halos
based
on a pure boson star solution
(soliton) are usually not successful, especially in the case of large DM
halos. Indeed, the mechanism
of gravitational cooling \cite{seidel94} 
typically leads to a ``core-halo''
structure with a
quantum core
(soliton)  in its ground state surrounded by an extended halo 
resulting from the quantum interferences of the excited states. This
``core-halo'' structure  has been evidenced in numerical simulations of the
Schr\"odinger-Poisson
equations
\cite{ch2,ch3,schwabe,mocz,moczSV,veltmaat,moczprl,moczmnras,veltmaat2} and can
be heuristically explained by an adaptation   of Lynden-Bell's 
statistical theory of violent relaxation \cite{lb,csr} as discussed in Ref.
\cite{wignerPH}. 
The quantum core may solve the core-cusp problem \cite{moore} of the cold dark
matter (CDM)
model and the halo, which is similar to an
isothermal or a Navarro-Frenk-White (NFW) profile, accounts for
the flat rotation curves of the
galaxies.  This type of core-halo
configurations leads to more realistic models of DM halos than a pure BEC
solution and
has been the subject of an intensive research over the last few years (see,
e.g., an exhaustive list of references in \cite{jeansapp} and in the recent
reviews
\cite{srm,rds,chavanisbook,marshrevue,leerevue,niemeyer,ferreira,huirevue}).

In the nonrelativistic limit, the KGE equations reduce to the
Schr\"odinger-Poisson equations in the noninteracting limit  or to the
Gross-Pitaevskii-Poisson (GPP) equations when the bosons interact through a
$|\varphi|^4$ potential.\footnote{The nonrelativistic limit of the KGE
equations
is discussed in \cite{abrilph,playa,chavmatos}  for a complex SF and in
\cite{phi6,tunnel} for a real SF. In the case of
a
complex SF, the potential that appears in the GP equation is the same as the
potential that appears in the KG equation. In the case of a real SF, they are
usually different (see Appendices \ref{sec_ra} and \ref{sec_rrsf} for a 
detailed discussion).} 
Using the Madelung \cite{madelung} transformation, one can introduce a
hydrodynamic representation of these equations in the form of compressible Euler
equations with an additional quantum potential.  The complete mass-radius
relation of nonrelativistic self-gravitating BECs with repulsive or attractive
self-interactions was obtained in \cite{prd1,prd2} either exactly by solving the
GPP equations numerically or analytically (approximately) by using variational
methods based on the minimization of energy at fixed mass and employing  a
Gaussian
ansatz for the wave function. For noninteracting bosons, one obtains the
mass-radius relation $M=9.95\, \hbar^2/(Gm^2R_{99})$ \cite{membrado,prd2}, showing
that the
radius of the star decreases as its mass increases (similarly to the 
mass-radius relation $M=1.49\times 10^{-3}\, h^6/[G^3m^3(\mu H)^5R^3]$ of
nonrelativistic white dwarf
stars \cite{chandrabook}). For a repulsive self-interaction, the mass-radius
relation is
modified at high masses. In the TF regime, corresponding to $M\rightarrow
+\infty$, the star is equivalent to a polytrope of index $n=1$, and the radius
of the BEC tends to a minimum value $R_{\rm TF}=\pi(a_s\hbar^2/Gm^3)^{1/2}$
\cite{tkachev,maps,leekoh,goodman,arbey,bohmer,prd1} independent of its mass.
These latter results were extended in the
general
relativistic regime by Chavanis and Harko \cite{chavharko} using a hydrodynamic
approach (see also \cite{mlbec} for a complementary discussion). This leads to
the
concept of general relativistic BEC stars with a maximum mass $M_{\rm max}^{\rm
GR}=0.307\, \hbar c^2\sqrt{a_s}/(Gm)^{3/2}$  \cite{chavharko}  equivalent to the
one found in Ref. \cite{colpi}. Chavanis and Harko \cite{chavharko} suggested
that, because of their superfluid core, neutron stars could be considered as BEC
stars. Indeed, neutrons could form Cooper pairs and behave as bosons of mass
$2m_n$
(where $m_n$ is the neutron mass). Since the maximum mass of BEC stars depends
on the self-interaction parameter, it can be larger than the Oppenheimer-Volkoff
limit $M_{\rm OV}=0.376\, (\hbar c/G)^{3/2}/m_n^2=0.7\, \, M_{\odot}$
obtained when the neutron star is modeled as an ideal Fermi gas.  This could
explain certain observations of  neutron stars with a mass $\sim 2\, M_{\odot}$
\cite{lp}, which cannot be explained with the Oppenheimer-Volkoff  model.

On the other hand, bosons can have an attractive $\frac{\lambda}{4\hbar
c}|\varphi|^4$ ($\lambda<0$)
self-interaction.  This is the case in particular for QCD axions with a mass
$m\sim 10^{-4}\, {\rm eV}/c^2$ and a
negative scattering length $a_s\sim -5.8\times 10^{-53}\, {\rm m}$
(corresponding to a 
self-interaction constant $\lambda\sim -7.39\times
10^{-49}$ and a decay constant $f=5.82\times 10^{10}\, {\rm GeV}$). Axions are
hypothetical pseudo-Nambu-Goldstone
bosons of the
Peccei-Quinn \cite{pq} phase transition associated with a $U(1)$ symmetry that
solves the strong charge parity (CP) problem of quantum chromodynamics (QCD).
They are described by a real SF ${\varphi}$ with a cosine self-interaction potential. Axions are possible DM candidates \cite{kc}. Their role in cosmology
has
been first
investigated in Refs. \cite{preskill,abbott,dine,davis}. The cosmological
evolution
of axions was considered by
Hogan and Rees \cite{rees} and Kolb
and Tkachev \cite{kt,kt2,kt3,kt2apj}. In
the early Universe, self-gravity between axions
can be neglected so they are governed by the (relativistic) sine-Gordon equation in an expanding background. Because
of the attractive self-interaction, axions can form miniclusters of mass $M_{\rm
axiton}\sim 10^{-12}\, M_{\odot}$ and size $R_{\rm
axiton}\sim 10^{9}\, {\rm m}$ called axion miniclusters
\cite{rees} or axitons \cite{kt2}. Tkachev \cite{tkachev,tkachevrt} took
self-gravity into account and considered the
possibility to form axion stars by Jeans instability.\footnote{He introduced
the names ``gravitationally bound axion condensates''
\cite{tkachev} and ``axionic Bose stars'' \cite{tkachevrt}, becoming later
``axion stars''.} He assumed a
repulsive $\frac{\lambda}{4\hbar
c}|\varphi|^4$  ($\lambda>0$)  self-interaction between axions and found a maximum
mass $M_{\rm max}^{\rm GR}\sim \sqrt{\lambda}M_p^3/m^2$ (see footnote 7).
However, when the self-interaction is attractive ($\lambda<0$), the equilibrium
state
of axion stars results from the balance between the gravitational attraction,
the attractive
self-interaction and the repulsive quantum potential. There is an equilibrium
state only below a maximum mass $M_{\rm max}^{\rm NR}=1.012\, \hbar/\sqrt{Gm
|a_s|}$ or $M_{\rm max}^{\rm NR}=5.073\,
M_P/\sqrt{|\lambda|}$ which was first identified by Chavanis
\cite{prd1,prd2,chavanisbook}.  This is the
maximum mass of
dilute axion stars \cite{mg16B}. Note that this maximum mass is a purely
nonrelativistic
result, contrary to the maximum mass of boson and fermion stars discussed
above which is due to general relativity. For dilute QCD axion
stars,
one finds $M_{\rm max}^{\rm NR}=6.46\times 10^{-14}\, M_{\odot}$ 
and a corresponding radius $R_{99}^*=227\, {\rm
km}$ which are of the order  of the mass and size of asteroids (by comparison
the Kaup mass is 
$M_{\rm max}^{\rm GR}=8.46\times
10^{-7}\, M_{\odot}$ and the Kaup radius is $R_{*}^{\rm GR}=1.19\times 10^{-2}\, {\rm
m}$).  This leads to
the
notion of ``axteroids''. For ULAs, we can have a much larger
maximum mass, of the order of galactic masses. Its precise
value depends, however, on the values of $m$ and $a_s$ which are not well
known. Taking $m=2.92\times 10^{-22}\, {\rm eV}/c^2$ and $a_s=-3.18\times
10^{-68}\, {\rm fm}$ (corresponding to 
$\lambda=-1.18\times 10^{-96}$ and $f=1.34\times 10^{17}\,
{\rm GeV}$) predicted in \cite{jeansapp}, we get $M_{\rm
max}^{\rm
NR}=5.10\times 10^{10}\,
M_{\odot}$ and
$R_{99}^*=1.09\, {\rm pc}$ (by comparison the Kaup mass is $M_{\rm max}^{\rm GR}=2.90\times
10^{11}\, M_{\odot}$ and the Kaup radius is $R_*^{\rm GR}=0.132\, {\rm pc}$). For
$M<M_{\rm
max}^{\rm NR}$ the
mass-radius
relation displays two branches of solutions \cite{prd1,prd2}. There are two
possible equilibrium
states for the same mass. The equilibrium states with $R>R_*$ are stable (energy
minima) and the equilibrium states with $R<R_*$ are unstable (energy maxima).
The stability of axion stars can be determined by
applying the Poincar\'e turning point argument \cite{prd1,prd2}.
The stable solutions define the branch of dilute axion stars
and the mass $M_{\rm max}^{\rm NR}$  represents their maximum mass
(the unstable solutions correspond to nongravitational
BECs) \cite{prd1,prd2}. For $M>M_{\rm max}^{\rm NR}$ there is no
equilibrium state and the axion star collapses \cite{bectcoll}.  
This leads to (i) a bosenova with the emission of relativistic
axions if we take special relativity into account \cite{tkachevprl}, (ii) a
black hole if the conditions where general relativity prevails are fulfilled
\cite{helfer,moss}, or (iii)  the formation of a dense axion star
\cite{braaten} 
if we take
into account  higher order terms in the expansion
of the self-interaction potential like, e.g., a repulsive $|\varphi|^6$
self-interaction
\cite{ebycollapse,phi6} that can stabilize the star against gravitational
collapse.\footnote{Visinelli
{\it et al.} \cite{visinelli} and Eby {\it et al.}
\cite{elssw} argue
that relativistic effects are crucial on the branch of dense axion stars while
self-gravity is negligible. As a
result, dense axion stars correspond to ``pseudobreathers'', ``oscillons''
or ``axitons'' which are
described by the sine-Gordon equation (they can be
viewed as the 3D version of usual 1D ``breathers'' \cite{breather}). For a real
SF, these
objects are known to
be unstable due to
particle number changing process such as the $3\rightarrow 1$ process and to 
decay via emission of relativistic axions. Visinelli
{\it et al.} \cite{visinelli} and Eby {\it et al.} \cite{elssw} argue that the
decay
timescale is much
shorter than any cosmological timescale so that dense axion stars are not
physically relevant. This
conclusion is, however, contested
by Braaten and Zhang \cite{braatenrevue} who argue that dense axion stars can
be long-lived. Note that dense axion stars described by a
{\it complex} SF (``axion boson stars'' \cite{guerra})  would be stable in the
relativistic regime because of charge conservation.} The
axion star can also fragment into
several
stable pieces (axion ``drops'') of mass $M'<M_{\rm max}$ \cite{davidson,cotner}
thereby preventing its complete collapse.  We refer to
\cite{braatenrevue,visinelliREVUE} and to the
introduction of \cite{tunnel} for recent reviews on axion stars and for an
exhaustive list of references.

\begin{table*}[t]
\centering
\begin{tabular}{|c|c|c|c|c|c|c|c|c|}
\hline
  & $M_{\rm max}$ & $R_*$
& ${\cal C}_{\rm max}$ & $R_*/R_S$ \\
\hline
White dwarfs \cite{chandra31} &  $3.10\, M_P^3/(\mu H)^2$ & $0$ &
$\infty$& $0$ \\
\hline
Neutron stars \cite{ov} &  $0.384\, M_P^3/m^2$ & $3.35\,
(\hbar^3/Gm^4c)^{1/2}$ & $0.114$ & $4.37$\\
\hline
Mini boson (mini soliton) stars \cite{kaup,rb} &  $0.633\, M_P^2/m$ & $6.03\,
\hbar/mc$ &
$0.105$ & $4.76$ \\
\hline
Massive boson stars \cite{colpi,chavharko} & 
$0.0612\,\sqrt{\lambda}M_P^3/m^2$ &
$0.383\,\sqrt{\lambda}(\hbar^3/Gm^4c)^{1/2}$
& $0.160$ & $3.13$ \\
\hline
Soliton star \cite{leecomment,tdlee,sss,fss}  &  $M_P^4/m^3$
& 
&  & \\
\hline
Oscillatons (real SF) \cite{ssreal,alcubierre} &  $0.606\, M_P^2/m$ & &
&\\
\hline
Dilute axion stars \cite{prd1,prd2} &  $5.07\,M_P/\sqrt{|\lambda|}$ & $1.10\,
\sqrt{|\lambda|}(\hbar^3/Gm^4c)^{1/2}$  & $4.61\, (m/M_P)^2/|\lambda|$ &\\
\hline
\end{tabular}
\label{table2}
\caption{Maximum mass
$M_{\rm max}$ of
different types of fermion and boson stars.  It is 
interesting to note that all
the scalings $M_P$, $M_P^2/m$, $M_P^3/m^2$ and  $M_P^4/m^3$ are represented. We
have also indicated the minimum radius $R_*$ and the maximum compactness ${\cal
C}_{\rm max}=GM_{\rm max}/R_* c^2$ of the star. The compactness of a
Schwarzschild
black hole is ${\cal C}_{\rm S}=1/2$. The Buchdahl inequality for a barotropic
relativistic star imposes ${\cal
C}\le 4/9=0.444$. The ratio between the star radius $R$ and the Schwarzschild
radius $R_{S}=2GM/c^2$ is $R/R_S=Rc^2/(2GM)=1/(2{\cal C})$. It is restricted by
the Buchdahl  inequality $R\ge (9/8)R_S$ so that a barotropic relativistic star
cannot be
a black hole.}
\end{table*}

These results have  been applied to DM halos made of ULAs. BECDM halos,
also called fuzzy dark matter (FDM) halos or SFDM halos,
typically have a core-halo structure made of a quantum core (soliton) in its
ground state surrounded by an approximately isothermal envelope
(atmosphere) arising from the
quantum interferences of excited states \cite{ggpp,moczSV,modeldm,wignerPH}.
This core-halo
structure results from a process of gravitational cooling \cite{seidel94} and
violent relaxation \cite{lb,csr}. The
results given in Refs. \cite{prd1,prd2}
describe
the ground state of a self-gravitating BEC so they apply either to
the ``minimum halo'' of mass $(M_h)_{\rm min}\sim 10^8\, M_{\odot}$ which is a
completely condensed
object without envelope (atmosphere) or to
the quantum core $M_c$ of large DM halos of mass $M_h\ge (M_h)_{\rm min}$. A
general expression of the core mass-halo
mass relation $M_c(M_h)$ for BECDM halos with an arbitrary
self-interaction has been obtained in \cite{modeldm,mcmh,mcmhbh,jeansapp,prsmv}
from thermodynamical considerations (see also
\cite{ch3,veltmaat,mocz,egg,bbbs} for other
justifications). The
mass $M_c$ of the quantum core increases with the halo mass $M_h$.  For
noninteracting bosons and for bosons with a repulsive self-interaction, it
can be shown that the core
mass $M_c$ is always much smaller than the maximum mass $M_{\rm max}^{\rm GR}$ set by
general relativity so the quantum core cannot collapse towards a
supermassive black hole (SMBH) \cite{mcmhbh,jeansapp,prsmv}. For bosons with an
attractive self-interaction,  the core mass $M_c$ could reach the maximum mass
$M_{\rm max}^{\rm NR}$ of Ref. \cite{prd1} in sufficiently large DM halos, and
collapse \cite{mcmhbh,jeansapp,prsmv}. The outcome of the collapse in that case
would be a dense axion
``star'' (soliton),\footnote{By an abuse of langage, we will sometimes use the
term axion ``star'' to designate the quantum core of axionic DM halos.} a black
hole, a bosenova or axion drops
\cite{braaten,davidson,cotner,bectcoll,ebycollapse,tkachevprl,helfer,phi6,
visinelli,moss,braatenrevue,elssw}.
 The conditions for this
collapse require, however, stronger
self-interactions ($f<10^{15}\, {\rm GeV}$) that those  ($10^{16}\,
{\rm
GeV}\le f\le 10^{18}\, {\rm
GeV}$) commonly
allowed by particle
physics and cosmology (see \cite{jeansapp,mg16B} for more details). Furthermore,
since $f\ll M_P$ the collapse leads to a dense axion star or to a bosenova, not
to
a SMBH. These
results can also find applications in the context of ``inflaton
clusters'' and ``inflaton
stars'' that could form in the very early universe \cite{mhe,ne,ene,esne}. 
There is a complete analogy between inflaton clusters and 
BECDM halos so that most of the results obtained for BECDM halos 
(core-halo solution, core mass-radius relation, core mass-halo mass relation...) can be
exported to the context of inflaton clusters. In particular, if the SF has an attractive
self-interaction, 
``inflaton stars'' can be stable only below the maximum mass $M_{\rm max}^{\rm
NR}$ of Ref.
\cite{prd1}. Padilla {\it et al.} \cite{padilla} argued that,
above that
critical mass, the inflaton star collapses
and forms a black hole. That could be a new mechanism to form primordial black
holes (PBHs) during the phase of reheating following inflation. However,
according to
the results of Refs.
\cite{tkachevprl,braaten,phi6}, it is also possible that the
collapse of inflaton stars leads to a bosenova or a dense inflaton star rather
than a black hole. Primordial black hole of mass $\sim 1\, {\rm g}$ can 
form only in sufficiently massive inflaton clusters
of mass $M_h\sim 10^{14}\, {\rm g}$ and they rapidly evaporate on a timescale
$10^{-30}\, {\rm s}$. We suggest that bosenova and dense inflaton stars may
occur in
less massive inflaton clusters if the bosons have a sufficiently attractive
self-interaction.
In addition to the quantum core of DM halos \cite{mcmh,mcmhbh,jeansapp,prsmv}
and inflaton clusters \cite{padilla}, other applications
of the
maximum mass of self-gravitating BECs with an attractive
self-interaction (like dilute axion stars) \cite{prd1,prd2} have
been discussed in Refs. \cite{dkr,schia,cdlmn,gpw,ebyapp,lcl,jain,ja}.

In this paper, we provide a simple approximate analytical expression of the
maximum mass of relativistic self-gravitating BECs with an arbitrary
$|\varphi|^4$ 
self-interaction.
This expression interpolates between the general relativistic 
maximum mass of noninteracting bosons stars, the general relativistic 
maximum mass of bosons stars with a repulsive self-interaction in the
TF limit, and the nonrelativistic maximum mass of dilute
axion stars with an attractive self-interaction. It therefore connects the
different 
expressions obtained in the literature reviewed above (see Table \ref{table2}
for a summary).
We
obtain the general structure  of our formula from simple considerations
and determine the numerical
coefficients in order to recover the exact asymptotic expressions of the maximum
mass in particular limits. We also show that the predictions from the Gaussian
ansatz are in good agreement with the exact values. As a result, our formula
should provide a relevant
approximation of the maximum mass of relativistic boson stars for any value
(positive or negative) of the self-interaction parameter.

The paper is organized as follows. In Sec. \ref{sec_sgbec}, we
recall the basic equations describing nonrelativistic self-gravitating BECs. In
Sec. \ref{sec_exact}, we discuss the exact mass-radius relation of
nonrelativistic self-gravitating BECs obtained by solving the GPP equations
numerically. In Sec. \ref{sec_pa}, we show that we can obtain an analytical
approximation of the mass-radius relation from an $f$-ansatz. We
determine the coefficients of this relation so as to recover the exact results
in particular limits. In Sec. \ref{sec_er}, we recall
the expression of the exact maximum mass of general relativistic BECs  obtained
by solving the KGE equations numerically. In Secs. \ref{sec_is} and 
\ref{sec_rq}, we obtain an analytical approximation of the maximum mass of 
self-gravitating BECs as a function of the scattering length $a_s$ of the
bosons for a
repulsive or an attractive self-interaction. We determine the coefficients of
this relation so as to recover the exact results in the noninteracting limit
($a_s=0$), in the TF limit (for
$a_s>0$), and in the nonrelativistic  limit (for $a_s<0$). In Sec.
\ref{sec_summ},
we summarize the main results of
our study. In Sec. \ref{sec_bhb}, we re-express our results in terms of the
dimensionless self-interaction constant $\lambda$ or in terms of the axion decay
constant $f$
and we discuss whether the collapse of the BEC star above the maximum mass leads
to a black hole or a bosenova.  In Sec. \ref{sec_cmhm}, we
apply our
results to DM halos and inflaton clusters and discuss whether the solitonic
core of these systems can become unstable in realistic situations. We conclude
in Sec.
\ref{sec_conclusion}. The
Appendices provide complements to our main results. In particular, we
develop interesting
analogies between self-gravitating BECs and models of extended elementary
particles.

\section{Nonrelativistic self-gravitating BECs}
\label{sec_sgbec}

In this section, we recall basic results applying to nonrelativistic
self-gravitating BECs at $T=0$ described by the GPP equations (see Refs.
\cite{prd1,prd2,ggpp} for more details).

\subsection{Gross-Pitaevskii-Poisson equations}
\label{sec_gpp}

We assume that DM is made of bosons (like the axion) in the form of BECs at
$T=0$.  We use a nonrelativistic approach based on Newtonian gravity.  The
evolution
of the wave function $\psi({\bf r},t)$ of a self-gravitating BEC is governed by
the
GPP equations
\begin{eqnarray}
\label{gpp1}
i\hbar \frac{\partial\psi}{\partial
t}=-\frac{\hbar^2}{2m}\Delta\psi+\frac{4\pi a_s\hbar^2}{m^2}|\psi|^2\psi+m\Phi\psi,
\end{eqnarray}
\begin{equation}
\label{gpp2}
\Delta\Phi=4\pi G |\psi|^2,
\end{equation}
where $\Phi({\bf r},t)$ is the gravitational potential and $m$ is the mass of
the
bosons.\footnote{The derivation
of the GPP equations (\ref{gpp1}) and (\ref{gpp2}) from the  KGE
equations is discussed in Appendices \ref{sec_ra} and \ref{sec_rrsf} and
references therein.} The mass density of the BEC is $\rho({\bf
r},t)=|\psi|^2$. The first term in Eq. (\ref{gpp1}) is the
kinetic term which accounts for the Heisenberg uncertainty
principle. The second term takes into account the self-interaction of the
bosons via a $|\psi|^4$ potential [see Eq. (\ref{nrl1})]
\begin{equation}
\label{gpp2b}
V(|\psi|^2)=\frac{2\pi a_s \hbar^2}{m^3}|\psi|^4,
\end{equation}
where  $a_s$ is the scattering
length of the bosons. The interaction between the bosons is repulsive when
$a_s>0$ and
attractive when $a_s<0$. The third term accounts for the
self-gravity of the BEC.

The GPP equations conserve the mass
\begin{eqnarray}
\label{gpp6}
M=\int |\psi|^2\, d{\bf r}
\end{eqnarray}
and the energy
\begin{equation}
\label{gpp7}
E_{\rm tot}=\frac{\hbar^2}{2m^2}\int |\nabla\psi|^2\, d{\bf r}+\frac{2\pi a_s\hbar^2}{m^3}\int
|\psi|^{4} \, d{\bf r}+\frac{1}{2}\int |\psi|^2 \Phi\, d{\bf r},
\end{equation}
which is the sum of the kinetic energy $\Theta$, the internal energy
$U=\int V(|\psi|^2)\, d{\bf r}$ and
the gravitational energy $W$ (i.e. $E_{\rm tot}=\Theta+U+W$).

\subsection{Madelung transformation}
\label{sec_mad}

Writing the wave function as
\begin{equation}
\label{mad1}
\psi({\bf r},t)=\sqrt{{\rho({\bf r},t)}} e^{iS({\bf r},t)/\hbar},
\end{equation}
where $\rho({\bf r},t)$ is the mass density  and $S({\bf r},t)$ is the action,
and making the Madelung \cite{madelung} transformation
\begin{equation}
\label{mad2}
\rho({\bf
r},t)=|\psi|^2\qquad {\rm and} \qquad {\bf u}=\frac{\nabla S}{m},
\end{equation}
where ${\bf u}({\bf r},t)$ is the velocity field, the GPP
equations (\ref{gpp1}) and  (\ref{gpp2}) can be written under the form of
hydrodynamic equations
\begin{equation}
\label{mad3}
\frac{\partial\rho}{\partial t}+\nabla\cdot (\rho {\bf u})=0,
\end{equation}
\begin{equation}
\label{mad3b}
\frac{\partial S}{\partial t}+\frac{(\nabla S)^2}{2m}+Q+\frac{4\pi
a_s\hbar^2}{m^2}\rho+m
\Phi=0,
\end{equation}
\begin{equation}
\label{mad4}
\frac{\partial {\bf u}}{\partial t}+({\bf u}\cdot
\nabla){\bf
u}=-\frac{1}{m}\nabla
Q-\frac{1}{\rho}\nabla P-\nabla\Phi,
\end{equation}
\begin{equation}
\label{mad5}
\Delta\Phi=4\pi G \rho,
\end{equation}
where
\begin{equation}
\label{mad6}
Q=-\frac{\hbar^2}{2m}\frac{\Delta
\sqrt{\rho}}{\sqrt{\rho}}=-\frac{\hbar^2}{4m}\left\lbrack
\frac{\Delta\rho}{\rho}-\frac{1}{2}\frac{(\nabla\rho)^2}{\rho^2}\right\rbrack
\end{equation}
is the quantum potential taking into account the Heisenberg uncertainty
principle, 
\begin{eqnarray}
\label{mad10q}
h(\rho)=V'(\rho)=\frac{4\pi a_s\hbar^2}{m^3}\rho
\end{eqnarray}
is the enthalpy
and
\begin{eqnarray}
\label{mad10}
P=\rho V'(\rho)-V(\rho)=\rho^2 \left\lbrack
\frac{V(\rho)}{\rho}\right\rbrack'=\frac{2\pi
a_s\hbar^2}{m^3}\rho^{2}
\end{eqnarray}
is the pressure arising from the self-interaction of the
bosons $V(\rho)=2\pi a_s\hbar^2\rho^{2}/m^3$ (see Appendix \ref{sec_ra}). This
quadratic equation of state is a
particular polytropic equation of
state $P=K\rho^{\gamma}$ of index
$\gamma=2$ and polytropic constant $K={2\pi a_s\hbar^2}/{m^3}$. 
The hydrodynamic equations
(\ref{mad3})-(\ref{mad5}) are called the
quantum Euler-Poisson equations. Equation (\ref{mad3}) is
the
continuity equation, Eq. (\ref{mad3b}) is the
quantum Hamilton-Jacobi (or Bernoulli) equation, Eq. (\ref{mad4}) is the
quantum Euler equation, and Eq. (\ref{mad5}) is the Poisson equation. In the TF
limit  where the
quantum potential $Q$
can be neglected (formally $\hbar=0$),\footnote{We note that
$\hbar$ appears in the quantum potential $Q$ and in the self-interaction
constant $g=4\pi a_s\hbar^2/m^3$. The TF limit (corresponding to
$\hbar\rightarrow 0$
with fixed $4\pi a_s\hbar^2/m^3$)
amounts to neglecting $Q$ but not $g$.} they
become
equivalent to the classical
Euler-Poisson equations for a barotropic gas
\cite{bt}.

The quantum Euler equations conserve the mass
\begin{eqnarray}
\label{mad11}
M=\int \rho\, d{\bf r}
\end{eqnarray}
and the energy 
\begin{eqnarray}
\label{mad12}
E_{\rm tot}=\Theta_c+\Theta_Q+U+W,
\end{eqnarray}
which is the sum of the classical kinetic energy
\begin{eqnarray}
\label{mad12a}
\Theta_c=\int \rho\frac{{\bf u}^2}{2}\, d{\bf r},
\end{eqnarray}
the quantum kinetic energy
\begin{eqnarray}
\label{mad12b}
\Theta_Q=\frac{\hbar^2}{8m^2}\int \frac{(\nabla\rho)^2}{\rho}\, d{\bf
r}=\frac{1}{m}\int\rho Q\, d{\bf r},
\end{eqnarray}
the internal energy
\begin{equation}
\label{mad12c}
U=\int V(\rho)\, d{\bf r}=\int
\rho\int^{\rho}\frac{P(\rho')}{{\rho'}^2}\,
d\rho'\, d{\bf r}=\int \frac{2\pi a_s
\hbar^2}{m^3}\rho^2\, d{\bf r},
\end{equation}
and the gravitational energy
\begin{eqnarray}
\label{mad12d}
W=\frac{1}{2}\int \rho \Phi\, d{\bf r}.
\end{eqnarray}
At equilibrium, the classical (macroscopic) kinetic energy vanishes and we get
\begin{eqnarray}
\label{mad12e}
E_{\rm tot}=\Theta_Q+U+W.
\end{eqnarray}

\subsection{Equilibrium states}
\label{sec_eqs}

A stationary solution of GPP equations is of the form
\begin{eqnarray}
\label{dm1}
\psi({\bf r},t)=\phi({\bf r})e^{-iEt/\hbar},
\end{eqnarray}
where $\phi({\bf r})=\sqrt{\rho({\bf r})}$ and $E$ are real. 
Substituting Eq. (\ref{dm1}) into Eqs.  (\ref{gpp1}) and  (\ref{gpp2}), we
obtain the eigenvalue problem
\begin{eqnarray}
\label{dm2}
-\frac{\hbar^2}{2m}\Delta\phi
+\frac{4\pi a_s\hbar^2}{m^2}\phi^3
+m\Phi\phi=E\phi,
\end{eqnarray}
\begin{equation}
\label{dm3}
\Delta\phi=4\pi G \phi^2,
\end{equation}
determining the eigenfunctions $\phi_n({\bf r})$ and the eigenvalues $E_n$. For the
fundamental mode (the one with the lowest energy) the  wavefunction $\phi(r)$ is
spherically symmetric and has no node so that the density profile decreases
monotonically with the radial distance.  Dividing Eq. (\ref{dm2}) by $\phi$ and
using $\rho=\phi^2$, we obtain the identity
\begin{eqnarray}
\label{dm4}
Q+\frac{4\pi a_s\hbar^2}{m^2}\rho+m\Phi=E,
\end{eqnarray}
which can also be obtained from the quantum Hamilton-Jacobi (or Bernoulli)
equation (\ref{mad3b}) by setting $S=-Et$.

Equivalent results can be obtained from the hydrodynamic equations 
(\ref{mad3})-(\ref{mad5}). 
Indeed, the condition of quantum
hydrostatic equilibrium,  corresponding to a
steady state of the quantum Euler equation (\ref{mad4}), reads
\begin{eqnarray}
\label{dm5}
\frac{\rho}{m}\nabla
Q+\nabla P+\rho\nabla\Phi={\bf 0}.
\end{eqnarray}
Dividing Eq. (\ref{dm5}) by $\rho$ and integrating the resulting expression with
the help of Eq. (\ref{mad10}), we recover Eq. (\ref{dm4}) where $E$ appears as a
constant of integration. On the other hand, combining Eq. (\ref{dm5}) with the
Poisson equation (\ref{mad5}), we obtain the
fundamental differential
equation of quantum
hydrostatic equilibrium
\begin{eqnarray}
\label{dm7}
\frac{\hbar^2}{
2m^2}\Delta
\left (\frac{\Delta\sqrt{\rho}}{\sqrt{\rho}}\right
)-\frac{4\pi a_s\hbar^2}{m^3}\Delta\rho=4\pi
G\rho.
\end{eqnarray}
This equation describes the balance between the quantum potential taking into
account the Heisenberg uncertainty principle, the pressure due to the
self-interaction of the bosons,  and the self-gravity.

These results can also be obtained from an energy principle. Indeed, one can
show (see Appendix B of \cite{jeansapp}) that (i) an equilibrium state of the
GPP equations is an extremum of energy $E_{\rm tot}$ at fixed mass $M$ and that
(ii) an equilibrium state is stable if, and only if, it is a
minimum of energy at fixed mass. We are led therefore to considering the
minimization problem
\begin{eqnarray}
\label{dm8}
\min\quad \lbrace E_{\rm tot}\quad | \quad M\quad {\rm fixed}\rbrace.
\end{eqnarray}
Writing the variational problem for the first variations (extremization
problem) as
\begin{eqnarray}
\label{dm9}
\delta E_{\rm tot}-\frac{\mu}{m}\delta M=0,
\end{eqnarray}
where $\mu$ (global chemical potential) is a Lagrange multiplier taking into
account the mass constraint, we obtain ${\bf u}={\bf 0}$ and
\begin{eqnarray}
\label{dm10}
Q+\frac{4\pi a_s\hbar^2}{m^2}\rho+m\Phi=\mu.
\end{eqnarray}
This relation is equivalent to Eq. (\ref{dm4}) provided that we make the
identification
$E=\mu$.
Therefore, the eigenenergy $E$ coincides with the global chemical
potential $\mu$. Equation
(\ref{dm10}) is also equivalent to the condition of quantum hydrostatic
equilibrium (\ref{dm5}). Therefore,  an extremum of energy at fixed mass is an
equilibrium state of the
GPP equations. Furthermore,  among all
possible equilibria, only minima of energy
at
fixed mass are dynamically stable with respect to the GPP equations (maxima or
saddle points are linearly unstable). The
stability of an equilibrium state can be established by studying the sign
of the second variations of energy or by solving an equation of pulsations.
These methods are equivalent and lead to a complicated eigenvalue problem (see
Appendix B of \cite{jeansapp}). The
stability of an equilibrium state can also be directly established, without
having to solve
an eigenvalue problem, by plotting the
series
of equilibria and using the Poincar\'e
\cite{poincare,katzpoincare} turning point criterion applied to the curve
$\mu(M)$, the Wheeler
\cite{htww} theorem applied to the curve  $M(R)$, or the Whitney theorem
\cite{whitney}  applied to the curve $E_{\rm tot}(M)$ (see Refs.
\cite{prd1,prd2,phi6} for
a specific application of these methods to the case of axion
stars).

\section{Exact  mass-radius relation of nonrelativistic self-gravitating BECs}
\label{sec_exact}

The fundamental equation (\ref{dm7}) of quantum hydrostatic equilibrium for a
self-gravitating BEC has been solved  numerically (exactly) in our previous
paper \cite{prd2} for an arbitrary $|\psi|^4$
self-interaction (repulsive or attractive). The nodeless solution describes a 
compact gravitational quantum
object (soliton/BEC) in its ground state. From
this solution we have determined the exact mass-radius relation of
nonrelativistic BEC
stars (see Figs. \ref{M-R-chi-pos-part1} and \ref{M-R-chi-neg-part1} below).
Here, we recall some exact results obtained in particular limits
that will be useful in the following.

\subsection{Noninteracting bosons}
\label{sec_nib}

For noninteracting bosons ($a_s=0$), the equation of quantum hydrostatic equilibrium (\ref{dm7}) reduces to
\begin{eqnarray}
\frac{\hbar^2}{
2m^2}\Delta
\left (\frac{\Delta\sqrt{\rho}}{\sqrt{\rho}}\right
)=4\pi
G\rho.
\end{eqnarray}
It can be solved numerically to obtain the density profile. The mass-radius
relation is 
given by \cite{membrado,prd2}
\begin{eqnarray}
\label{dm12}
M=9.95\frac{\hbar^2}{Gm^2R_{99}},
\end{eqnarray}
where $R_{99}$ represents the radius containing $99\%$ of the mass (the density
profile extends to infinity so it has not a compact support). The mass decreases
as the radius increases. The equilibrium states are all
stable.

{\it Remark:} The scaling of the mass-radius relation (\ref{dm12}) can be
understood by writing that the radius $R$ of the BEC is of the order of the de 
Broglie wavelength of the bosons $\lambda_{\rm dB}=\hbar/(mv)$ constructed with
the virial velocity $v\sim (GM/R)^{1/2}$.

\subsection{Bosons with a repulsive self-interaction}
\label{sec_rep}

\begin{figure}[!h]
\begin{center}
\includegraphics[clip,scale=0.3]{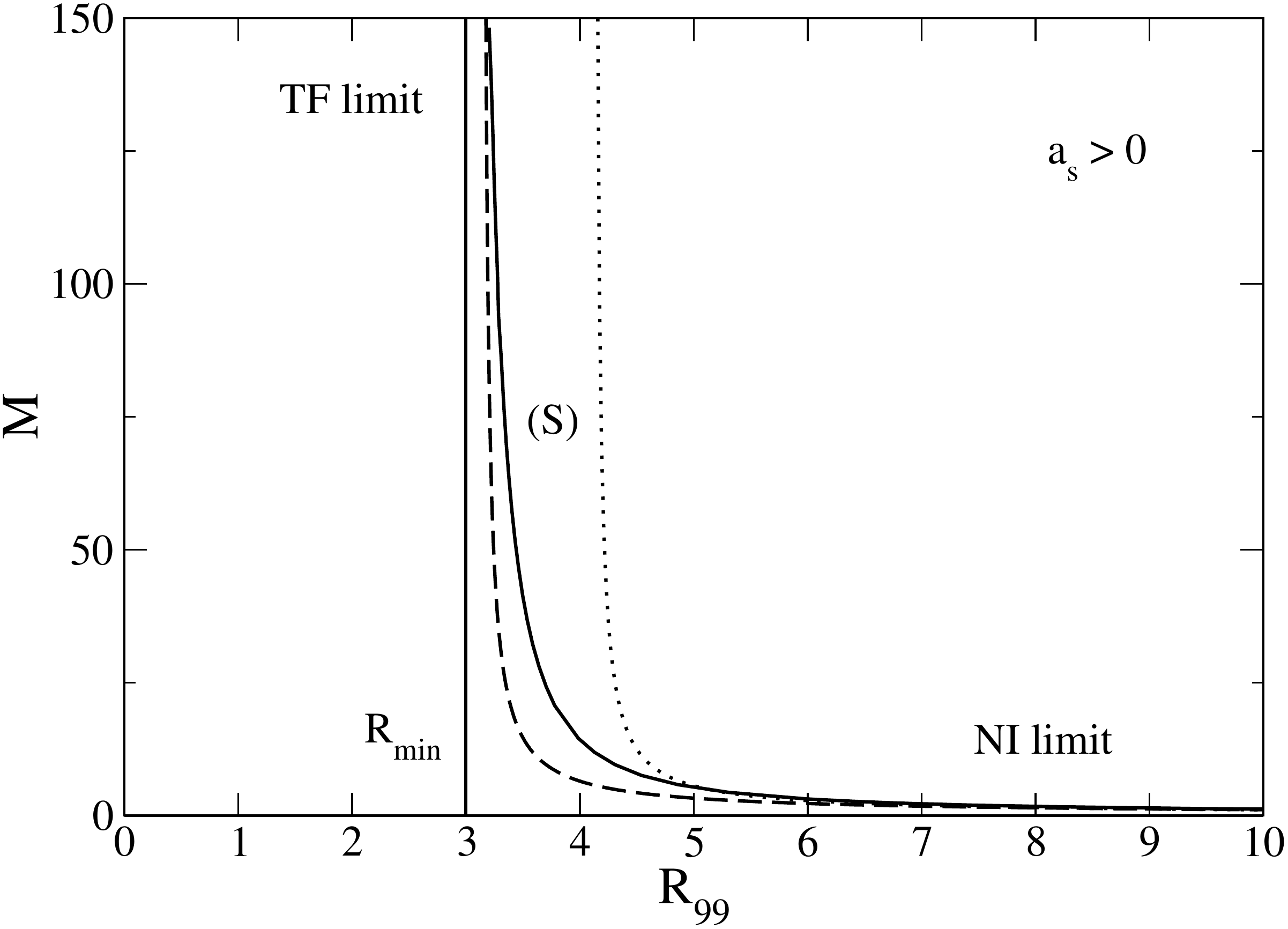}
\caption{Mass-radius relation of self-gravitating BECs with $a_s>0$ (full line:
exact result \cite{prd2}; dotted line: Gaussian ansatz \cite{prd1}; dashed line:
fit
from Eq. (\ref{pa1}) with $a=9.946$ and $b=\pi$). The mass is normalized by
$M_a=\hbar/\sqrt{Gma_s}$ and the radius by $R_a=(a_s\hbar^2/Gm^3)^{1/2}$.}
\label{M-R-chi-pos-part1}
\end{center}
\end{figure}

For bosons with a repulsive self-interaction ($a_s>0$), 
the exact mass-radius relation is represented in Fig. \ref{M-R-chi-pos-part1}.
The mass decreases as the radius increases. In the TF limit $M\gg M_a=
\hbar/\sqrt{Gma_s}$ where 
the quantum potential can be neglected, the system is equivalent to a polytrope
of index $\gamma=2$.  The equation of quantum hydrostatic equilibrium
(\ref{dm7}) reduces to
\begin{eqnarray}
\label{dm7tf}
\Delta\rho+\frac{Gm^3}{a_s\hbar^2}\rho=0.
\end{eqnarray}
This equation is equivalent to the Lane-Emden equation 
of index $n=1$  \cite{chandrabook}. It can
be solved analytically leading to a density profile of the form\footnote{This
analytical solution was first given by Ritter \cite{ritter} in the context of
self-gravitating polytropic spheres. It was previously used by Laplace
\cite{laplace} to model the earth interior (see footnote 10 in
\cite{jeansapp}).}
\begin{eqnarray}
\label{mg18}
\rho(r)=\frac{\rho_0 R_{\rm TF}}{\pi r}\sin \left (\frac{\pi r}{R_{\rm
TF}}\right
).
\end{eqnarray}
In the TF limit, the equilibrium states have a unique radius given
by \cite{tkachev,maps,leekoh,goodman,arbey,bohmer,prd1}
\begin{eqnarray}
\label{dm15}
R_{\rm TF}=\pi\left (\frac{a_s\hbar^2}{Gm^3}\right )^{1/2},
\end{eqnarray}
which is independent of their mass $M$. This is the minimum radius of
self-gravitating BECs with a repulsive self-interaction. In the noninteracting
(NI) limit $M\ll M_a=\hbar/\sqrt{Gma_s}$ and
$R\gg R_{\rm TF}$, we recover Eq. (\ref{dm12}).

{\it Remark:} Let us recall some basic results valid in the TF limit \cite{prd1} that we shall need later.  The central density of the self-gravitating BEC is determined by its mass according to
\begin{eqnarray}
\label{rhoM}
\rho_0=\frac{\pi M}{4R_{\rm TF}^3}.
\end{eqnarray}
Its total energy is
\begin{eqnarray}
\label{Etot}
E_{\rm tot}=-\frac{GM^2}{2R_{\rm TF}}.
\end{eqnarray}
When slightly displaced from its equilibrium configuration, the BEC oscillates 
with a pulsation which is of the order of the inverse dynamical time (see
\cite{prd1} for more precise results)
\begin{eqnarray}
\label{tdyn}
t_D\sim \frac{1}{\sqrt{G\rho_0}}\sim \left (\frac{R_{\rm TF}^3}{GM}\right )^{1/2}.
\end{eqnarray}

\subsection{Bosons with an attractive self-interaction}
\label{sec_att}

\begin{figure}[!h]
\begin{center}
\includegraphics[clip,scale=0.3]{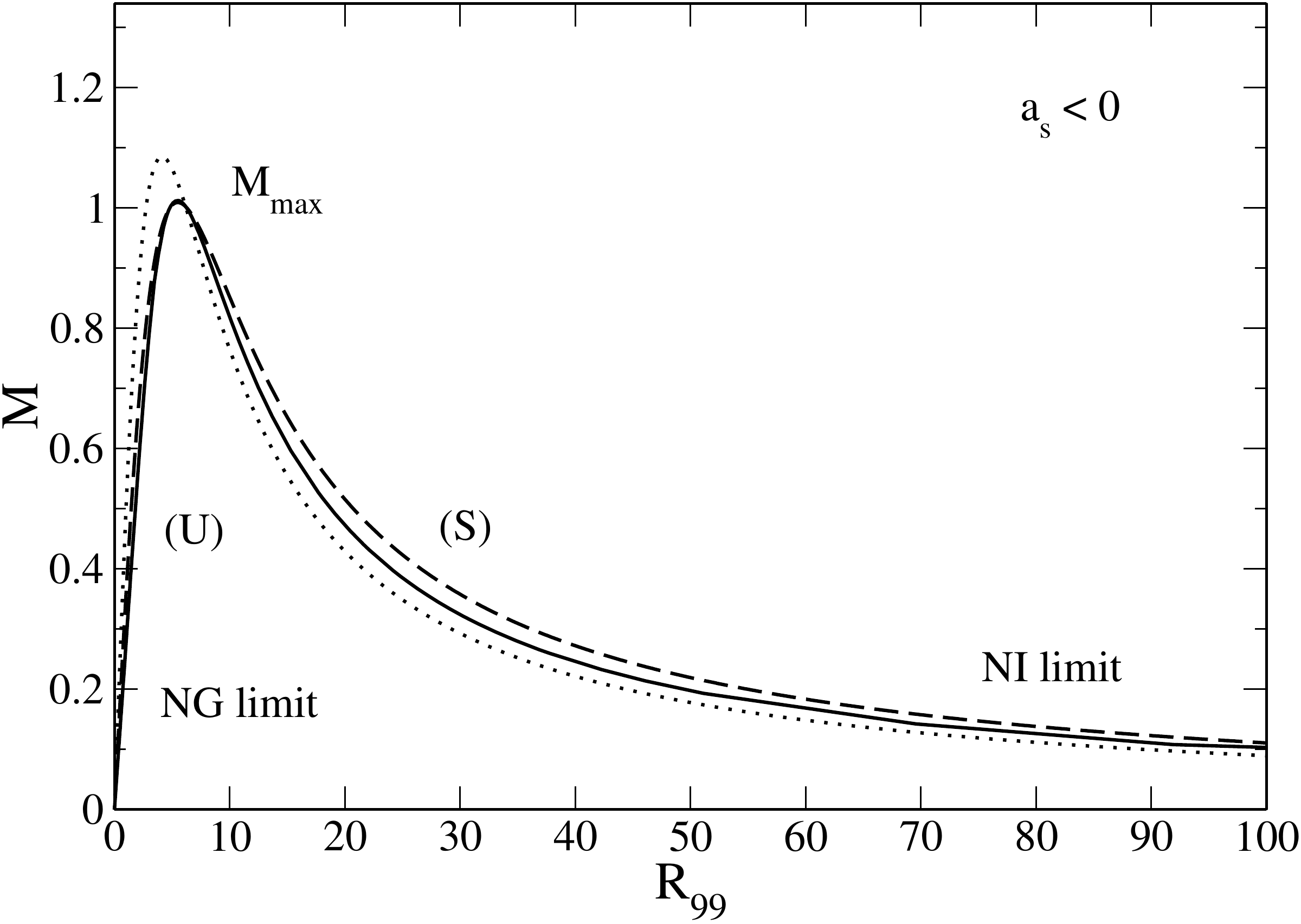}
\caption{Mass-radius relation of self-gravitating BECs with $a_s<0$ (full line:
exact result \cite{prd2}; dotted line: Gaussian ansatz \cite{prd1}; dashed line:
fit
from Eq. (\ref{pa1}) with $a=11.1$ and $b=5.5$). The mass is normalized by
$M_a=\hbar/\sqrt{Gm|a_s|}$ and the radius by
$R_a=(|a_s|\hbar^2/Gm^3)^{1/2}$.}
\label{M-R-chi-neg-part1}
\end{center}
\end{figure}

For bosons with an attractive self-interaction ($a_s<0$), 
the exact mass-radius relation is represented in
Fig. \ref{M-R-chi-neg-part1}. The mass increases as the radius increases,
reaches a maximum value \cite{prd1,prd2} 
\begin{eqnarray}
\label{dm16}
M_{\rm max}^{\rm NR}=1.012\, \frac{\hbar}{\sqrt{Gm|a_s|}}
\end{eqnarray}
at
\begin{eqnarray}
\label{dm17}
R_{99}^*=5.5\, \left (\frac{|a_s|\hbar^2}{Gm^3}\right )^{1/2}.
\end{eqnarray}
and decreases. $M_{\rm max}^{\rm NR}$ is the maximum mass of dilute axion stars \cite{mg16B}.
There is no equilibrium state with $M>M_{\rm max}^{\rm NR}$. In
that case, the BEC is
expected to collapse \cite{bectcoll}. The outcome of the collapse (dense
axion star, black hole, bosenova, axion drops...) is discussed in
\cite{braaten,davidson,cotner,bectcoll,ebycollapse,tkachevprl,helfer,phi6,
visinelli,
moss}. For $M<M_{\rm max}$ there are two possible equilibrium states with the same mass. The equilibrium
states with $R>R_{99}^*$ are stable and 
the equilibrium states with $R<R_{99}^*$ are
unstable. This can be shown by using the Poincar\'e criterion, the
Wheeler theorem, the Whitney theorem, or by investigating the sign of the
squared pulsation
\cite{prd1,prd2,phi6}. We note that the maximum mass is connected to the minimum stable radius by
\begin{eqnarray}
\label{dm17r}
M_{\rm max}^{\rm NR}=5.57\, \frac{\hbar^2}{Gm^2R_{99}^*},
\end{eqnarray}
which presents the same scaling as Eq. (\ref{dm12}).

In the nongravitational (NG) limit $M\ll M_{\rm max}^{\rm NR}$ and $R\ll
R_{99}^*$, the  equation of quantum hydrostatic equilibrium (\ref{dm7})
reduces to 
\begin{eqnarray}
-\frac{\hbar^2}{
2m}\frac{\Delta\sqrt{\rho}}{\sqrt{\rho}}+\frac{4\pi a_s\hbar^2}{m^2}\rho=E.
\end{eqnarray}
This equation is equivalent to the ordinary (nongravitational) GP equation with an attractive self-interaction. It can be solved numerically to obtain the density profile. The
mass-radius
relation is given by (see, e.g., \cite{prd2})
\begin{eqnarray}
\label{dm19}
M=0.275\, \frac{m R_{99}}{|a_s|}.
\end{eqnarray}
These equilibrium states are unstable. In the NI limit  $M\ll M_{\rm max}^{\rm NR}$ and $R\gg R_{99}^*$, we
recover Eq. (\ref{dm12}). These equilibrium states are stable.

\section{Approximate  mass-radius relation of nonrelativistic self-gravitating BECs from
the $f$-ansatz}
\label{sec_pa}

In Ref. \cite{prd1}, using a Gaussian ansatz for the wave function, we have
obtained an approximate analytical expression of the mass-radius relation of
self-gravitating BECs. In Ref. \cite{jeansapp} we have shown that the form of
this relation is independent of the precise shape of the wave function. The
shape of the wave function just determines the coefficients entering in this
relation. We have then proposed to determine these coefficients by matching the
asymptotic expressions of the analytical (approximate) mass-radius relation with
the asymptotic expressions of the exact (numerical) mass-radius relation
obtained in \cite{prd2}. We briefly recall this procedure below.

\subsection{$f$-ansatz}
\label{sec_sim}

Stable BEC stars (axion stars or the quantum core of BECDM halos)
correspond to minima of energy $E_{\rm tot}$ at fixed mass $M$. We can obtain an
approximate analytical form of the mass-radius relation by making an ansatz for
the wave function.\footnote{Here, we restrict ourselves to the equilibrium
state, so we just need to make an ansatz for the density profile. See Sec. 8
of \cite{ggpp} for a more general study.}  To be as
general as possible, we consider an ansatz of the form (that we call
$f$-ansatz)
\begin{eqnarray}
\rho({\bf r},t)=\frac{M}{R(t)^3}f\left\lbrack \frac{\bf r}{R(t)}\right \rbrack,
\label{mra1}
\end{eqnarray}
where $f({\bf x})$ is an arbitrary (physical) function. We impose $\int f({\bf
x})\,
d{\bf x}=1$ to satisfy the normalization condition (or the conservation of
mass). On the other
hand, the gravitational potential can be determined from the Poisson equation
(\ref{mad5}). Using Eq. (\ref{mra1}) we obtain
\begin{eqnarray}
\Phi({\bf r},t)=\frac{GM}{R(t)}g\left\lbrack \frac{\bf r}{R(t)}\right \rbrack,
\label{mra4}
\end{eqnarray}
where $g({\bf x})$ is the solution of
\begin{eqnarray}
\Delta g=4\pi f({\bf x}).
\label{mra5}
\end{eqnarray}
We can now use the ansatz (\ref{mra1})-(\ref{mra5}) to determine 
the different functionals that appear in the energy from Eq. (\ref{mad12}). We
find
\begin{eqnarray}
\Theta_Q=\sigma\frac{\hbar^2M}{m^2R^2}\quad {\rm with}\quad
\sigma=\frac{1}{8}\int \frac{(\nabla f)^2}{f}\, d{\bf x},
\label{mra7}
\end{eqnarray} 
\begin{eqnarray}
U=\zeta\frac{2\pi
a_s\hbar^2M^2}{m^3R^3}\quad {\rm
with}\quad \zeta=\int f^{2}({\bf x})\, d{\bf x},
\label{mra9}
\end{eqnarray}
\begin{equation}
W=-\nu \frac{GM^2}{R}\quad {\rm with}\quad \nu=-\frac{1}{2}\int f({\bf x})g({\bf
x})\, d{\bf x}.
\label{mra8}
\end{equation}
If we use a Gaussian ansatz $f({\bf
x})=\frac{1}{\pi^{3/2}}e^{-x^2}$, the values 
of the coefficients are  $\sigma_{\rm G}=3/4$, $\zeta_{\rm
G}=1/(2\pi)^{3/2}$, and  $\nu_{\rm G}=1/\sqrt{2\pi}$
\cite{prd1}.

With the ansatz from Eq. (\ref{mra1}) the total energy can be
written as
\begin{eqnarray}
E_{\rm tot}(R)=\sigma\frac{\hbar^2M}{m^2R^2}-\nu
\frac{GM^2}{R}+\zeta\frac{2\pi
a_s\hbar^2M^2}{m^3R^3}.
\label{mra11}
\end{eqnarray}
At equilibrium, the condition $E_{\rm min}'(R)=0$ (extremum of energy) gives the
mass-radius relation\footnote{As shown in \cite{prd1,ggpp,bectcoll} the
mass-radius
relation can also be obtained from the equilibrium virial theorem or from the
Lagrangian formalism.}
\begin{eqnarray}
-2\sigma\frac{\hbar^2M}{m^2R^3}+\nu \frac{GM^2}{R^2}-6\pi\zeta
\frac{a_s\hbar^2M^2}{m^3R^{4}}=0
\label{pul3}
\end{eqnarray}
or, equivalently,
\begin{equation}
M=\frac{\frac{2\sigma}{\nu}\frac{\hbar^2}{Gm^2R}}{1-\frac{6\pi\zeta}{\nu}\frac{
a_s\hbar^2 } { Gm^3R^2 } }.
\label{mrga}
\end{equation}
The BEC is stable provided that $E_{\rm tot}''(R)>0$ which corresponds  to the
requirement that the equilibrium state is a minimum of
energy  or, equivalently, that the squared pulsation is positive \cite{prd1}.

We see that the form of the analytical mass-radius 
relation is independent of the ansatz. Indeed, it is always given by
\begin{equation}
\label{pa1}
M=\frac{a\frac{\hbar^2}{Gm^2R}}{1-b^2\frac{a_s\hbar^2}{Gm^3R^2}},
\end{equation}
where only the values of the coefficients $a$ and $b$ depend on the ansatz. 
Following Ref. \cite{jeansapp}, we shall determine the coefficients $a$ and $b$
so as to recover the exact
mass-radius relation in some particular limits. We finally note that the
mass-radius
relation can be written under the normalized form 
\begin{equation}
\label{pa1b}
\frac{M}{M_s}=\frac{\frac{R_{s}}{R}}{1\mp\left (\frac{R_{s}}{R}\right
)^2}
\end{equation}
with $R_s=b(|a_s|\hbar^2/Gm^3)^{1/2}$ and $M_s=(a/b)\hbar/\sqrt{Gm|a_s|}$ (the
upper sign corresponds to a repulsive self-interaction and the lower sign to
an attractive self-interaction).

{\it Remark:} With the Gaussian ansatz, we get $a_G^*=2\sigma_G/\nu_G=3.76$ and
$b_G^*=(6\pi\zeta_G/\nu_G)^{1/2}=1.73$.
However, below, we shall identify the radius $R$
with $R_{99}$, not with the radius $R$ of the $f$-ansatz defined in Eq.
(\ref{mra1}). Since
$R_{99}=2.38167\, R$ with the Gaussian ansatz \cite{prd1}, we obtain
$a_G=2.38167\,
a_G^*=8.96$ and $b_G=2.38167\,
b_G^*=4.12$ to be compared with the more exact values of $a$ and
$b$ found below [see Eqs. (\ref{pa5u}) and (\ref{pa5v})].

\subsection{Noninteracting bosons}
\label{sec_panb}

For noninteracting bosons ($a_s=0$), the mass-radius relation from Eq.
(\ref{pa1}) reduces to
\begin{equation}
\label{pa3}
M=a\frac{\hbar^2}{Gm^2R}.
\end{equation}
If we identify $R$ with the radius $R_{99}$ containing $99\%$ of the mass and
compare Eq. (\ref{pa3}) with the exact mass-radius relation of noninteracting
self-gravitating BECs from Eq. (\ref{dm12}), we get $a=9.946$.

{\it Remark:} The density profile of a noninteracting
self-gravitating BEC (soliton) is often fitted by the empirical
profile introduced by Schive {\it et al.} \cite{ch2,ch3}. In Ref. \cite{modeldm}
we have shown that a Gaussian profile \cite{prd1}, which is simpler, also fits
the soliton quite well up to the halo radius (see Fig. 2 of \cite{modeldm}). A
Gaussian profile with the mass-radius relation from Eq. (\ref{pa1}) may
also provide a convenient approximation of the density profile of
self-gravitating BECs with repulsive or attractive self-interaction. 

\subsection{Repulsive self-interaction}
\label{sec_parb}

For bosons with a repulsive self-interaction ($a_s>0$), in the TF limit ($\hbar\rightarrow
0$ with fixed $g=4\pi a_s\hbar^2/m^3$), the mass-radius
relation from Eq. (\ref{pa1}) reduces to
\begin{equation}
\label{pa5}
R=b\left (\frac{a_s\hbar^2}{Gm^3}\right )^{1/2}.
\end{equation}
If we identify $R$ with 
the radius at which the density vanishes and compare Eq. (\ref{pa5}) with the
exact radius of self-gravitating BECs in the TF limit from Eq. (\ref{dm15}), we
get $b=\pi$. On the other
hand, in the noninteracting limit, we recover the
result from Eq. (\ref{pa3}) leading to $a=9.946$. We shall adopt these values of
$a$ and $b$ in the repulsive case (see Fig. \ref{M-R-chi-pos-part1} for a
comparison with the exact result). Therefore, we take
\begin{equation}
\label{pa5u}
a=9.946,\qquad b=\pi\qquad ({\rm repulsive}).
\end{equation}

\subsection{Attractive self-interaction}
\label{sec_paab}

For bosons with an attractive  self-interaction ($a_s<0$), the mass-radius
relation from Eq. (\ref{pa1}) displays 
a maximum mass 
\begin{equation}
\label{pa8}
M_{\rm max}=\frac{a}{2b} \frac{\hbar}{\sqrt{Gm|a_s|}}\qquad {\rm at}\qquad
R_*=b\left (\frac{|a_s|\hbar^2}{Gm^3}\right )^{1/2}.
\end{equation}
They are connected by
\begin{eqnarray}
\label{dm17rj}
M_{\rm max}=\frac{a}{2} \frac{\hbar^2}{Gm^2R_*}.
\end{eqnarray}
If we identify $R_*$ with the radius $(R_*)_{99}$ 
containing $99\%$ of the mass and compare Eq. (\ref{pa8}) with the exact values
of the maximum mass and of the corresponding  radius from Eqs. (\ref{dm16}) and
(\ref{dm17}), we get $b=5.5$ and $a/2b=1.012$, leading to $a=11.1$. We shall
adopt these values in the attractive case (see Fig. \ref{M-R-chi-neg-part1} for
a comparison with the exact result). Therefore, we take
\begin{equation}
\label{pa5v}
a=11.1,\qquad b=5.5 \qquad ({\rm attractive}).
\end{equation}
We note that the value  $a=11.1$
obtained from the maximum mass is relatively close to  the value $a=9.946$
obtained in the noninteracting limit (see Sec. \ref{sec_panb}). In the
nongravitational
limit, the mass-radius relation from Eq. (\ref{pa1}) reduces to
\begin{equation}
\label{pa9}
M=\frac{a}{b^2} \frac{mR}{|a_s|}.
\end{equation}
The value $a/b^2=0.367$ obtained from the maximum mass is relatively close to
the exact value $0.275$ from Eq.
(\ref{dm19}). This is a consistency check.

\section{Exact maximum mass of boson stars due to general relativity}
\label{sec_er}

In this section, we recall the expression of the maximum mass of general
relativistic BECs (boson stars)  at $T=0$ described
by the KGE equations. Above that maximum mass, the system collapses towards a black hole.

\subsection{Noninteracting bosons}
\label{sec_erni}

The maximum mass and the minimum radius of a noninteracting boson star 
set by general relativity are \cite{kaup,rb}
\begin{equation}
\label{er1}
M_{\rm max}^{\rm GR}=0.633\, \frac{\hbar c}{Gm},\qquad R_{*}^{\rm
GR}=6.03\, \frac{\hbar}{mc}.
\end{equation}
They satisfy the relation
\begin{equation}
\label{er2}
R_{*}^{\rm
GR}=9.53\, \frac{GM_{\rm
max}}{c^2}. 
\end{equation}
The compactness at the maximum mass is ${\cal C}\equiv GM_{\rm max}/R_*^{\rm
GR}c^2=0.105$.\footnote{We have adopted the value of the radius
$R_{*}^{\rm
GR}=6.03\, {\hbar}/{mc}$ given by Seidel and Suen \cite{ss90}. This is the
radius containing $95\%$ of the mass. The radius containing $99\%$ of the
mass is larger, implying a smaller compactness. For example, Choi {\it et al.}
\cite{choi} consider the
radius containing $99\%$ of the mass and find a maximum compactness ${\cal
C}\simeq 0.08$ (see also \cite{babe}). This yields $R_{*}^{\rm
GR}=7.91\, {\hbar}/{mc}$.} These scalings can be 
obtained as explained in Appendix B.2
of \cite{prd1} (see also Secs. \ref{sec_is} and \ref{sec_rq}). The Kaup radius
is of the order of the Compton wavelength of the boson.

\subsection{Repulsive self-interaction in the TF limit}
\label{sec_ertf}

The maximum mass and the minimum radius of a boson star with a repulsive
$|\varphi|^4$ self-interaction  in the TF limit set by
general relativity are \cite{colpi,chavharko}
\begin{equation}
\label{er3}
M_{\rm max}^{\rm GR}=0.307\, \frac{\hbar c^2\sqrt{a_s}}{(Gm)^{3/2}},\qquad
R_{*}=1.92\, \left
(\frac{a_s\hbar^2}{Gm^3}\right )^{1/2}.
\end{equation}
They satisfy the relation
\begin{equation}
R_{*}^{\rm GR}=6.25\, \frac{GM_{\rm
max}}{c^2}. \label{er4}
\end{equation}
The compactness at the maximum mass is ${\cal C}\equiv GM_{\rm max}/R_*^{\rm
GR}c^2=0.16$ \cite{chavharko}. These
scalings
can be obtained as explained in Appendix B.3
of \cite{prd1} (see also Sec. \ref{sec_is}). The critical radius $R_*$ is 
of the same order as the minimum radius $R_{\rm TF}$ of a nonrelativistic
self-interacting BEC in the TF approximation (see Sec. \ref{sec_rep}).

{\it Remark:} In the TF limit, a self-interacting complex SF is equivalent to a
fluid with a barotropic equation of state $P(\epsilon)$ determined by the
potential $V(|\varphi|^2)$ (see Appendix
\ref{sec_ra}). In the case of a
repulsive $|\varphi|^4$
self-interaction [see Eq. (\ref{pq2})], the equation of state is given
by Eq. (\ref{rtf7qw}). The mass-radius relation of
the boson star, and its maximum mass (\ref{er3}),  may therefore be obtained by
solving
the Oppenheimer-Volkoff  equation of hydrostatic equilibrium, as done in Ref.
\cite{chavharko},
instead of solving the KGE equations, as done in Ref. \cite{colpi}.

\section{Relativistic corrections as we approach the Schwarzschild
radius}
\label{sec_is}

\begin{figure}[!h]
\begin{center}
\includegraphics[clip,scale=0.3]{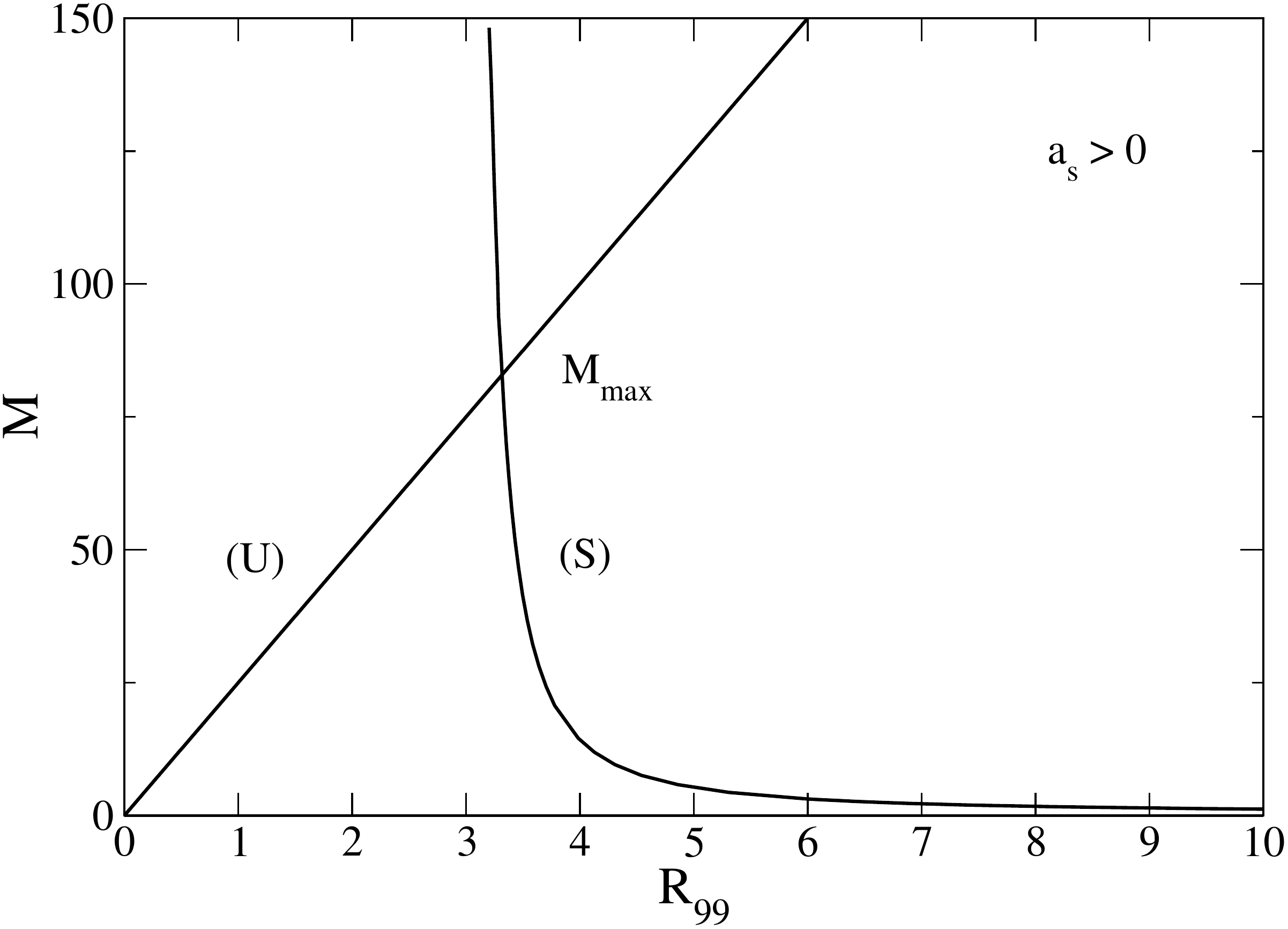}
\caption{Graphical construction determining the maximum mass of
self-gravitating BECs with $a_s>0$ due to general relativity. We use the
normalization of Fig. \ref{M-R-chi-pos-part1}. The maximum mass
$M_{\rm max}$ is qualitatively obtained by taking the intersection between the
mass-radius relation $M(R)$ of Newtonian BECs and the Schwarzschild line
$R=kGM/c^2$ (in scaled variables, its slope is $kGm/|a_s| c^2$).}
\label{M-R-chi-pos-part1mm}
\end{center}
\end{figure}

\begin{figure}[!h]
\begin{center}
\includegraphics[clip,scale=0.3]{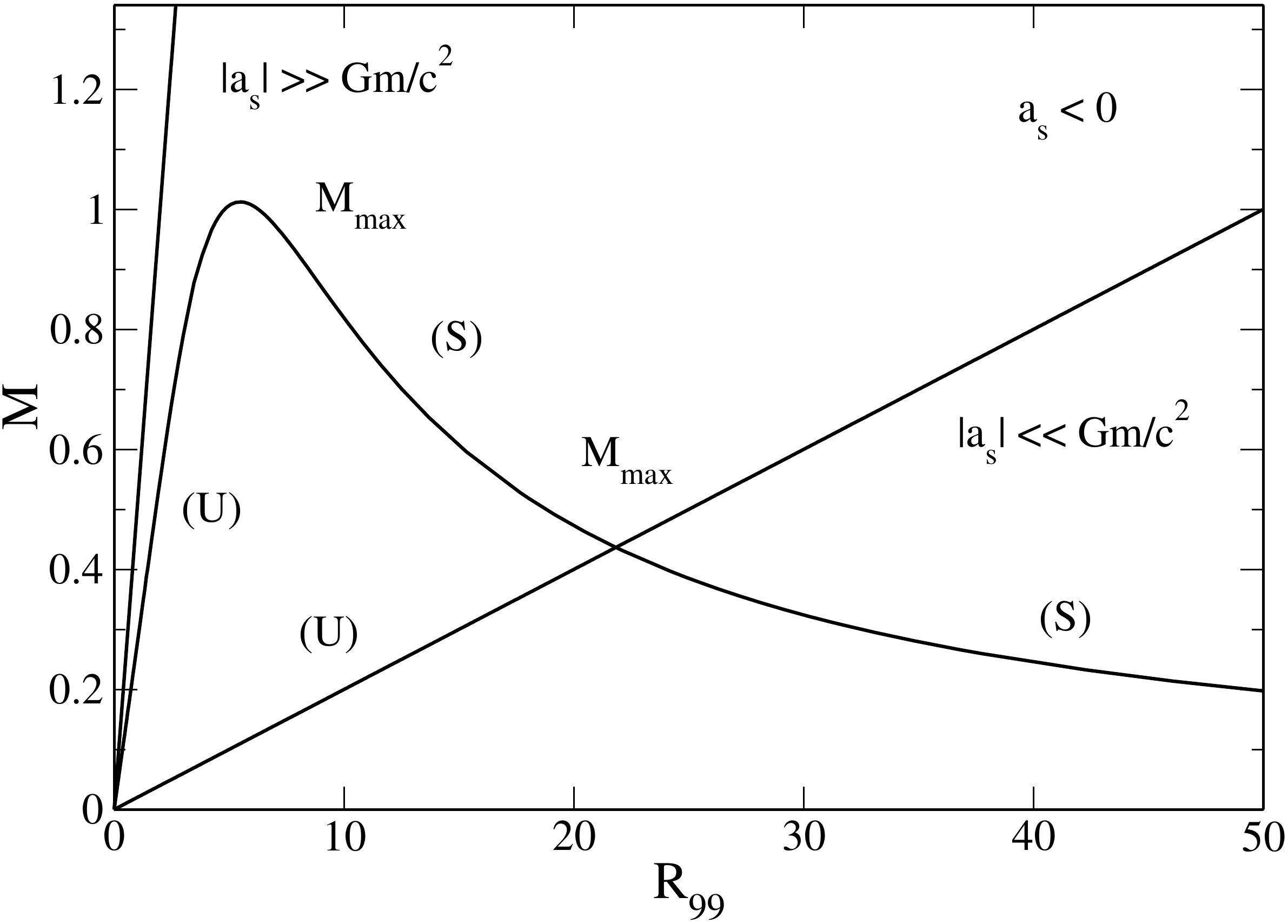}
\caption{Same as Fig. \ref{M-R-chi-pos-part1mm} for $a_s<0$. An intersection
exists only when $|a_s|c^2/Gm$ is sufficiently small (see the main text for
details).
When $|a_s|c^2/Gm$ is large, the maximum mass is given by the
nonrelativistic limit (\ref{dm16}) or by relativistic corrections in the quantum
potential (see Sec. \ref{sec_rq}).}
\label{M-R-chi-neg-part1mm}
\end{center}
\end{figure}

A relativistic star becomes dynamically unstable when its
radius approaches the
Schwarzschild radius $R_S=2GM/c^2$ \cite{chandra64}. The condition $R\sim R_S$ 
combined 
with the nonrelativistic mass-radius relation of the star determines the order
of magnitude of its maximum mass $M_{\rm max}^{\rm GR}$ due to
general relativity (see Appendix B of \cite{prd1}).  Let us apply this argument
to BEC stars (see Figs. \ref{M-R-chi-pos-part1mm} and
\ref{M-R-chi-neg-part1mm} for an illustration).

Substituting $R=kGM/c^2$ with $k\sim 1$ into Eq. (\ref{pa1}), we obtain after
simplification
\begin{equation}
M_{\rm max}=\left (\frac{a}{k}\right )^{1/2}\frac{\hbar
c}{Gm}\sqrt{1+\frac{b^2}{ak}\frac{a_s c^2}{Gm}}.
\label{is1}
\end{equation}
This equation is of the form
\begin{equation}
M_{\rm max}=A\frac{\hbar c}{Gm}\sqrt{1+B\frac{a_s c^2}{Gm}}
\label{is2}
\end{equation}
with $A=(a/k)^{1/2}$ and $B=b^2/ak$. The radius corresponding to the maximum mass is
\begin{equation}
R_*=\frac{kGM_{\rm max}}{c^2}.
\label{is3}
\end{equation}
Substituting Eqs. (\ref{is1}) and (\ref{is2}) into Eq. (\ref{is3}), we get
\begin{equation}
R_*=\sqrt{ka}\frac{\hbar }{mc}\sqrt{1+\frac{b^2}{ak}\frac{a_s c^2}{Gm}}
\label{is4}
\end{equation}
and
\begin{equation}
R_{*}=kA\frac{\hbar}{mc}\sqrt{1+B\frac{a_s c^2}{Gm}}.
\label{is5}
\end{equation}
If we define the compactness of the relativistic star at the maximum mass by
\begin{equation}
{\cal C}=\frac{GM_{\rm max}}{R_*c^2},
\label{is6}
\end{equation}
we obtain
\begin{equation}
{\cal C}=\frac{1}{k}.
\label{is7}
\end{equation}
By construction, our approximations assume 
that the compactness is independent of the value of $a_s$. Of course, this
is not rigorously true. However, this is a reasonable approximation as we can
see by considering the exact values of the compactness of boson stars
in two extreme limits (see Sec. \ref{sec_er}),
namely the noninteracting limit (${\cal C}=0.105$) and the TF
limit (${\cal C}=0.16$).

We shall now determine the values of $A$ and $B$ in Eq. (\ref{is2}) in 
order to match the exact results from Secs. \ref{sec_erni} and \ref{sec_ertf}.

\subsection{Noninteracting limit}
\label{sec_isni}

In the noninteracting limit ($a_s=0$), Eqs. (\ref{is1})-(\ref{is5}) reduce to
\begin{equation}
M_{\rm max}=\left (\frac{a}{k}\right )^{1/2}\frac{\hbar c}{Gm}=A\frac{\hbar
c}{Gm}
\label{is8}
\end{equation}
and
\begin{equation}
R_{*}=\sqrt{ka}\frac{\hbar}{mc}=kA\frac{\hbar}{mc}.
\label{is9}
\end{equation}
This returns the maximum mass (\ref{er1}) of mini boson stars
\cite{kaup}.
Comparing Eqs.
(\ref{is8}) and (\ref{is9}) with
Eqs. (\ref{er1}) and (\ref{er2}) we get $A=0.633$ and $k=1/{\cal C}=9.53$.

{\it Remark:} With these values of $A$ and $k$ we obtain $a=kA^2=3.82$ instead
of the value
$a=9.946$ (or $a_G=8.96$) computed in
Sec. \ref{sec_pa}.

\subsection{TF limit}
\label{sec_istf}

For boson stars with a repulsive self-interaction in the TF limit
($\hbar\rightarrow 0$ with fixed $g=4\pi a_s\hbar^2/m^3$),  Eqs.
(\ref{is1})-(\ref{is5})
reduce to
\begin{equation}
M_{\rm max}=\frac{b}{k}\left (\frac{a_s\hbar^2 c^4}{G^3m^3}\right
)^{1/2}=A\sqrt{B}\left (\frac{a_s\hbar^2 c^4}{G^3m^3}\right )^{1/2}\label{is10}
\end{equation}
and
\begin{equation}
R_{*}=b\left (\frac{a_s\hbar^2}{Gm^3}\right )^{1/2}=kA\sqrt{B}\left
(\frac{a_s\hbar^2}{Gm^3}\right )^{1/2}.
\label{is11}
\end{equation}
This returns the maximum mass (\ref{er3}) of massive boson stars
\cite{colpi,chavharko}. Note
that
$R_*$ is
independent of $k$. Comparing Eqs.
(\ref{is10}) and (\ref{is11}) with Eqs. (\ref{er3}) and (\ref{er4})
we get $A\sqrt{B}=0.307$ and $k=1/{\cal C}=6.25$. Taking $A=0.633$ from Sec.
\ref{sec_isni} we get $B=0.235$.

{\it Remark:} With these values of $A$ and $k$ we obtain  $a=kA^2=2.50$
and $b=kA\sqrt{B}=1.92$ instead of the values
$a=9.946$ and $b=\pi$ (or $a_G=8.96$ and $b_G=4.12$) computed in Sec.
\ref{sec_pa}.

\subsection{Interpolation formulas}
\label{sec_isint}

Using the previous results, we can obtain simple interpolation formulas for
the maximum mass and minimum radius of boson stars. We shall take
\begin{equation}
\label{is12}
A=0.633,\qquad B=0.235, \qquad k=6.25.
\end{equation}
In this manner, the asymptotic expressions of the
maximum mass are exact. The value of the radius is also
exact  in the
TF limit. By contrast, the value of the radius is not exact (but
approximately correct) for
noninteracting bosons because of the slow change of compactness with the
self-interaction which is not taken into account in our approach. We propose
therefore the interpolation formulas
\begin{equation}
M_{\rm max}=0.633\, \frac{\hbar c}{Gm}\sqrt{1+0.235\, \frac{a_s c^2}{Gm}},
\label{is13}
\end{equation}
\begin{equation}
R_{*}=3.96\, \frac{\hbar}{mc}\sqrt{1+0.235\frac{a_s c^2}{Gm}}.
\label{is14}
\end{equation}
We note that these expressions are defined only for $a_s>-4.25\, Gm/c^2$ (see
Fig. \ref{M-R-chi-neg-part1mm}). This
suggests that the present treatment is only reliable for positive values of the
scattering length.  This is because the maximum mass of boson stars with an
attractive self-interaction ($a_s<0$) is essentially a nonrelativistic result
\cite{prd1}. Therefore, relativistic corrections in the maximum mass do not come
from the fact that the radius of the star approaches the Schwarzschild radius
but rather from relativistic corrections arising in the quantum
potential as discussed  in Sec. \ref{sec_rq} below.  

For $a_s\rightarrow 0$, Eqs. (\ref{is13}) and (\ref{is14}) reduce to
\begin{equation}
M_{\rm max}\simeq 0.633\, \frac{\hbar c}{Gm}\left (1+0.118\, \frac{a_s
c^2}{Gm}\right ),
\label{is15}
\end{equation}
\begin{equation}
R_{*}\simeq 3.96\, \frac{\hbar}{mc}\left (1+0.118\frac{a_s c^2}{Gm}\right ).
\label{is16}
\end{equation}

\subsection{Improved interpolation formulas}

We can improve the preceding interpolation formulas by allowing the parameter
$B$ to be different in the expressions (\ref{is2}) and (\ref{is5})
determining the maximum mass and the corresponding radius. We write
\begin{equation}
M_{\rm max}=A\frac{\hbar c}{Gm}\sqrt{1+B\frac{a_s c^2}{Gm}},
\label{uk0}
\end{equation}
\begin{equation}
R_{*}=A'\frac{\hbar}{mc}\sqrt{1+B'\frac{a_s c^2}{Gm}},
\label{uk1}
\end{equation}
and we determine the constants $A$, $B$, $A'$ and $B'$ so as to reproduce the
exact asymptotic behaviors from Eqs. (\ref{er1}) and (\ref{er3}). In this
manner we get
\begin{equation}
\label{uk2}
A=0.633,\quad B=0.235, \quad A'=6.03,\quad B'=0.101,
\end{equation}
leading to
\begin{equation}
M_{\rm max}=0.633\frac{\hbar c}{Gm}\sqrt{1+0.235\frac{a_s c^2}{Gm}},
\label{uk3}
\end{equation}
\begin{equation}
R_{*}=6.03\frac{\hbar}{mc}\sqrt{1+0.101\frac{a_s c^2}{Gm}}.
\label{uk4}
\end{equation}
This gives a maximum compactness 
\begin{equation}
{\cal C}=0.105\, \frac{\sqrt{1+0.235\frac{a_s c^2}{Gm}}}{\sqrt{1+0.101\frac{a_s
c^2}{Gm}}},
\label{uk5}
\end{equation}
depending on $a_s$, and going from ${\cal C}=0.105$ in the noninteracting limit
$a_s=0$ to ${\cal C}=0.160$ in the TF limit $a_s\rightarrow +\infty$.

\section{Relativistic corrections in the quantum potential}
\label{sec_rq}

\subsection{Relativistic Hamiltonian}

In the weak gravity limit, the relativistic Hamiltonian is
given by $H=\int_0^{+\infty}T_0^0 4\pi r^2\, dr$ where the time-time
component $T_0^0$ of the energy-momentum tensor is given by Eq. (\ref{kgp14b}).
If we introduce the pseudo wave function $\psi({\bf
r},t)=e^{-iEt/\hbar}\phi({\bf
r})$ where $E$ includes relativistic corrections, one can show
\cite{croon,choi} that the Hamiltonian decomposes into $H=Mc^2+H_{\rm
NR}+H_{\rm R}$ where the first term is the rest mass energy (with $M=Nm$), the
second term is the nonrelativistic Hamiltonian and the third term is the
relativistic correction to the kinetic energy (or quantum potential). From Eq.
(\ref{kgp14b}) we see
that the kinetic energy is equal to
\begin{eqnarray}
H_{\rm kin}=\frac{\hbar^2}{2m^2}\int \left
(1+\frac{2\Phi}{c^2}\right )|\nabla\psi|^2\, d{\bf r}.
\label{rqm2}
\end{eqnarray}
Consequently, the  relativistic correction to the kinetic
energy is
\begin{eqnarray}
H_{\rm R}=\frac{\hbar^2}{m^2c^2}\int \Phi|\nabla\psi|^2\, d{\bf r}.
\label{rqm1}
\end{eqnarray}
Using the hydrodynamic representation of the SF (see Sec.
\ref{sec_mad})
and ignoring the rest-mass term which is just a constant,  the total energy
taking into account
relativistic corrections in the quantum potential is $E_{\rm tot}=E_{\rm
NR}+E_{\rm R}$, where $E_{\rm NR}$ is given by Eq. (\ref{mad12}) and 
\begin{eqnarray}
E_{\rm R}=\frac{\hbar^2}{m^2c^2}\int (\nabla \sqrt{\rho})^2\Phi\, d{\bf r}.
\label{rq1}
\end{eqnarray}
Using the $f$-ansatz of Sec. \ref{sec_pa}, we obtain
\begin{eqnarray}
E_{\rm R}=\frac{\hbar^2GM^2}{m^2c^2R^3}\int (\nabla \sqrt{f})^2 g\, d{\bf x}.
\label{rq2}
\end{eqnarray}
Since the value of the integral is negative, we can write
\begin{equation}
E_{\rm R}=-\chi\frac{\hbar^2GM^2}{m^2c^2R^3}\quad {\rm
with}\quad \chi=-\int (\nabla \sqrt{f})^2 g\, d{\bf
x}>0.
\label{rq3}
\end{equation}
If we use a Gaussian ansatz, we get
$\chi_G\simeq 0.997$ (see Appendix
\ref{sec_g}).

Combining Eqs. (\ref{mra11}) and (\ref{rq3}), we find that the total energy
is
\begin{equation}
E_{\rm tot}(R)=\sigma\frac{\hbar^2M}{m^2R^2}-\nu \frac{GM^2}{R}+\zeta
\frac{2\pi a_s\hbar^2M^2}{m^3R^{3}}-\chi\frac{\hbar^2GM^2}{m^2c^2R^3}.
\label{rq4}
\end{equation}
Remarkably, the scaling of the relativistic 
correction $E_{\rm R}$ is the same as the scaling of the internal energy $U$
arising from the self-interaction of the bosons (for a $|\varphi|^4$
self-interaction). As a result, the total energy of a relativistic BEC star
can be rewritten as
\begin{equation}
E_{\rm tot}(R)=\sigma\frac{\hbar^2M}{m^2R^2}-\nu \frac{GM^2}{R}+\zeta
\frac{2\pi a_s\hbar^2M^2}{m^3R^{3}}\left (1-\kappa \frac{Gm}{a_s c^2}\right )
\label{rq5}
\end{equation}
with 
\begin{eqnarray}
\kappa=\frac{\chi}{2\pi\zeta}.
\label{rq6}
\end{eqnarray}
If we use a  Gaussian ansatz, we get $\kappa_G=2.50$. We see that the results of
the nonrelativistic study \cite{prd1} remain valid provided that we make
the substitution
\begin{eqnarray}
a_s\rightarrow a_s\left (1-\kappa \frac{Gm}{a_s c^2}\right )\equiv a_s^*.
\label{rq7}
\end{eqnarray}
In particular, according to Eqs. (\ref{pa1}) and (\ref{rq7}), the mass-radius
relation of relativistic BEC stars is
\begin{equation}
\label{rq8}
M=\frac{a\frac{\hbar^2}{Gm^2R}}{1-b^2\frac{a_s\hbar^2}{Gm^3R^2}\left (1-\kappa \frac{Gm}{a_s c^2}\right )
}.
\end{equation}
Since we have used scaled variables,
Figs. \ref{M-R-chi-pos-part1} and \ref{M-R-chi-neg-part1} remain the same. We
just have to replace $a_s$ by $a_s^*$ in the normalization. There
is a critical scattering length
\begin{eqnarray}
(a_s)_c=\kappa \frac{Gm}{c^2}=\kappa\frac{r_S}{2},
\label{rq9}
\end{eqnarray}
where $r_S=2Gm/c^2$ is the Schwarzschild  (gravitational) radius of
the bosons
\cite{abrilphas}. When $a_s=(a_s)_c$, the effective scattering length
vanishes ($a_s^*=0$). When $a_s\ge (a_s)_c$ (i.e., $a_s^*\ge 0$),
the mass-radius relation is monotonic like in Fig. \ref{M-R-chi-pos-part1}.
When $a_s<(a_s)_c$ (i.e.,
$a_s^*<0$), it displays a maximum mass like in Fig. \ref{M-R-chi-neg-part1}.
Therefore, when relativistic corrections
are taken into account in the quantum potential, we find the existence of a
maximum mass not only when $a_s<0$ but also when $0\le a_s<\kappa Gm/c^2$. Using
Eqs. (\ref{pa8}) and (\ref{rq7}), the maximum mass and the corresponding radius
are given by 
\begin{eqnarray}
\label{rq10}
M_{\rm max}=\frac{a}{2b}\, \frac{\hbar}{\sqrt{Gm\left |a_s-\kappa
\frac{Gm}{c^2}\right |}},
\end{eqnarray}
\begin{eqnarray}
\label{rq11}
R_*=b\, \left (\frac{\left |a_s-\kappa \frac{Gm}{c^2}\right | \hbar^2}{Gm^3}\right )^{1/2}.
\end{eqnarray}
They can be written as
\begin{eqnarray}
\label{rq12}
M_{\rm max}=C\, \frac{\hbar}{\sqrt{Gm\left |a_s-\kappa \frac{Gm}{c^2}\right |}}
\end{eqnarray}
\begin{eqnarray}
\label{rq13}
R_*=D\, \left (\frac{\left |a_s-\kappa \frac{Gm}{c^2}\right | \hbar^2}{Gm^3}\right )^{1/2}.
\end{eqnarray}
with $C=a/2b$ and $D=b$.

{\it Remark:} Using the values $a=11.1$ and $b=5.5$ of Eq. (\ref{pa5v}) we
obtain
$C=1.012$ and $D=5.5$. Using the values $a_G=8.96$ and $b_G=4.12$ of the
Gaussian ansatz we find that $C=1.09$ and $D=4.12$.

\subsection{Noninteracting limit}
\label{sec_rqni}

In the noninteracting limit ($a_s=0$), Eqs. (\ref{rq10})-(\ref{rq13})
reduce to
\begin{eqnarray}
\label{dm16d}
M_{\rm max}=\frac{a}{2b\sqrt{\kappa}}\, \frac{\hbar
c}{Gm}=\frac{C}{\sqrt{\kappa}}\, \frac{\hbar c}{Gm}
\end{eqnarray}
\begin{eqnarray}
\label{dm17d}
R_*=b\sqrt{\kappa}\, \frac{\hbar}{m c}=D\sqrt{\kappa}\, \frac{\hbar}{m c}
\end{eqnarray}
\begin{eqnarray}
\label{dm17e}
R_*=\frac{2b^2\kappa}{a}\frac{GM_{\rm max}}{c^2}=\frac{D\kappa}{C}\frac{GM_{\rm
max}}{c^2}.
\end{eqnarray}
This returns the maximum mass (\ref{er1}) of mini boson stars
\cite{kaup}.  The
compactness
at the maximum mass is ${\cal C}=a/(2b^2\kappa)=C/(D\kappa)$.
Comparing Eqs. (\ref{dm16d})-(\ref{dm17e}) with Eqs. (\ref{er1}) and 
(\ref{er2}) we get $C/\sqrt{\kappa}=0.633$ and $D\sqrt{\kappa}=6.03$.

{\it Remark:}  Using the values $a_G=8.96$, $b_G=4.12$ and $\kappa_G=2.50$ of
the Gaussian ansatz we find that $M_{\rm max}=0.688\, {\hbar
c}/{Gm}$, $R_*=6.51 \, {\hbar}/{m c}$ and ${\cal C}=0.106$ in good
agreement with the exact
values $M_{\rm max}=0.633\, {\hbar c}/{Gm}$, $R_*=6.03 \, {\hbar}/{m c}$
and ${\cal C}=0.105$ \cite{kaup}. This is remarkable because these predictions
are obtained without
numerical calculation. In particular, the compactness is predicted almost
exactly. We also find that the relativistic mass-radius relation is 
given by [see Eq. (\ref{rq8}) with $a_s=0$] 
\begin{equation}
\label{rq8zero}
M=\frac{a\frac{\hbar^2}{Gm^2R}}{1+b^2\kappa \frac{\hbar^2}{m^2c^2R^2}}.
\end{equation}
Its graphical representation can be deduced from Fig. \ref{M-R-chi-neg-part1} by
taking
$M_a=\hbar
c/(\sqrt{\kappa}Gm)$ and $R_a=\sqrt{\kappa}\hbar/mc$ (corresponding to
$a_s\rightarrow a_s^*=-\kappa Gm/c^2$). 

\subsection{Attractive self-interaction}
\label{sec_rqatt}

For bosons with an attractive self-interaction ($a_s<0$), the expression
(\ref{rq12}) of the maximum mass is defined for all $a_s$. When $a_s\rightarrow
-\infty$, we obtain
\begin{equation}
\label{pa8b}
M_{\rm max}=\frac{a}{2b}\frac{\hbar}{\sqrt{Gm|a_s|}}=C
\frac{\hbar}{\sqrt{Gm|a_s|}},
\end{equation}
\begin{equation}
\label{pa8c}
R_*=b\left (\frac{|a_s|\hbar^2}{Gm^3}\right )^{1/2}=D\left
(\frac{|a_s|\hbar^2}{Gm^3}\right )^{1/2}.
\end{equation}
This returns the maximum mass (\ref{dm16}) of nonrelativistic
dilute axion stars \cite{prd1}. Comparing Eqs.
(\ref{pa8b}) and (\ref{pa8c}) with Eqs. (\ref{dm16}) and (\ref{dm17}) we get
$C=1.012$ and $D=5.5$. Combined with the results of Sec.
\ref{sec_rqni}, we see that we cannot satisfy all the constraints (we have four
equations for three unknowns). The idea is to privilege exact asymptotic results
for the maximum mass with respect to the radius. Therefore, we will take
$C=1.012$ and $C/\sqrt{\kappa}=0.633$. This  gives $\kappa=2.56$ in good
agreement with the value $\kappa_{\rm G}=2.50$ obtained from the Gaussian
ansatz. Then, from the
results of Sec. \ref{sec_rqni}, we get $D=6.03/\sqrt{\kappa}=3.77$ and from the
results of this section $D=5.5$. The disagreement between these two values is
not too strong. In the following we shall adopt $D=5.5$.

{\it Remark:} Using the values $a_G=8.96$ and
$b_G=4.12$ of the
Gaussian ansatz we get $M_{\rm max}=1.09\, {\hbar}/{\sqrt{Gm|a_s|}}$,
$R_*=4.12 \, \sqrt{|a_s|\hbar^2/Gm^3}$ and ${\cal C}=0.264\, Gm/|a_s|c^2$  in
good agreement
with the exact
values $M_{\rm max}=1.012\, {\hbar}/{\sqrt{Gm|a_s|}}$,
$R_*=5.5 \, \sqrt{|a_s|\hbar^2/Gm^3}$ and ${\cal C}=0.184\, Gm/|a_s|c^2$
\cite{prd1}. This is remarkable because these predictions are obtained without
numerical calculation. In particular, the maximum mass is predicted almost
exactly.

\subsection{Repulsive self-interaction}

For bosons with a repulsive self-interaction ($a_s>0$), 
the expression (\ref{rq12}) of the maximum mass is defined only for
$a_s<\kappa Gm/c^2$. In particular, it does not return the maximum 
mass of massive boson stars \cite{colpi,chavharko} from Eq.
(\ref{er3}) in the TF limit (i.e. when $a_s\rightarrow +\infty$). Indeed, the
maximum mass from Eq. (\ref{rq12})
diverges when $a_s\rightarrow \kappa Gm/c^2$ and the mass-radius relation from
Eq. (\ref{rq8}) does not display a maximum mass when $a_s>\kappa Gm/c^2$. This
means that relativistic corrections of strong gravity as we approach the
Schwarzschild radius, besides the relativistic correction $E_{\rm R}$ of weak
gravity in the quantum
potential, are important in that
case as discussed in Sec. \ref{sec_is}.

\subsection{Interpolation formulas}

Using the previous results, we can obtain simple interpolation formulas for
the maximum mass and minimum radius of boson stars. We shall take
\begin{equation}
C=1.012,\qquad D=5.5, \qquad  \kappa=2.56.
\label{pa8d}
\end{equation}
In this manner, the asymptotic expressions of the maximum mass for $a_s=0$ and
$a_s\rightarrow -\infty$ are 
exact. The asymptotic expression of the corresponding radius is also exact for
$a_s\rightarrow -\infty$ while it is only approximate for  $a_s=0$. We propose
therefore the interpolation formulas
\begin{eqnarray}
\label{dm16f}
M_{\rm max}=1.012\, \frac{\hbar}{\sqrt{Gm\left |a_s-2.56 \frac{Gm}{c^2}\right
|}},
\end{eqnarray}
\begin{eqnarray}
\label{dm17h}
R_*=5.5\, \left (\frac{\left |a_s-2.56 \frac{Gm}{c^2}\right | \hbar^2}{Gm^3}\right )^{1/2}.
\end{eqnarray}
We note that these expressions are defined only for $a_s<2.56\,
Gm/c^2$.\footnote{The bound $a_s<2.56\,
Gm/c^2$ obtained when we consider relativistic corrections to the quantum
potential is ``antisymmetric'' with respect to the bound $a_s>-4.25\,
Gm/c^2$ found in Sec. \ref{sec_is} when we consider the criterion based on
the
Schwarzschild radius. This shows that the two approaches are complementary to
each other.} This
suggests that the 
present treatment is only reliable for  negative
values of the scattering
length. 
This is because the maximum mass of boson stars with a  repulsive
self-interaction ($a_s\ge 0$) is a general relativistic result which is
essentially due to the fact that the radius of the system approaches the
Schwarzschild
radius as shown in Sec. \ref{sec_is}. It is not directly due to relativistic
corrections in the
quantum potential.

For $a_s\rightarrow 0$, Eqs. (\ref{dm16f}) and (\ref{dm17h}) reduce to
\begin{eqnarray}
\label{dm16g}
M_{\rm max}\simeq 0.633\, \frac{\hbar c}{Gm}\left (1+0.195\, \frac{a_s
c^2}{Gm}\right ),
\end{eqnarray}
\begin{eqnarray}
\label{dm17i}
R_*\simeq 8.8\, \frac{\hbar}{m c}\left (1-0.195\, \frac{a_s c^2}{Gm}\right ).
\end{eqnarray}
Comparing Eqs. (\ref{dm16g}) and (\ref{dm17i}) with Eqs. (\ref{is15}) and
(\ref{is16}), we see that the expressions coincide approximately for the maximum
mass while they strongly differ for the radius because the values of the
radius at $a_s=0$ and the signs in front of the
correction $a_s c^2/Gm$ are not the same. This is why we have privileged the
behavior
of $R_*$ at $a_s\rightarrow \pm\infty$ rather than at
$a_s=0$.

{\it Remark:} Using the values $a_G=8.96$, $b_G=4.12$ and $\kappa_G=2.50$ of
the Gaussian ansatz to evaluate the coefficients in Eqs. (\ref{rq10}) and
(\ref{rq11}), we get
\begin{eqnarray}
\label{rq12ga}
M_{\rm max}=1.09\, \frac{\hbar}{\sqrt{Gm\left |a_s-2.50 \frac{Gm}{c^2}\right
|}},
\end{eqnarray}
\begin{eqnarray}
\label{rq13ga}
R_*=4.12\, \left (\frac{\left |a_s-2.50 \frac{Gm}{c^2}\right |
\hbar^2}{Gm^3}\right )^{1/2}.
\end{eqnarray}
We stress that these results are obtained from the Gaussian ansatz without
numerical calculation. They are in good agreement with the interpolation
formulas  (\ref{dm16f}) and (\ref{dm17h}) which rely on the exact asymptotic
expressions of the maximum mass and minimum radius obtained numerically.

\subsection{Improved interpolation formulas}

We can improve the preceding interpolation formulas by allowing the parameter
$\kappa$ to be different in the expressions (\ref{rq12}) and (\ref{rq13})
determining the maximum mass and the corresponding
radius. We write
\begin{eqnarray}
\label{ru0}
M_{\rm max}=C\, \frac{\hbar}{\sqrt{Gm\left |a_s-\kappa \frac{Gm}{c^2}\right
|}},
\end{eqnarray}
\begin{eqnarray}
\label{ru1}
R_*=D\, \left (\frac{\left |a_s-\kappa' \frac{Gm}{c^2}\right |
\hbar^2}{Gm^3}\right )^{1/2},
\end{eqnarray}
and we determine the constants $C$, $D$, $\kappa$ and $\kappa'$ so as to
reproduce the exact
asymptotic behaviors from Eqs.  (\ref{dm16}), (\ref{dm17}) and (\ref{er1}). In
this manner, we get
\begin{equation}
\label{ru2}
C=1.012,\quad D=5.5, \quad \kappa=2.56,\quad \kappa'=1.20,
\end{equation}
leading to
\begin{eqnarray}
\label{ru3}
M_{\rm max}=1.012\, \frac{\hbar}{\sqrt{Gm\left |a_s-2.56 \frac{Gm}{c^2}\right
|}},
\end{eqnarray}
\begin{eqnarray}
\label{ru4}
R_*=5.5\, \left (\frac{\left |a_s-1.20 \frac{Gm}{c^2}\right |
\hbar^2}{Gm^3}\right )^{1/2}.
\end{eqnarray}
This gives a maximum compactness 
\begin{equation}
{\cal C}=\frac{0.184\, \frac{Gm}{c^2}}{\sqrt{\left (a_s-2.56
\frac{Gm}{c^2}\right )\left (a_s-1.20 \frac{Gm}{c^2}\right )}},
\label{ru5}
\end{equation}
depending on $a_s$, and going from ${\cal C}=0.105$ in the noninteracting limit
$a_s=0$ to
${\cal C}\sim 0.184\, Gm/|a_s|c^2$ in the nonrelativistic limit
$a_s\rightarrow -\infty$.

\section{Summary of the main results}
\label{sec_summ}

\begin{figure}[!h]
\begin{center}
\includegraphics[clip,scale=0.3]{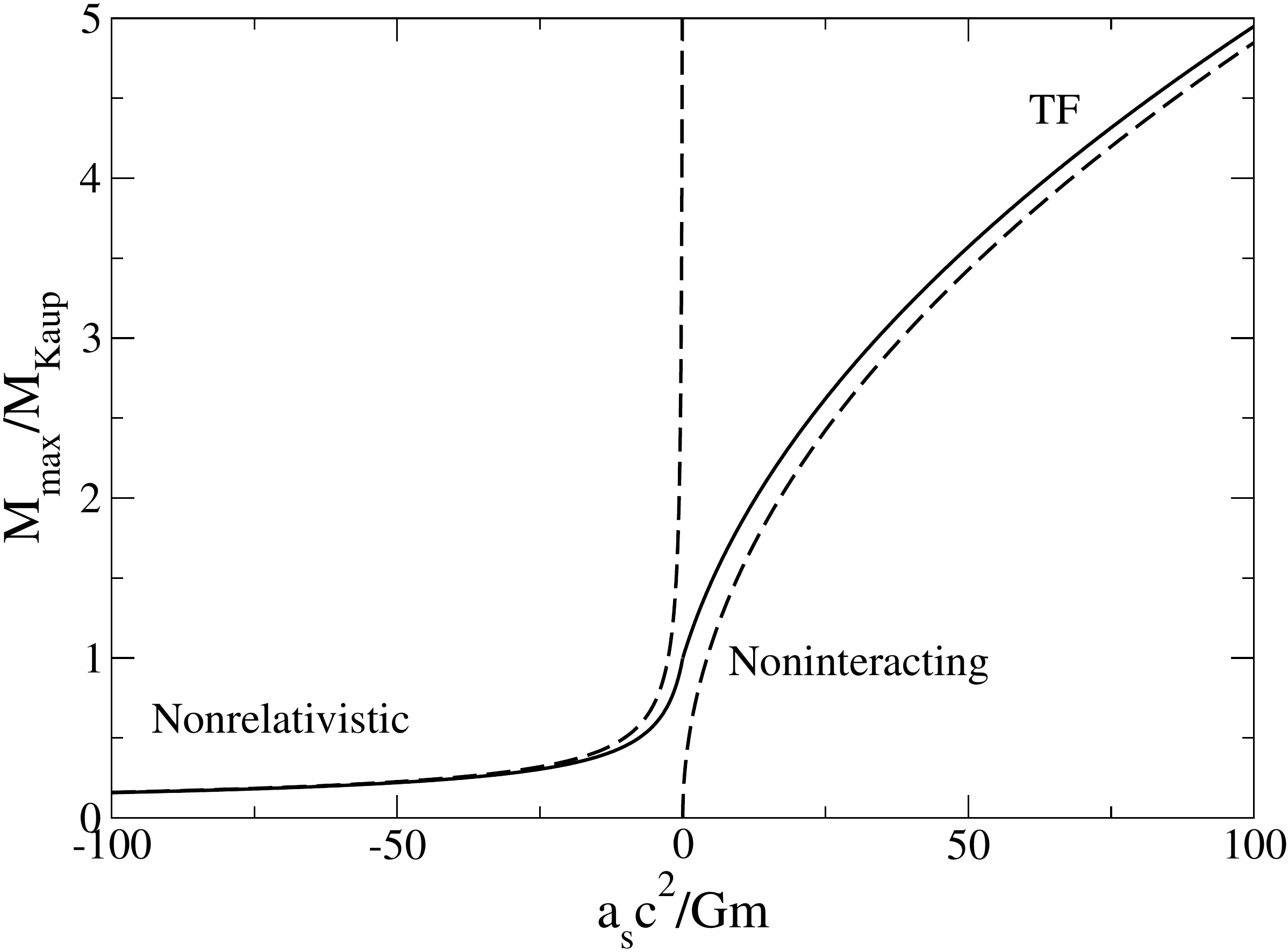}
\caption{Maximum mass $M_{\rm max}$ of boson stars normalized by the Kaup mass
$M_{\rm Kaup}$
corresponding to noninteracting bosons ($a_s=0$)  \cite{kaup,rb} as a function
of the scattering
length $a_s$ of the bosons normalized by their semi-Schwarzschild radius
$Gm/c^2$. The
dashed lines correspond to the TF limit
($a_s\rightarrow +\infty$) \cite{colpi,chavharko} and to the nonrelativistic
limit ($a_s\rightarrow
-\infty$) \cite{prd1}.}
\label{masse}
\end{center}
\end{figure}

\begin{figure}[!h]
\begin{center}
\includegraphics[clip,scale=0.3]{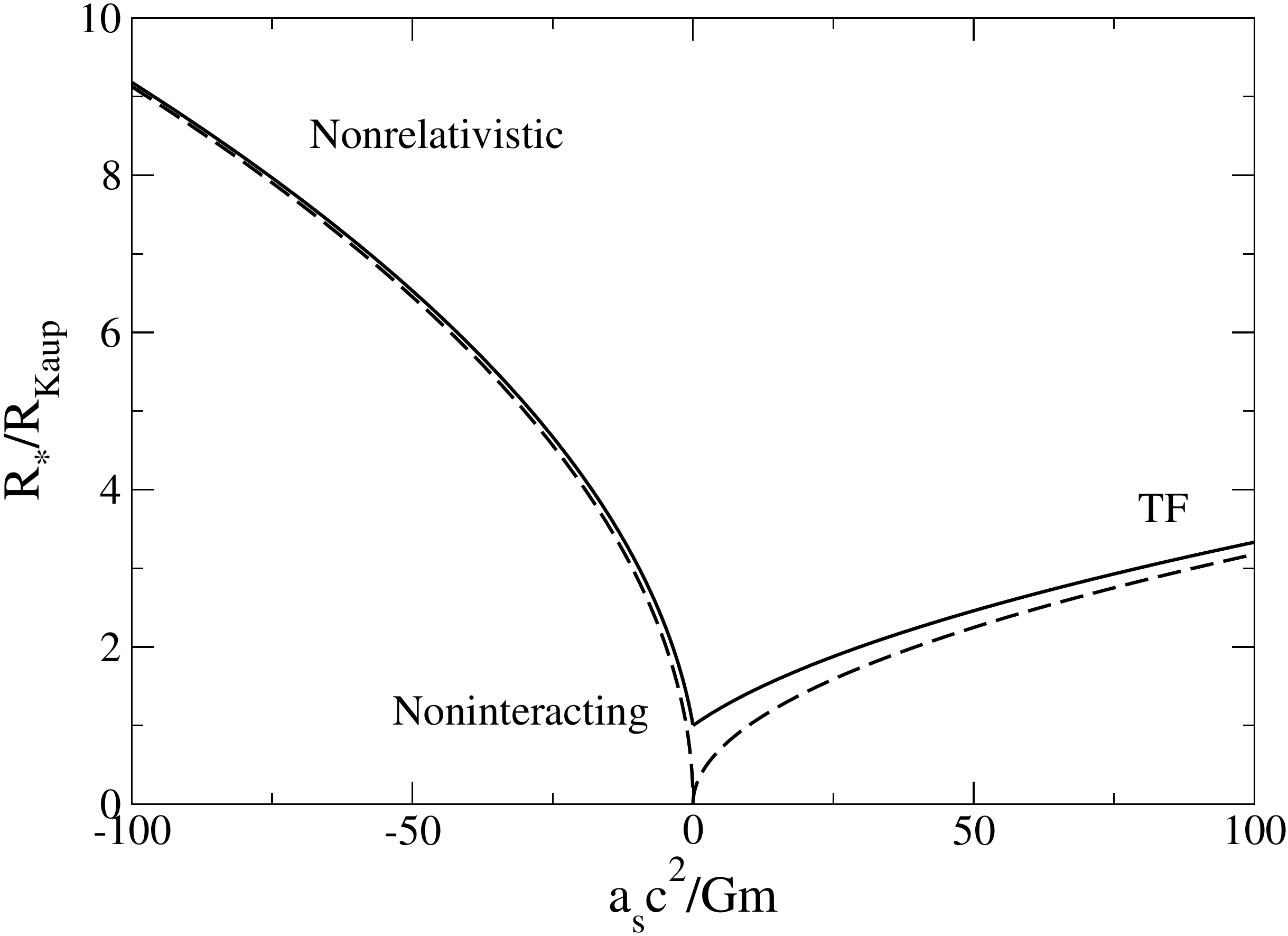}
\caption{Minimum radius $R_{*}$ of boson stars normalized by the Kaup radius
$R_{\rm Kaup}$ as
a function of the scattering
length $a_s$ of the bosons normalized by their semi-Schwarzschild radius
$Gm/c^2$.}
\label{rayon}
\end{center}
\end{figure}

\begin{figure}[!h]
\begin{center}
\includegraphics[clip,scale=0.3]{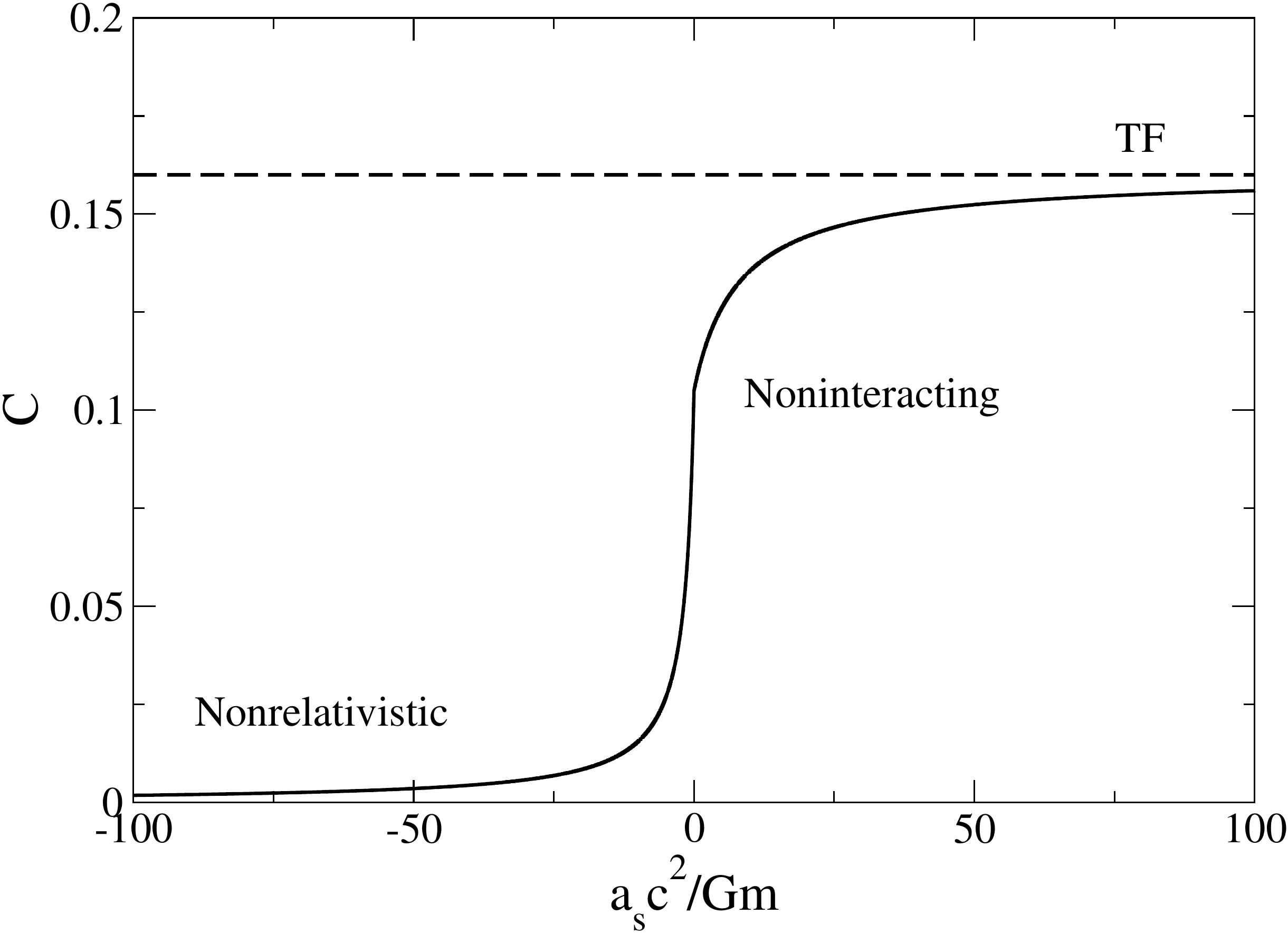}
\caption{Maximum compactness ${\cal
C}=GM_{\rm max}/R_* c^2$ of boson stars as
a function of the scattering
length $a_s$ of the bosons normalized by their semi-Schwarzschild radius
$Gm/c^2$. The noninteracting limit ($a_s=0$) corresponds to mini boson stars
(${\cal C}=0.105$) \cite{kaup,rb}, the TF limit ($a_s\rightarrow +\infty$)
corresponds to massive
boson stars (${\cal C}=0.16$)  \cite{colpi,chavharko}, and the nonrelativistic
limit
($a_s\rightarrow -\infty$) corresponds to dilute axion stars
(${\cal C}\rightarrow 0$) \cite{prd1,prd2,mg16B}.}
\label{compactness}
\end{center}
\end{figure}

In this section, we summarize the results
obtained previously for boson stars with repulsive or attractive
self-interaction. The maximum mass, minimum radius and maximum compactness are
plotted as a function of the scattering length in Figs. \ref{masse}-\ref{compactness}.

For boson stars with a repulsive self-interaction ($a_s\ge 0$), we have obtained
the
following formulas  [see Eqs. (\ref{uk3}) and (\ref{uk4})]
\begin{equation}
M_{\rm max}=0.633\, \frac{\hbar c}{Gm}\sqrt{1+0.235\, \frac{a_s c^2}{Gm}},
\label{e1}
\end{equation}
\begin{equation}
R_{*}=6.03\, \frac{\hbar}{mc}\sqrt{1+0.101\frac{a_s c^2}{Gm}},
\label{e1b}
\end{equation}
or, equivalently,
\begin{equation}
M_{\rm max}=0.307\, \frac{\hbar c^2\sqrt{a_s}}{(Gm)^{3/2}}\sqrt{1+4.25\,
\frac{Gm}{a_s c^2}},
\label{e1j}
\end{equation}
\begin{equation}
R_{*}=1.92\, \left
(\frac{a_s\hbar^2}{Gm^3}\right )^{1/2}\sqrt{1+9.90\,
\frac{Gm}{a_s c^2}}.
\label{e1bj}
\end{equation}
These expressions
interpolate between the
maximum mass and minimum radius [see Eq. (\ref{er1})] of noninteracting bosons
stars  \cite{kaup,rb} and the
maximum mass  and minimum radius [see Eq. (\ref{er3})]  of bosons stars with a
strong repulsive self-interaction (TF limit)
\cite{colpi,chavharko}. The maximum mass is due to general relativity in the
strong gravity regime. We expect
these approximate results to be more 
accurate for $a_s\gg r_S=Gm/c^2$ than for $a_s\ll r_S$.
We see that a repulsive self-interaction increases the maximum mass and the
minimum radius of noninteracting boson stars. Therefore, a
repulsive self-interaction delays the collapse. Eqs. (\ref{e1j}) and
(\ref{e1bj}) also give the quantum corrections to the TF approximation used to
obtain  Eq. (\ref{er3}) \cite{colpi,chavharko}. We see that quantum corrections
 increase the maximum mass and minimum radius with respect to the TF
approximation. The maximum compactness ${\cal C}=GM_{\rm max}/R_* c^2$ can be
obtained as a function of the scattering length $a_s$ from Eqs.
(\ref{e1})-(\ref{e1bj}) [see Eq. (\ref{uk5})]. It goes from
${\cal
C}=0.105$  in the
noninteracting case ($a_s=0$) to ${\cal C}=0.16$ 
in the TF limit ($a_s\rightarrow +\infty$).

For boson stars with an attractive self-interaction  ($a_s\le 0$), like axions,
we
have obtained the following  formulas [see Eqs. (\ref{ru3}) and
(\ref{ru4})] 
\begin{eqnarray}
\label{dm16fq}
M_{\rm max}=0.633\, \frac{\hbar c}{Gm}\frac{1}{\sqrt{1+0.391\frac{|a_s|
c^2}{Gm}}},
\end{eqnarray}
\begin{eqnarray}
\label{dm17hq}
R_*=6.03\, \frac{\hbar}{mc}\sqrt{1+0.833\frac{|a_s| c^2}{Gm}},
\end{eqnarray}
or, equivalently,
\begin{eqnarray}
\label{dm16i}
M_{\rm max}=1.012\, \frac{\hbar}{\sqrt{Gm |a_s|}}\frac{1}{\sqrt{1+2.56
\frac{Gm}{|a_s|c^2}}},
\end{eqnarray}
\begin{eqnarray}
\label{dm17k}
R_*=5.5\, \left (\frac{ |a_s|\hbar^2}{Gm^3}\right )^{1/2}\sqrt{1+1.20
\frac{Gm}{|a_s|c^2}}. 
\end{eqnarray}
These expressions interpolate between
the general relativistic 
maximum mass and minimum radius [see Eq. (\ref{er1})] of noninteracting bosons
stars \cite{kaup,rb}  and the nonrelativistic maximum mass and
minimum radius  [see Eqs.  (\ref{dm16}) and (\ref{dm17})] of
boson stars with an attractive self-interaction \cite{prd1}. The maximum mass is
due to general relativity in the strong gravity regime when   $|a_s|\ll r_S$
and  to the attractive self-interaction $+$ general relativity in the weak
gravity regime when $|a_s|\ll r_S$.
We expect these
approximate results to be more 
accurate for $|a_s|\gg r_S=Gm/c^2$ than for $|a_s|\ll r_S$.
We see that an attractive self-interaction decreases the maximum mass and
increases the minimum
radius of noninteracting boson stars.  Therefore, an attractive 
self-interaction favors the collapse. Eqs. (\ref{dm16i}) and (\ref{dm17k}) also
give the relativistic corrections to the nonrelativistic results
from
Eqs. (\ref{dm16}) and (\ref{dm17}) \cite{prd1,prd2}. We see that
relativistic corrections reduce the maximum mass and increase the
minimum radius with respect to the nonrelativistic approximation. The maximum
compactness ${\cal C}=GM_{\rm max}/R_* c^2$ can be obtained as a function of the
scattering length $a_s$ from Eqs. (\ref{dm16fq})-(\ref{dm17k}) [see Eq.
(\ref{ru5})].
It goes from ${\cal C}=0.105$  in the noninteracting case ($a_s=0$) to ${\cal
C}=0$ in 
the nonrelativistic limit ($a_s\rightarrow -\infty$).

\section{Black hole or bosenova?}
\label{sec_bhb}

In the previous sections, we have expressed the maximum mass of boson stars in
terms of the
scattering length $a_s$ of the bosons. In the present section, we 
reformulate these results in terms of the dimensionless self-interaction
constant $\lambda$ or in terms of the axion decay constant $f$ (see Appendices
\ref{sec_ra} and \ref{sec_rrsf}). Then, we discuss whether the collapse
of the boson star above the maximum mass leads to   a black hole or a
bosenova. We introduce relevant transition scales separating these two regimes.

\subsection{Maximum mass}

The dimensionless self-interaction constant and the axion
decay constant  (when $a_s<0$) are defined by\footnote{Depending on whether we
consider a real or a complex SF, there may be a multiplicative factor $2/3$ in
the expression of $\lambda$ [see Eq. (\ref{sc3})]. We adopt here the
definitions from Eqs. (\ref{lambda}) and (\ref{f})  in order to be
consistent with our previous papers.}
\begin{eqnarray}
\frac{\lambda}{8\pi}=\frac{a_s m c}{\hbar},
\label{lambda}
\end{eqnarray}
\begin{eqnarray}
f=\left (\frac{\hbar c^3 m}{32\pi |a_s|}\right
)^{1/2}=\frac{mc^2}{2\sqrt{|\lambda|}}.
\label{f}
\end{eqnarray}

In the noninteracting limit, the maximum mass of boson stars due to general
relativity [see Eq. (\ref{er1})] can be written as
\begin{equation}
\label{er1wy}
M_{\rm max,NI}^{\rm GR}=0.633\, \frac{M_P^2}{m}.
\end{equation}
For $m\sim 1\, {\rm GeV/c^2}$ (nucleon mass) we get $M_{\rm
max,NI}^{\rm GR}\sim 8.46\times 10^{-20}\, M_{\odot}$ (mini boson star). The
maximum mass becomes comparable to the solar mass $M_{\rm max,NI}^{\rm GR}\sim
M_{\odot}$ for $m\sim 10^{-10}\,
{\rm eV/c^2}$.

On the other hand, for boson stars with a repulsive self-interaction, the
maximum mass due to general relativity  can be written, in the TF
approximation,  as [see Eq. (\ref{er3})]
\begin{eqnarray}
\label{mtf}
M_{\rm max,TF}^{\rm GR}=0.0612\,\sqrt{\lambda}\frac{M_P^3}{m^2}.
\end{eqnarray}
Following Colpi {\it et al.} \cite{colpi} we introduce the rescaled 
dimensionless self-interaction constant
\begin{equation}
\Lambda=\frac{\lambda}{4\pi} \frac{M_P^2}{m^2}
\label{bigl}
\end{equation}
in terms of which
\begin{eqnarray}
M_{\rm max,TF}^{\rm GR}=0.217\,\sqrt{\Lambda}\frac{M_P^2}{m}.
\label{mtfg}
\end{eqnarray}
The TF approximation is valid when
\begin{eqnarray}
\Lambda\gg 1 \qquad {\rm i.e.} \qquad \lambda\gg \left (\frac{m}{M_P}\right )^2.
\end{eqnarray}
Since $m\ll M_P$ in general, the TF approximation is valid even when
$\lambda\sim 1$ or smaller.\footnote{
Conversely, the self-interaction is negligible if
$\Lambda\ll 1$ i.e. $\lambda\ll (m/M_P)^2$. Since $m\ll M_P$, the
self-interaction can be neglected only if it is extraordinarily tiny.
For example, for a boson mass $m\sim 10^{-22}\, {\rm eV/c^2}$,
the self-interaction can be neglected only if $|\lambda|\ll 10^{-100}$ (!).} We
see
that when $\Lambda\gg 1$ the maximum mass $M_{\rm max,TF}^{\rm GR}\sim
\sqrt{\Lambda}{M_P^2}/{m}$ of self-interacting boson stars is much larger than
the maximum mass $M_{\rm max,NI}^{\rm GR}\sim {M_P^2}/{m}$ of noninteracting
boson
stars. They differ by a factor $\sqrt{\lambda}{M_P}/{m}$. Written under the
form $M_{\rm max,TF}^{\rm GR}\sim
\sqrt{\lambda}{M_P^3}/{m^2}$, we see that, when $\lambda\sim 1$, the maximum
mass of self-interacting boson stars is of the order of the Chandrasekhar
mass $M_{\rm Chandra}\sim {M_P^3}/{m^2}$ for fermion stars. For
$m\sim 1\, {\rm
GeV/c^2}$ ($\sim$ nucleon mass) the maximum mass $M_{\rm
max,TF}^{\rm GR}\sim M_{\odot}$ is of the order of the solar mass  (massive
boson stars).  For $m\sim 1\, {\rm MeV/c^2}$ ($\sim$ electron mass) the maximum
mass $M_{\rm max,TF}^{\rm GR}\sim 10^6\, M_{\odot}$  is of the order of the mass
of SMBHs in AGNs. For  $m\sim
100\, {\rm GeV/c^2}$ ($\sim$ Higgs mass) we get $M_{\rm
max,TF}^{\rm GR}\sim 10^{-4}\, M_{\odot}$. In all these examples, the TF
approximation is justified because $\lambda\sim 1\gg (m/M_P)^2\sim 10^{-38}$,
$10^{-44}$ and $10^{-34}$, respectively.

For boson stars with an arbitrary repulsive self-interaction, we
can write the maximum mass   [see Eq. (\ref{uk3})] as
\begin{equation}
M_{\rm max}=0.633\, \frac{M_P^2}{m}\sqrt{1+0.117\, \Lambda}.
\label{e1jj}
\end{equation}
When $\Lambda=0$ we recover Eq. (\ref{er1wy}) and when $\Lambda\gg 1$ we recover
Eq. (\ref{mtfg}).
Interestingly, our analytical formula (\ref{e1jj}) is consistent with the
formula
$M_{\rm max}=(2/\pi)\sqrt{1+\Lambda/8}M_P^2/m$
obtained by Mielke et Schunck \cite{sm2000,sm2003} from different
considerations.

For axion stars with an attractive self-interaction, the maximum mass due to
the self-interaction can be written, in the nonrelativistic
limit, as [see Eq. (\ref{dm16})] 
\begin{eqnarray}
\label{dm16l}
M_{\rm max}^{\rm NR}=5.07\, \frac{M_P}{\sqrt{|\lambda|}}=10.1\, \left
(\frac{f^2\hbar}{c^3m^2G}\right )^{1/2}
\end{eqnarray}
or, using Eq. (\ref{bigl}), as
\begin{eqnarray}
\label{dm16ll}
M_{\rm max}^{\rm NR}=1.43\, \frac{M_P^2}{m\sqrt{|\Lambda|}}.
\end{eqnarray}
The nonrelativistic limit is valid when
\begin{equation}
|\Lambda|\gg 1 \quad {\rm i.e.} \quad |\lambda|\gg \left (\frac{m}{M_P}\right
)^2
\end{equation}
Since $m\ll M_P$ in general, the nonrelativistic approximation is valid even
when
$|\lambda|\sim 1$ or smaller (see footnote 28 with $\Lambda$ replaced by
$|\Lambda|$). We see
that when $|\Lambda|\gg 1$ the maximum mass $M_{\rm max}^{\rm NR}\sim
{M_P^2}/{m\sqrt{|\Lambda|}}$ of dilute axion stars with an attractive
self-interaction is much smaller than
the mass $M_{\rm max,NI}^{\rm GR}\sim {M_P^2}/{m}$ of noninteracting boson
stars. They differ by a factor $|\lambda|^{-1/2} m/{M_P}$. Written under the
form $M_{\rm max}^{\rm NR}\sim
{M_P}/{\sqrt{|\lambda|}}$, we see that, when $|\lambda|\sim 1$, the maximum
mass of dilute axion stars is of the order of the Planck mass. For QCD axions
with  $m\sim 10^{-4}\, {\rm eV}/c^2$ and $a_s\sim
-5.8\times 10^{-53}\, {\rm m}$
(corresponding to $\lambda\sim -7.39\times
10^{-49}$ and $f=5.82\times 10^{10}\, {\rm GeV}$), we get
$M_{\rm max}^{\rm NR}=6.46\times 10^{-14}\, M_{\odot}$ 
and $R_{99}^*=227\, {\rm
km}$ (axteroids). For ULAs with $m=2.92\times 10^{-22}\, {\rm eV}/c^2$ and
$a_s=-3.18\times
10^{-68}\, {\rm fm}$ (corresponding to 
$\lambda=-1.18\times 10^{-96}$ and $f=1.34\times 10^{17}\,
{\rm GeV}$) predicted in \cite{jeansapp} we get $M_{\rm
max}^{\rm
NR}=5.10\times 10^{10}\,
M_{\odot}$ and
$R_{99}^*=1.09\, {\rm pc}$ (DM cores). In these two examples, the
nonrelativistic approximation is justified because $|\lambda|\gg
(m/M_P)^2\sim 10^{-64}$ and $10^{-100}$, respectively (the second
approximation is marginally valid).

For dilute axion stars with an arbitrary attractive self-interaction, we
can write the maximum mass [see Eq.
(\ref{ru3})] as
\begin{equation}
M_{\rm max}=0.633\, \frac{M_P^2}{m}\frac{1}{\sqrt{1-0.196\, \Lambda}}.
\label{e1g}
\end{equation}
When $\Lambda=0$ we recover Eq. (\ref{er1wy}) and when $|\Lambda|\gg 1$ we
recover Eq. (\ref{dm16ll}). We note that the  formula of  Mielke et Schunck
\cite{sm2000,sm2003} quoted above is not valid when $\Lambda<0$.

\subsection{Transition scales}

We introduce the
transition scales
\begin{equation}
(a_s)_t=\frac{2Gm}{c^2}=r_S,
\label{t1}
\end{equation}
\begin{equation}
\frac{\lambda_t}{16\pi}=\frac{Gm^2}{\hbar c}=\left (\frac{m}{M_P}\right )^2,
\label{t1b}
\end{equation}
\begin{equation}
f_t=\frac{M_P c^2}{8\sqrt{\pi}}\sim 10^{18}\, {\rm GeV}.
\label{t1c}
\end{equation}
We note that $(a_s)_t$ is of the order of the 
Schwarzschild radius of the boson $r_S=2Gm/c^2$, $\lambda_t$ is of the order of
the gravitational coupling constant $\alpha_g=Gm^2/\hbar c=(m/M_P)^2$ and $f_t$
is of the order of the Planck mass energy $M_P c^2\sim
10^{19}\, {\rm GeV}$ (note that it is independent of the mass $m$ of the boson).

\begin{figure}[!h]
\begin{center}
\includegraphics[clip,scale=0.3]{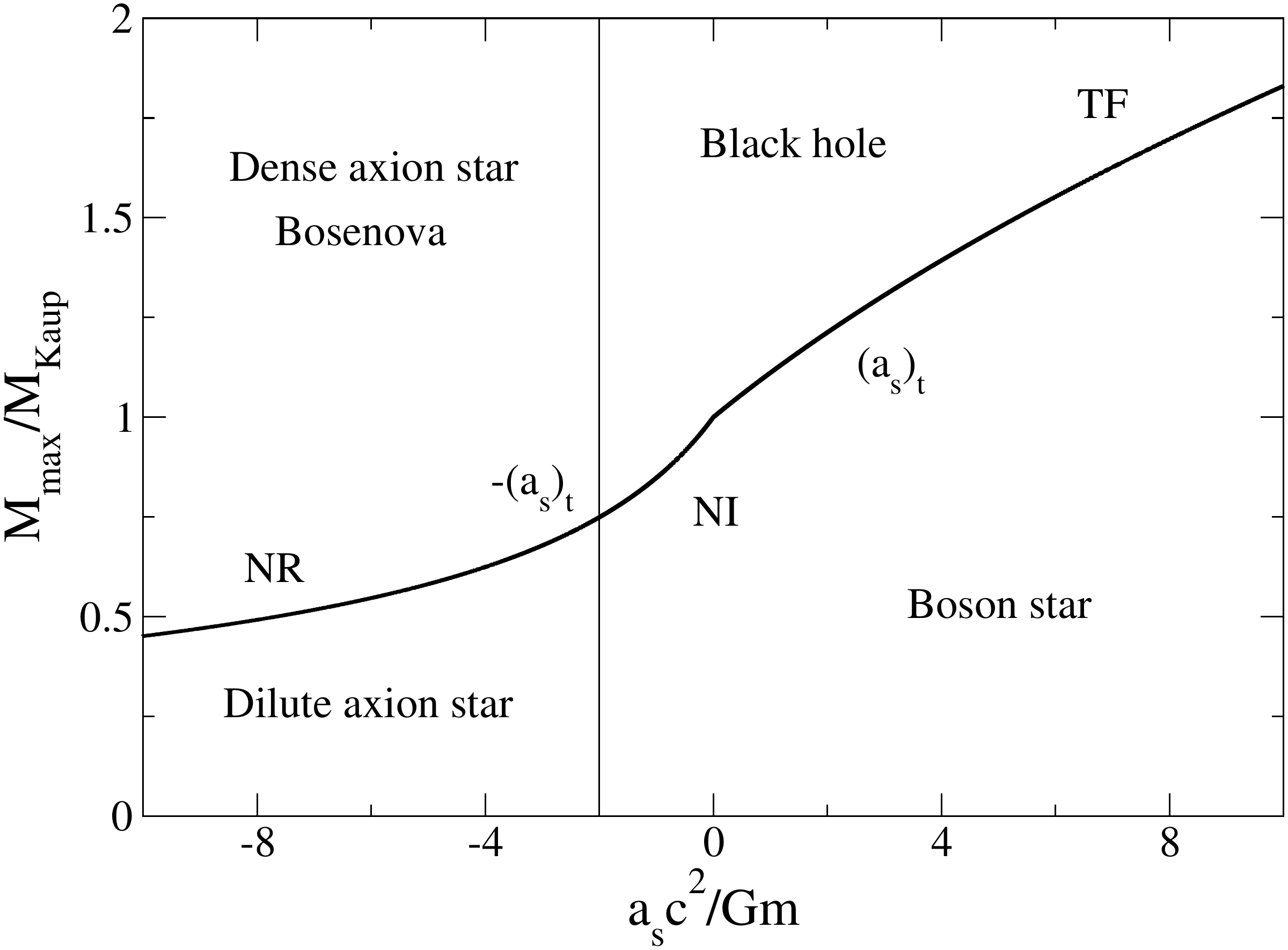}
\caption{Phase diagram of boson stars with a
$|\varphi|^4$ self-interaction (see text for details). We have represented the
maximum mass above which the boson star becomes unstable as a function of the
scattering length [see Eqs. (\ref{uk3}) and (\ref{ru3})]. This
maximum mass is $M_{\rm max,NI}^{\rm GR}$ when $a_s=0$, $M_{\rm max,TF}^{\rm GR}$ when $a_s\gg (a_s)_t$ (repulsive
case) and $M_{\rm max}^{\rm NR}$ when $|a_s|\gg (a_s)_t$ (attractive case).}
\label{massezoom}
\end{center}
\end{figure}

For a repulsive self-interaction ($a_s\ge 0$), the maximum mass $M_{\rm
max}^{\rm GR}$ is due to general
relativity. When $M<M_{\rm max}^{\rm GR}$ the boson star is stable and when
$M>M_{\rm max}^{\rm GR}$ it collapses towards a black hole. The
transition between the
noninteracting regime and the TF regime occurs when $M_{\rm max,NI}^{\rm
GR}\sim
M_{\rm max,TF}^{\rm
GR}$ giving $a_s\sim (a_s)_t$ (i.e.
$\lambda\sim \lambda_t$). When $a_s\ll
(a_s)_t$ (i.e. $\lambda\ll \lambda_t$) we are in the noninteracting regime and
when $a_s\gg (a_s)_t$ (i.e. $\lambda\gg \lambda_t$) we are in the TF regime. In
each case, when the mass overcomes  $M_{\rm max}^{\rm GR}$, a black hole is
formed.

For an attractive self-interaction, the instability may be due to general
relativity or to the self-interaction of the bosons. When
$a_s\simeq 0$ the maximum mass $M_{\rm
max}^{\rm GR}$ is due to general
relativity. When $M<M_{\rm max,NI}^{\rm GR}$ the boson star is stable and when
$M>M_{\rm max,NI}^{\rm GR}$ it collapses towards a black hole. When
$a_s\rightarrow
-\infty$ the maximum
mass $M_{\rm max}^{\rm NR}$ is due to the
attractive self-interaction. When
$M<M_{\rm max}^{\rm NR}$ the dilute axion star is stable and
when $M>M_{\rm max}^{\rm NR}$ it collapses and forms a dense axion
star (stabilized by, e.g., a $|\varphi|^6$ repulsive self-interaction) or
explodes in a
bosenova. The transition
between the noninteracting regime and the
nonrelativistic regime occurs when $M_{\rm max}^{\rm NR}\sim  
M_{\rm max,NI}^{\rm GR}$ giving $|a_s|\sim (a_s)_t$ (i.e. $|\lambda|\sim
\lambda_t$ and $f\sim f_t$). When $|a_s|\ll
(a_s)_t$ (i.e. $|\lambda|\ll \lambda_t$ or $f\gg f_t$) we are in the
noninteracting regime.  In that case, when the mass overcomes 
$M_{\rm max,NI}^{\rm
GR}$, a black hole is formed. When $|a_s|\gg (a_s)_t$ (i.e. $|\lambda|\gg
\lambda_t$ or
$f\ll f_t$) we are in the nonrelativistic regime.  In that case, when the mass
overcomes  $M_{\rm max}^{\rm NR}$, a dense axion star or a bosenova is formed.

A phase diagram presenting these different possibilities is shown in Fig.
\ref{massezoom}.

\subsection{Triple point}

\begin{figure}[!h]
\begin{center}
\includegraphics[clip,scale=0.3]{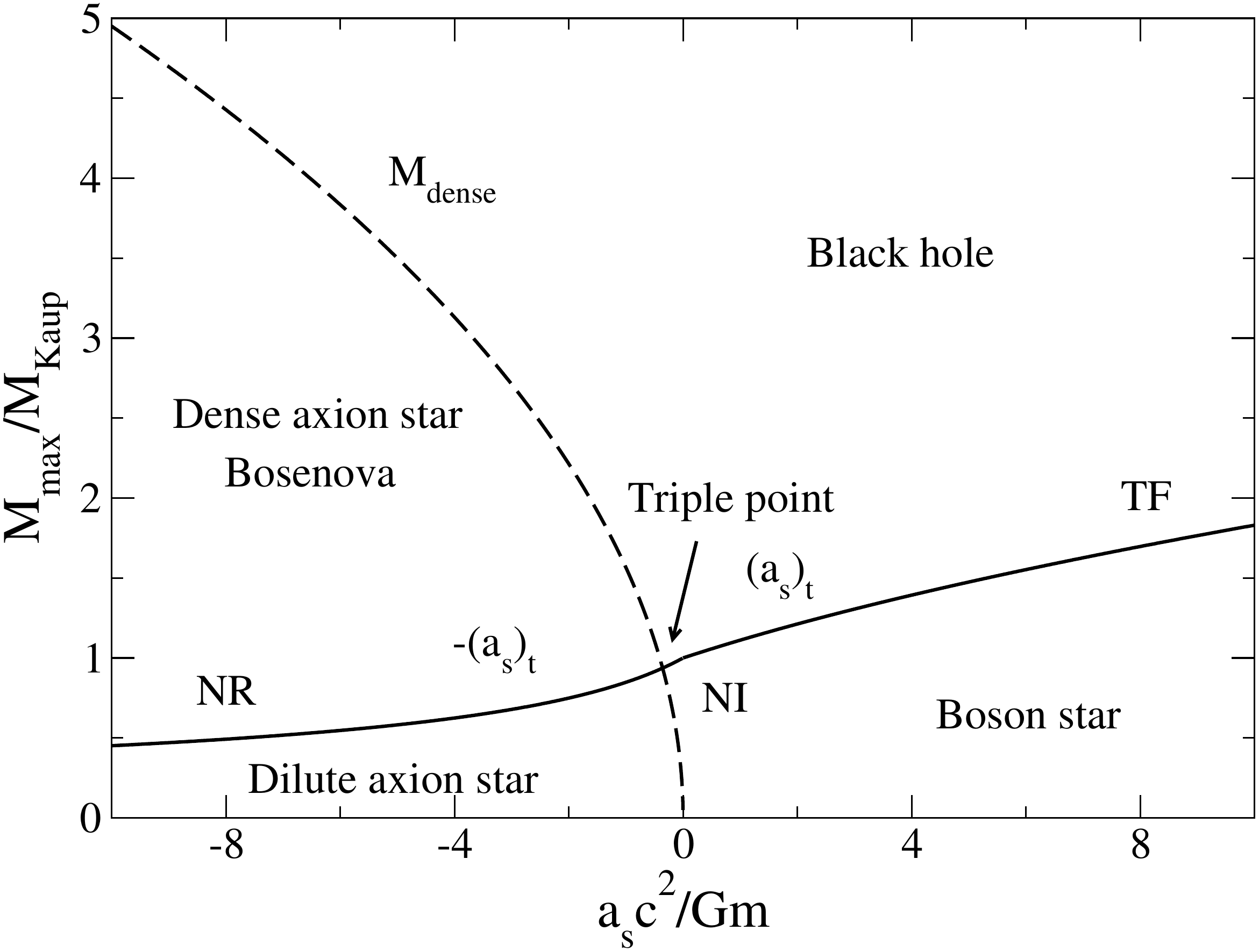}
\caption{Phase diagram of boson stars with a
$|\varphi|^4$ and a $|\varphi|^6$ self-interaction (see text for details). It
displays a triple point at $\sim
[-r_S,M_{\rm Kaup}]$ separating boson stars, black holes, and dense axion
stars or bosenova. With respect to Fig. \ref{massezoom} we have added the
mass $M_{\rm max,dense}^{\rm GR}$ above which dense axion stars ($a_s<0$) become
general relativistically
unstable and collapse towards a black hole (the corresponding curve is $M_{\rm
max,GR}^{\rm
dense}/M_{\rm Kaup}=1.56\, (|a_s|c^2/Gm)^{1/2}$). This
figure can be compared to Fig. 31 of \cite{phi6}.}
\label{triple}
\end{center}
\end{figure}

A repulsive  $|\varphi|^6$ self-interaction can stabilize a collapsing axion
star above the mass $M_{\rm max}^{\rm NR}$ and allow the formation of a dense
axion star.
However, a dense axion star can itself undergo a gravitational
instability of general relativistic origin above a maximum mass estimated to be
\cite{phi6,tunnel}:
\begin{eqnarray}
\label{dm16der}
M_{\rm max,dense}^{\rm GR}=0.991\, \left
(\frac{|a_s|\hbar^2c^4}{G^3m^3}\right )^{1/2}
\end{eqnarray}
i.e.
\begin{equation}
\label{dm16db}
M_{\rm max,dense}^{\rm GR}=0.0988\, \left
(\frac{\hbar^3c^7}{G^3f^2m^2}\right )^{1/2}=0.198\,
\sqrt{|\lambda|}\frac{M_P^3}{m^2}.
\end{equation}
This mass presents the same scaling as the maximum mass of a boson star in the
TF regime [see Eq. (\ref{er3})] except that, in the present case, $a_s<0$. We
note
that
$M_{\rm max,dense}^{\rm GR}\sim  
M_{\rm max,NI}^{\rm GR}$ when $|a_s|\sim
(a_s)_t$ (i.e. when $|\lambda|\sim
\lambda_t$ and $f\sim f_t$). This leads to a triple point at $\sim
[-r_S,M_{\rm Kaup}]$ separating boson
stars, black holes and dense axion stars or bosenova (see Fig.
\ref{triple}). This triple point was obtained numerically in \cite{helfer}
and reproduced analytically (qualitatively) in
\cite{phi6}. When $a_s>-(a_s)_t$ we expect to observe a boson star for
$M<M_{\rm max}^{\rm GR}$ and a black hole for $M>M_{\rm
max}^{\rm GR}$. When $a_s<-(a_s)_t$ we expect to observe a dilute axion star
for
$M<M_{\rm max}^{\rm NR}$,  a  dense axion star or a bosenova for
$M_{\rm max}^{\rm NR}<M<M_{\rm max,dense}^{\rm GR}$, and a black hole for
$M>M_{\rm
max,dense}^{\rm GR}$.

\section{Application to DM halos and inflaton
clusters: Core mass -- halo mass relation}
\label{sec_cmhm}

We now apply our results to DM halos and inflaton clusters.

The results of Secs. \ref{sec_sgbec}-\ref{sec_bhb} describe
the ground state of a self-gravitating BEC. In cosmology, they characterize a
``minimum halo'' of mass $(M_h)_{\rm min}$ which is a purely condensed object
without atmosphere.
Larger DM
halos of mass $M_h>(M_h)_{\rm min}$ have a
``core-halo'' structure
with a quantum core (soliton) in its ground state and an approximately
isothermal atmosphere which arises from the quantum interferences of excited
states.
This
core-halo structure has been explained in Refs.
\cite{ggpp,moczSV,modeldm,wignerPH} by the process of gravitational
cooling \cite{seidel94} and the theory of
violent relaxation \cite{lb,csr}. The
results of Secs. \ref{sec_sgbec}-\ref{sec_bhb} also
apply to
the quantum core of large DM halos. 
The mass $M_c$ of the quantum core increases
with
the halo mass $M_h$ (see Fig. \ref{mhmcnorm}). For noninteracting bosons and for
bosons with
a repulsive
self-interaction we may wonder if the core mass can reach the maximum mass
$M_{\rm max}^{\rm GR}$ [see Eqs. (\ref{er1}) and (\ref{er3})] set by general
relativity and collapse towards a black hole. For bosons with an attractive
self-interaction we may wonder if the core mass can reach the Newtonian maximum
mass $M_{\rm max}^{\rm NR}$  [see Eq. (\ref{dm16})] and collapse towards a
dense axion star (soliton) or a black hole, or
explode in a bosenova
\cite{braaten,cotner,bectcoll,ebycollapse,tkachevprl,helfer,phi6,visinelli,
moss}.

\subsection{General formalism}
\label{sec_gf}

In Refs. \cite{modeldm,mcmh,mcmhbh,jeansapp}, we have derived the core mass
-- halo mass relation of DM halos  (without or with the presence of a central
black hole) from a thermodynamic approach. We have
obtained a general relation $M_c(M_h)$ valid for noninteracting bosons as well
as for
bosons with a repusive or an attractive self-interaction (and
for fermions). To obtain this relation we have proceeded in three steps:

(i) We have first
shown
\cite{modeldm,mcmh} that the
maximization of the Lynden-Bell entropy (justified by the
theory of violent relaxation \cite{lb,wignerPH}) at fixed mass and energy leads
to the
``velocity
dispersion tracing'' relation according to which the velocity dispersion in the
core $v_c^2\sim GM_c/R_c$ is equal to the velocity dispersion in the
halo $v_h^2\sim GM_h/r_h$. This relation can be written as
\begin{equation}
\label{cmhm1}
v_c\sim v_h \qquad \Rightarrow \qquad \frac{M_c}{R_c}\sim \frac{M_h}{r_h}.
\end{equation}

(ii) To determine the core mass-radius relation $M_c(R_c)$ we have used a
Gaussian ansatz yielding \cite{prd1}
\begin{equation}
\label{cmhm2}
M_c=\frac{3.76\, \frac{\hbar^2}{Gm^2R_c}}{1-3\frac{
a_s\hbar^2}{Gm^3 R_c^2}}
\end{equation}
or, equivalently,
\begin{equation}
\label{cmhm2r}
R_c=1.87\ \frac{\hbar^2}{Gm^2M_c}\left (1\pm\sqrt{1+0.849\,
\frac{Gma_sM_c^2}{\hbar^2}}\right ).
\end{equation}

(iii) To determine the halo mass-radius relation $M_h(r_h)$ we have assumed
that the atmosphere is isothermal,\footnote{The fact that the
atmosphere should be isothermal is justified in Ref. \cite{wignerPH} by the
Lynden-Bell \cite{lb} theory of violent relaxation. An
(approximately) isothermal atmosphere seems to be validated by numerical
simulations showing density profiles decreasing as $r^{-2}$ \cite{moczSV}
and exhibiting a Maxwellian velocity distribution \cite{veltmaat,sgbne,esne}.}
and we have used the fact
that the surface density $\Sigma_0$ of the DM halos is universal. This
leads to the halo mass-radius relation \cite{modeldm}
\begin{equation}
\label{cmhm3}
M_h=1.76\,\Sigma_0 r_h^2.
\end{equation}

Combining
Eqs. (\ref{cmhm1})-(\ref{cmhm3}), we obtain the core mass -- halo mass relation
$M_c(M_h)$ under the form
\begin{equation}
\label{mcmhexp}
M_c=2.23\, \frac{\hbar\Sigma_0^{1/4}M_h^{1/4}}{G^{1/2}m}\left
(1+1.06\, \frac{a_s}{m}\Sigma_0^{1/2}M_h^{1/2}\right )^{1/2}.
\end{equation}
As we mentioned above, the ``minimum halo'' is a  purely condensed object
without
atmosphere. Writing $M_c=M_h$, meaning that the quantum core contains all the
mass (i.e. there is no isothermal halo around it), we obtain 
\begin{equation}
\label{minhexp}
(M_h)_{\rm min}^{3/2}=4.99\, \frac{\hbar^{2}\Sigma_0^{1/2}}{Gm^{2}}\left
(1+1.06\, \frac{a_s}{m}\Sigma_0^{1/2}(M_h)_{\rm min}^{1/2}\right ).
\end{equation}
This equation determines the mass $(M_h)_{\rm min}$ of the minimum halo as a
function of $m$ and $a_s$. Alternatively, for a given value of $(M_h)_{\rm min}$
deduced from the observations, Eq. (\ref{minhexp})
determines a constraint between $m$ and $a_s$.

\begin{figure}[!h]
\begin{center}
\includegraphics[clip,scale=0.3]{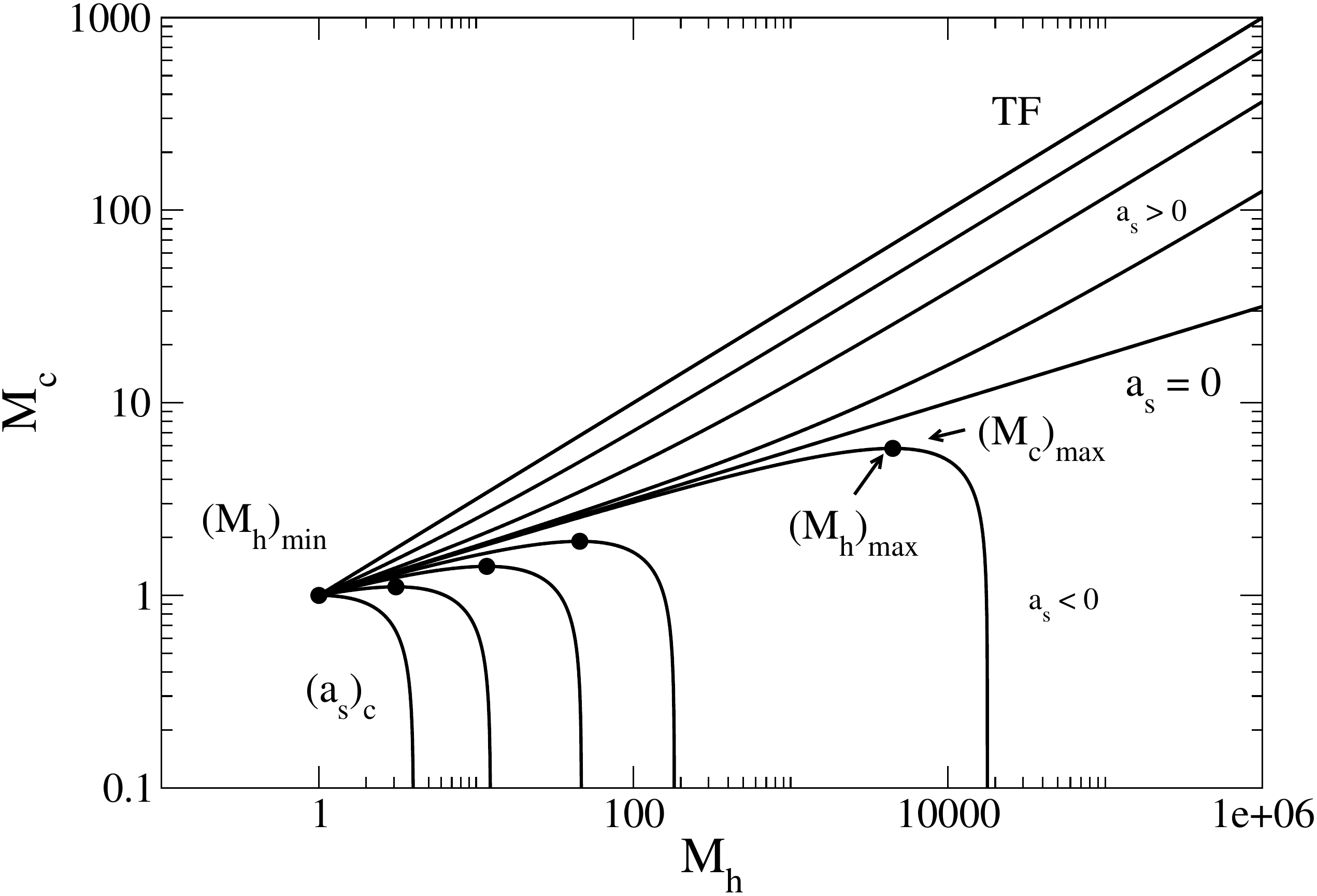}
\caption{Core mass $M_c$ as a function
of the halo mass $M_h$ for different
values of $a_s$ [see Eqs. (\ref{cmhm6}) and (\ref{cmhm7})]. The mass is
normalized by $(M_{h})_{\rm
min}$. We have highlighted the curve corresponding to the
noninteracting case
$a_s=0$ [see Eq. (\ref{cmhm9})], and the curve corresponding to the TF
limit
$a_s/a'_*\gg 1$ [see Eq. (\ref{cmhm11})]. When the self-interaction is
attractive ($a_s<0$), the quantum core becomes unstable when it
reaches the maximum mass $(M_c)_{\rm max}$ [see Eq. (\ref{cmhm13})]. This
occurs in a halo of mass $(M_h)_{\rm max}$ [see Eq. (\ref{cmhm14})].}
\label{mhmcnorm}
\end{center}
\end{figure}

Universal relations can be obtained by introducing an appropriate
normalization \cite{mcmh,mcmhbh,jeansapp}.
For a given
value of the minimum
halo mass $(M_h)_{\rm min}$  we introduce  the mass
scale
\begin{equation}
\label{cmhm4}
m_0=2.23\frac{\hbar\Sigma_0^{1/4}}{
G^ {1/2}(M_h)_{\rm min}^{3/4}}
\end{equation}
and the self-interaction scale
\begin{equation}
\label{cmhm5}
a'_*=2.11\frac{\hbar}{G^{
1/2}\Sigma_0^{1/4}(M_h)_{\rm min}^{5/4}}.
\end{equation}
With these
scales, the normalized DM particle mass -- scattering length relation
(\ref{minhexp}) can be written as 
\begin{equation}
\label{cmhm6}
\frac{a_s}{a'_*}=\left (\frac{m}{m_0}\right )^3-\frac{m}{m_0}
\end{equation}
and the normalized  core mass -- halo mass relation (\ref{mcmhexp}) can be
written
as
\begin{eqnarray}
\label{cmhm7}
\frac{M_c}{(M_h)_{\rm min}}&=&\frac{m_0}{m}\left (\frac{M_h}{(M_h)_{\rm
min}}\right )^{1/4}\nonumber\\
&\times&\sqrt{1+\frac{m_0}{m} \frac{a_s}{a'_*}
\left (\frac{M_h}{(M_h)_{\rm min}}\right )^{1/2}}.
\end{eqnarray}
This leads to the universal curves plotted in
Fig. \ref{mhmcnorm}. The only
input is the DM particle mass $m$ or, equivalently, its scattering length
$a_s$ (they are related to each other by Eq. (\ref{cmhm6})). The minimum halo
mass $(M_h)_{\rm min}$ obtained from the observations determines the scales
$m_0$ and
$a'_*$.

The above relations were derived in the context of DM
halos \cite{modeldm,mcmh,mcmhbh,jeansapp}. In that case, the
universal surface density infered from the observations is
$\Sigma_0=141\, M_{\odot}/{\rm pc}^2$ \cite{kormendy,spano,donato}. On the other
hand, it is an observational evidence that there is no DM halo with a mass
smaller than $M_h\sim 10^7-10^8\, M_{\odot}$. Taking a minimum halo
mass $(M_h)_{\rm min}=10^8\,
M_{\odot}$ to fix the ideas, we get $m_0=2.25\times 10^{-22}\, {\rm eV}/c^2$ and
$a'_*=4.95\times 10^{-62}\, {\rm fm}$. Then, for a given value of the scattering
length $a_s$, we obtain the boson mass $m$ from Eq. (\ref{cmhm6}) and the core
mass -- halo mass relation $M_c(M_h)$ from Eq. (\ref{cmhm7}). A detailed
discussion of the core mass -- halo mass relation of DM halos has been given in
\cite{modeldm,mcmh,mcmhbh,jeansapp}. For a
repulsive
self-interaction (or no self-interaction) it is found that, in realistic DM
halos ($M_h<10^{14}\, M_{\odot}$), the quantum core mass $M_c$ is always much
smaller than $M_{\rm
max}^{\rm GR}$. Therefore, it cannot collapse towards a black hole.\footnote{For
noninteracting 
bosons or for bosons with a repulsive self-interaction, we find that the core
mass can reach the critical mass (\ref{er1}) or (\ref{er3}) only in halos of
mass larger than $M_h\sim 10^{-2}c^4/G^2\Sigma_0\sim 10^{22}\, M_{\odot}$ which
are not realistic (the biggest DM halos have a mass  $M_h\sim 10^{14}\,
M_{\odot}$). Remarkably, this result is independent of the characteristics
(mass, scattering length...) of the DM particle (see Appendix C of
\cite{mcmhbh}). We conclude that the
quantum
core is always stable in practice and that it cannot collapse towards a
black hole. The fact that $M_c\ll M_{\rm max}^{\rm GR}$ also
justifies that we use a nonrelativistic approach.} For an
attractive self-interaction, and for values of $f$ in the range $10^{16}\, {\rm
GeV}\le f\le 10^{18}\, {\rm
GeV}$ predicted by particle physics and cosmology \cite{hui}, it is found that
the quantum core (soliton)
is always stable ($M_c<M_{\rm max}^{\rm NR}$) in  realistic DM
halos of mass $M_h<10^{14}\, M_{\odot}$.\footnote{For bosons
with an attractive self-interaction, we find that the quantum core mass
can reach the critical mass (\ref{dm16}) in realistic halos (of mass $M_h\le 
10^{14}\, M_{\odot}$) provided that the axion decay constant $f$ is smaller than
$f_c=4.22\times 10^{15}\, {\rm GeV}$. Remarkably, this result is independent of
the DM particle mass \cite{mcmh,jeansapp}.
However, such small
values of $f$ seem to be excluded by constraints from cosmology and particle
physics
which place $f$ in the range $10^{16}\, {\rm
GeV}\le f\le 10^{18}\, {\rm
GeV}$. We conclude that the quantum core is always stable in practice and that
it cannot collapse towards a dense axion star or a black hole, or explode in a
bosenova.} 
However, we must be careful about 
observational constraints on a DM particle whose precise nature has not yet
been established. If $f$ can be smaller than $10^{15}\, {\rm
GeV}$ (a possibility to consider), the soliton can become unstable in
sufficiently large realistic DM
halos and collapse. For example, for $m\sim 10^{-22}\, {\rm eV/c^2}$ and $f\sim
10^{14}\, {\rm GeV}$ (corresponding to $|\lambda|\sim 10^{-91}$) we find a
maximum core mass  $M_{\rm max}^{\rm NR}\sim 10^8\, M_{\odot}$. This value
is of the same order as the typical mass of the minimum halo $(M_h)_{\rm
min}\sim 10^8\, M_{\odot}$ implying that  the quantum core of all the DM halos
($M_h\ge (M_h)_{\rm
min}$) would be  unstable in that case.
Since $f\ll M_P
c^2\sim 10^{19}\,  {\rm GeV}$, the collapse of the quantum core leads
to a dense soliton or a bosenova, not to a black hole (see Sec. \ref{sec_bhb}).

A similar discussion has been given by Padilla {\it et al.}
\cite{prsmv} in the case of DM halos by using the theoretical formalism
described
previously 
\cite{modeldm,mcmh,mcmhbh,jeansapp}. Recently, Padilla {\it et al.}
\cite{padilla} applied the same formalism to inflaton
clusters
\cite{mhe,ne,ene,esne}. In particular, they considered the possibility that the
self-interaction between bosons is attractive and that the inflaton stars
collapse towards a black hole when they reach the maximum mass $M_{\rm max}^{\rm
NR}$ [see Eq. (\ref{dm16})]. Below, we complement their results and also treat
the case of a repulsive self-interaction.

The general results of Eqs. (\ref{cmhm1})-(\ref{cmhm7}) remain valid in the
case of inflaton clusters except for a
change of scales. We thus have to determine the relevant scales for this
problem. Niemeyer and Easther \cite{ne} adopt a boson mass $m=6.35\times
10^{-6}\, M_{P}$ based on the earlier work of
Ref. \cite{efg}. In their numerical simulations, Eggemeier {\it et
al.} \cite{ene} find that inflaton halos of mass $M_h\sim 20\, {\rm kg}$ and
radius $r_h\sim 10^{-20}\, {\rm m}$ form roughly $10^{-24}\, {\rm s}$ after the
big bang. They also observe clusters of mass $M_h\sim 0.01\, {\rm kg}$ and size
$r_h\sim 10^{-22}\, {\rm m}$.  From
these values one obtains the typical surface densities
$\Sigma\sim 2\times 10^{44}\, {\rm g/m^2}$ and $\Sigma\sim
10^{45}\, {\rm g/m^2}$ respectively. These results are consistent with a constant surface
density\footnote{If the surface density of the inflaton clusters is not
constant, we should replace $\Sigma_0$ by $M_h/(1.76 r_h^2)$ in Eqs.
(\ref{mcmhexp}) and (\ref{minhexp}). In that case, the core mass $M_c$ depends
on $M_h$ and $r_h$
individually. For simplicity, we will assume below that $\Sigma_0$ is constant.}
but its value  is considerably larger than in the case of DM
halos where $\Sigma_0=295\, {\rm g/m^2}$. To be specific we shall take
$\Sigma_0=10^{44}\, {\rm
g/m^2}$. Using Eq. (\ref{cmhm4}) with $m=m_0=6.35\times 10^{-6}\,
M_{P}=1.38\times 10^{-10}\, {\rm
g}$ we obtain
a minimum halo mass $(M_h)_{\rm min}=5.74\times 10^{-5}\, {\rm g}$. There
should be no inflaton cluster below this
threshold value. Then, Eq.
(\ref{cmhm5}) gives $a'_*=1.72\times
10^{-30}\, {\rm m}$ (corresponding to $\lambda'_*=17.0$ and
$f'_*=9.39\times 10^{12}\, {\rm GeV}$).

\subsection{Noninteracting bosons}
\label{sec_appni}

For noninteracting bosons, the core mass -- halo mass relation (\ref{mcmhexp})
reduces to
\begin{equation}
\label{cmhm9}
M_c=2.23\left (\frac{\hbar^4\Sigma_0 M_h}{G^2m^4}\right )^{1/4}.
\end{equation}
In the noninteracting limit, the boson mass is $m=m_0=6.35\times 10^{-6}\,
M_{P}$.
For an inflaton cluster of mass $M_h=1\, {\rm g}$ and radius $r_h=7.54\times 10^{-23}\, {\rm m}$, we
obtain a core mass $M_c=6.59\times 10^{-4}\, {\rm g}$ and a core
radius $R_c=1.32\times 10^{-25}\, {\rm m}$. This theoretical prediction 
is consistent with the numerical results of Eggemeier {\it et
al.} \cite{esne} who find an inflaton star of mass $M_c\sim
10^{-6}\, {\rm kg}$ in a cluster of mass $M_h\sim 10^{-3}\, {\rm kg}$.

The maximum mass and the minimum radius of a noninteracting boson star at $T=0$
set by general relativity are given by Eqs. (\ref{er1}) and (\ref{er2}).
For a boson of mass $m=6.35\times 10^{-6}\, M_{P}$, we
obtain $M_{\rm max,NI}^{\rm GR}=2.17\, {\rm g}$
and $R_{\rm min,NI}^{\rm GR}=1.53\times 10^{-29}\, {\rm m}$. The maximum mass is
much larger than the typical quantum core mass (inflaton star) of inflaton
clusters in the simulations of Eggemeier {\it et
al.} \cite{ene,esne} ($M_c\ll
M_{\rm max,NI}^{\rm GR}$). Therefore, a
nonrelativistic approach is justified in most inflaton clusters. According
to Eq. (\ref{cmhm9}) the mass
of the
soliton becomes equal to the maximum mass ($M_c=M_{\rm max,NI}^{\rm GR}$)  in an
inflaton
cluster of
mass
\begin{equation}
\label{cmhm9g}
(M_h)_{\rm max}^{\rm NI}=6.49\times 10^{-3} \frac{c^4}{G^2\Sigma_0}=1.18\times
10^{14}\, {\rm g}.
\end{equation}
Remarkably, this expression is independent of the boson mass (see Appendix C of
\cite{mcmhbh}). Above that mass, the inflaton star collapses
towards a black hole. This may
provide a new mechanism for PBH formation during reheating
\cite{padilla}.\footnote{We note, however, that very massive
inflaton clusters are required
to form PBHs. Such massive inflaton clusters may not be very numerous (see the
much smaller typical masses obtained in the
simulations of Eggemeier {\it et
al.} \cite{ene,esne}).
Therefore, the quantity of PBHs produced by this mechanism may be small.
Furthermore, PBHs of mass $m\sim 1\, {\rm g}$ rapidly evaporate by Hawking
\cite{hawking} radiation on a timescale $t_{\rm evap}\sim G^2M^3/\hbar c^4\sim
10^{-30}\, {\rm s}$.}  Inflaton stars  in less
massive inflaton
clusters are stable and do not form PBHs.

{\it Remark:} An inflaton cluster would collapse towards a black hole as a whole
if
$r_h<2GM_h/c^2$. Using the mass-radius relation from Eq. (\ref{cmhm3}), this criterion gives $M_h>0.142\, c^4/(G^2\Sigma_0)=2.57\times
10^{15}\, {\rm g}$. This critical mass is one order of magnitude larger than the
critical mass  $(M_h)_{\rm max}^{\rm NI}$ from Eq. (\ref{cmhm9g}) 
above which its core  collapses. This suggests that the quantum core
(soliton) collapses first.

\subsection{Repulsive self-interaction in the TF limit}

For bosons with a repulsive self-interaction in the TF limit, the core mass --
halo mass relation (\ref{mcmhexp}) reduces to 
\begin{equation}
\label{cmhm11}
M_c=2.30\left (\frac{\hbar^2\Sigma_0 a_s M_h}{Gm^3}\right )^{1/2}.
\end{equation}
In the
TF limit $a_s\gg a'_*$, the ratio $a_s/m^3$ is given by 
$a_s/m^3=a'_*/m_0^3=0.654\, {\rm m/g^3}$ [see Eq. (\ref{cmhm6})] (for  $m=6.35\times 10^{-6}\, M_{P}$ this
corresponds to $\lambda=17.0$ and $a_s=1.73\times 10^{-30}\, {\rm m}$).  For an
inflaton cluster of mass $M_h=1\, {\rm g}$, we
obtain a core mass $M_c=7.60\times 10^{-3}\, {\rm g}$
and a
core radius $R_c=1.04\times 10^{-24}\, {\rm m}$.

The maximum mass and the minimum radius of a self-interacting boson star at
$T=0$ in the TF limit set by
general relativity are given by Eqs. (\ref{er3}) and (\ref{er4}).
For a ratio $a_s/m^3=0.654\, {\rm m/g^3}$, we
obtain $M_{\rm
max,TF}^{\rm GR}=1.36\times 10^5\, {\rm g}$
and $R_{\rm min,TF}^{\rm GR}=6.34\times 10^{-25}\, {\rm m}$. The maximum mass is
much larger than the typical quantum core mass of inflaton
clusters ($M_c\ll M_{\rm max,TF}^{\rm GR}$). Therefore, a
nonrelativistic approach is justified in most inflaton clusters.
According to Eq. (\ref{cmhm11}) the
mass
of the
soliton
would be equal to the maximum mass ($M_c=M_{\rm max,TF}^{\rm GR}$)  in an
inflaton
cluster of
mass 
\begin{equation}
\label{cmhm11k}
(M_h)_{\rm max}^{\rm TF}=0.0178 \frac{c^4}{G^2\Sigma_0}=3.23\times 10^{14}\,
{\rm g}.
\end{equation}
Remarkably, this expression is independent of the ratio $a_s/m^3$ (see Appendix
C of
\cite{mcmhbh}). The comments made at the end of Sec.
\ref{sec_appni} also
apply to the present situation. There is, however, one important
difference. The mass of the PBHs formed by this mechanism is much larger than
in the noninteracting case (this is because self-interacting boson stars
\cite{colpi,chavharko} are more massive than noninteracting boson stars
\cite{kaup,rb}). As a result, their evaporation time is longer. A PBH of mass
$m\sim 10^5\, {\rm g}$ evaporates on a 
timescale $t_{\rm evap}\sim G^2M^3/\hbar c^4\sim
10^{-15}\, {\rm s}$. This timescale is still short but could be
significantly increased
(since it depends on the cubic power of $M$) if the self-interaction is
larger.

\subsection{Attractive self-interaction}
\label{sec_asi}

For bosons with an attractive self-interaction, the core mass -- halo mass
relation presents a maximum  (see Fig. \ref{mhmcnorm})  when the core
mass
reaches the
critical value\footnote{Recall that we use here the values of the Gaussian
ansatz in order to be consistent with the general formalism of Sec.
\ref{sec_gf}.}
\begin{eqnarray}
\label{cmhm13}
(M_c)_{\rm max}=1.085 \frac{\hbar}{\sqrt{G m |a_s|}}&=&10.9 \left
(\frac{f^2\hbar}{c^3m^2G}\right
)^{1/2}\nonumber\\
&=&5.44\frac{M_{P}}{\sqrt{|\lambda|}}
\end{eqnarray}
at which it becomes unstable and collapses \cite{prd1,bectcoll}. The collapse of
the core, leading to a dense
axion star (soliton), a black hole (provided that
$M_c>M_{\rm max,dense}^{\rm GR}$) or a bosenova
\cite{braaten,cotner,bectcoll,ebycollapse,tkachevprl,helfer,phi6,visinelli,
moss}, occurs in a  DM halo of mass 
\begin{equation}
\label{cmhm14}
(M_h)_{\rm max}=0.223 \frac{m^2}{a_s^2\Sigma_0}=2255 \frac{f^4}{\hbar^2
c^6\Sigma_0}=141\, \frac{m^4c^2}{\Sigma_0\hbar^2\lambda^2}.
\end{equation}
From Eqs. (\ref{cmhm13}) and (\ref{cmhm14}) we obtain
the relation
\begin{equation}
\label{cmhm13b}
(M_c)_{\rm max}=1.58\left (\frac{\hbar^4\Sigma_0 (M_h)_{\rm max}}{G^2m^4}\right
)^{1/4},
\end{equation}
presenting the same scaling as Eq. (\ref{cmhm9}). Therefore,  in Fig.
\ref{mhmcnorm}, the curve $(M_c)_{\rm max}[(M_h)_{\rm max}]$ connecting the
bullets is close to the curve $M_c(M_h)$ corresponding to $a_s=0$. We note that
the maximum
halo
mass $(M_h)_{\rm max}$ depends only on $f$ while the  maximum core mass
$(M_c)_{\rm
max}$ depends on $f$ and $m$.  On the other hand, the  maximum core mass
$(M_c)_{\rm
max}$ depends only on $\lambda$ while the maximum
halo
mass $(M_h)_{\rm max}$ depends on $\lambda$ and $m$.

According to Eq. (\ref{cmhm14}), the value of $f$ below which the quantum
core of an inflaton halo of mass $M_h=1\, {\rm g}$ collapses is $f_{\rm
crit}=1.53\times 10^{14}\,
{\rm GeV}$ (for  $m=6.35\times 10^{-6}\, M_{P}$ this
corresponds to $\lambda_{\rm crit}=-6.42\times 10^{-2}$ and $(a_s)_{\rm
crit}=-6.50\times 10^{-33}\, {\rm m}$). In
that case, the critical core mass is $(M_c)_{\rm max}=4.68\times 10^{-4}\, {\rm
g}$.\footnote{It is comparable to the core mass corresponding to $a_s=0$ on
account of the remark following Eq. (\ref{cmhm13b}). On the other hand, for
$m=6.35\times
10^{-6}\,
M_{P}$ and $|\lambda|\sim 1$ we have $f\sim 3.17\times 10^{-6}\,
M_{P}c^2$ [see Eq. (\ref{f})]. According to Eq. (\ref{cmhm14}), the
quantum core becomes unstable in a halo of mass $(M_h)_{\rm max}=4.13\times
10^{-3}\, {\rm g}$. The critical core mass is then $(M_c)_{\rm max}=5.44\,
M_P=1.18\times 10^{-4}\, {\rm
g}$.} Since $f_{\rm crit}\ll f_t\sim M_P
c^2\sim 10^{19}\, {\rm GeV}$
(i.e. $|\lambda_{\rm crit}|\gg \lambda_t\sim (m/M_P)^2\sim 4.03\times 10^{-11}$
and $|(a_s)_{\rm crit}|\gg 2Gm/c^2\sim 2.05\times 10^{-40}\, {\rm m}$) and
$(M_c)_{\rm max}\ll M_{\rm
max,dense}^{\rm GR}=2.71\times
10^4\, {\rm g}$, the collapse of the
quantum core leads to a dense axion ``star'' (soliton)
\cite{braaten,ebycollapse,phi6}
or
a bosenova \cite{tkachevprl}, not to a black hole \cite{helfer,moss} (see
Sec. \ref{sec_bhb}). The possibility to form dense inflation
stars and bosenova has not been
considered by Padilla {\it et al.} \cite{padilla}. 
The critical decay constant $f_{\rm crit}$ becomes of the order of the Planck
scale
$M_P
c^2$ (or $|\lambda_{\rm crit}|\sim (m/M_P)^2$ and $|(a_s)_{\rm crit}|\sim
2Gm/c^2$), allowing the formation of PBHs \cite{padilla},
in much more massive
inflaton clusters of mass  $M_h\sim 10^{14}\,
{\rm g}$ or larger. This essentially
returns the
results of Sec. \ref{sec_appni} because for $f\sim M_P c^2$ we
are in the noninteracting limit. For $f\lesssim M_P c^2$, the attractive
self-interaction may facilitate the formation of PBHs  in
sufficiently massive inflaton clusters.

{\it Remark:}  Equations (\ref{cmhm1})-(\ref{cmhm7}) are valid in the
nonrelativistic limit. When $a_s\le 0$ we can easily extend our results to the
relativistic regime by making the substitution $a_s\rightarrow a_s-\kappa
Gm/c^2$ [see Eq. (\ref{rq7})] with $\kappa_{\rm G}=2.50$. In that
case, the maximum mass $(M_c)_{\rm max}$ in Fig. \ref{mhmcnorm} is given by Eq.
(\ref{ru3}). It returns 
 $M_{\rm max}^{\rm NR}$ [see Eq. (\ref{dm16})] when $|a_s|\gg
(a_s)_t$
and  $M_{\rm max, NI}^{\rm GR}$ [see
Eq. (\ref{er1})] when $|a_s|\ll (a_s)_t$. If we make the substitution from Eq.
(\ref{rq7}) in Eq.
(\ref{cmhm14})
and take $a_s=0$ (which is equivalent to directly taking $a_s=-\kappa Gm/c^2$
in Eq. (\ref{cmhm14})) we get $(M_h)_{\rm max}=0.0357\,
c^4/(G^2\Sigma_0)=6.47\times 10^{14}\, {\rm g}$
(we have used $\kappa_G=2.50$). This value can be compared to the value
$M_h=6.49\times 10^{-3}
c^4/(G^2\Sigma_0)=1.18\times 10^{14}\, {\rm g}$ obtained in Sec.
\ref{sec_appni}. On the other hand,  for
$f\sim f_t\sim M_Pc^2/8\sqrt{\pi}$ (see
Sec. \ref{sec_bhb}) we find that $(M_h)_{\rm max}\sim 0.0558\,
c^4/(G^2\Sigma_0)$. This result qualitatively agrees with the value obtained in
the noninteracting case.

\subsection{Virial mass}
\label{sec_vm}

Following our previous works \cite{modeldm,mcmh,mcmhbh,jeansapp}, we
have defined the halo mass
$M_h$ and the halo radius $r_h$ such that $r_h$ represents the distance at
which the central density is divided by $4$. However, some authors
\cite{ch3,prsmv,padilla} use another definition of
the halo mass  and
halo radius. They introduce the virial mass $M_v$ and the virial radius $r_v$
through the relation
\begin{eqnarray}
\label{sc1}
M_v=\frac{4}{3}\pi \rho_{200} r_v^3,
\end{eqnarray}
where $\rho_{200}=200\rho_b$ is $200$ times the background density. Using
\begin{eqnarray}
\label{sc2}
\frac{GM_v}{r_v}\sim \frac{GM_h}{r_h},
\end{eqnarray}
in consistency with Eq. (\ref{cmhm1}), and combining this relation with Eqs.
(\ref{cmhm3}) and
(\ref{sc1}),
we obtain 
\begin{eqnarray}
\label{sc3}
M_h\sim \frac{1}{1.76\, \Sigma_0}\left ( \frac{4}{3}\pi
\rho_{200}\right )^{2/3} M_v^{4/3}.
\end{eqnarray}
The relation between the halo mass $M_h$ and the  virial mass $M_v$
exhibits the scaling $M_h\propto M_v^{4/3}$ \cite{modeldm}. We can use this
relation to express the previous results in terms of $M_v$ instead of $M_h$.
In particular, we find $M_c\propto M_v^{1/3}$ for noninteracting bosons (in
agreement with \cite{ch3}) and 
$M_c\propto M_v^{2/3}$ for bosons with a repulsive self-interaction  in the TF
limit \cite{mcmh}.

\section{Conclusion}
\label{sec_conclusion}

In this paper, by using simple considerations, we have obtained general 
approximate analytical expressions for the maximum mass and the minimum radius of 
relativistic self-gravitating Bose-Einstein condensates at $T=0$ with repulsive
or attractive $|\varphi|^4$ self-interaction  (see Sec. \ref{sec_summ} for a
summary).

For boson stars with a repulsive self-interaction, our analytical  expressions
[see Eqs. (\ref{e1})-(\ref{e1bj})]
interpolate between the
maximum mass and minimum radius [see Eq. (\ref{er1})] of noninteracting bosons
stars  \cite{kaup,rb} and the
maximum mass  and minimum radius [see Eq. (\ref{er3})]  of bosons stars with a
strong self-interaction (TF limit)
\cite{colpi,chavharko}. The noninteracting regime is valid for $a_s\ll
r_S=2Gm/c^2$ or $\lambda\ll (m/M_P)^2$  and the TF regime is valid for $a_s\gg
r_S=2Gm/c^2$ or $\lambda\gg (m/M_P)^2$. In all cases, the maximum mass has a 
general relativistic origin, i.e., it is due to the fact that the radius of the
star approaches the Schwarzschild radius (strong gravity). 
Above the maximum mass, there is no
equilibrium state and the boson star collapses towards a black hole.

For boson stars with an attractive self-interaction (axion stars), our
analytical expressions [see Eqs. (\ref{dm16fq})-(\ref{dm17k})]
 interpolate between the general relativistic 
maximum mass and minimum radius [see Eq. (\ref{er1})] of noninteracting bosons
stars \cite{kaup,rb}  and the nonrelativistic maximum mass and
minimum radius  [see Eqs.  (\ref{dm16}) and (\ref{dm17})] of 
boson stars with a strong self-interaction \cite{prd1}. When $|a_s|\ll
r_S=2Gm/c^2$, $|\lambda|\ll (m/M_P)^2$ or $f\gg M_Pc^2$ (noninteracting regime),
the maximum mass has a general relativistic origin (strong gravity). Above that 
mass the boson star collapses towards a black hole. When $|a_s|\gg
r_S=2Gm/c^2$, $|\lambda|\gg (m/M_P)^2$ or $f\ll M_Pc^2$ (nonrelativistic
regime),  the maximum mass is essentially due to  the attractive
self-interaction of the bosons and to relativistic corrections in the
kinetic energy (or quantum potential) of the BEC (weak
gravity). Above that mass, the dilute
axion star collapses towards a dense axion star or explodes in a bosenova (it
can collapse towards a black hole only if it has a very large
mass $M>M_{\rm max,dense}^{\rm GR}$ given by Eqs. (\ref{dm16der})
and (\ref{dm16db})).

We have confirmed the existence of a triple point at $(a_s,M)\sim
(-r_S,M_{\rm Kaup})$  separating boson stars, black holes, and dense axion
stars or bosenova. We have considered applications of these results to DM halos
and inflaton clusters. We have shown that the quantum core of DM halos is stable
in general (it would collapse towards a black hole in DM halos of mass
$M_h>10^{14}\,
M_{\odot}$ which are not realistic  and it would form a dense axion star or a
bosenova in realistic
halos of mass $M_h<10^{14}\, M_{\odot}$ provided that $f<10^{15}\, {\rm GeV}$
which seems to be excluded by constraints from cosmology and particle
physics).\footnote{We have, however, considered the possibility that these
constraints may be by-passed and that the quantum core of DM halos may be
unstable.} We have also
discussed the possibility to form PBHs in inflaton clusters
\cite{padilla}. We have shown that, for noninteracting bosons of mass $m\sim
10^{-5}\, M_{P}$, the quantum core (inflaton star) of an inflaton cluster can
collapse towards a black hole only if the inflaton cluster is sufficiently
massive, of the order of $M_h\sim 10^{14}\, {\rm
g}$. In that case, we form a PBH of mass $\sim 1\, {\rm g}$.  In
less massive inflaton clusters, the quantum core is stable and no black hole can
form. We have mentioned
that the amount of PBHs formed by this mechanism may not be very large and
that PBHs of mass $\sim 1\, {\rm g}$ quickly evaporate ($t_{\rm evap}\sim
10^{-30}\, {\rm s}$). For bosons with a repulsive self-interaction, the mass of
the PBHs is larger and their evaporation time longer (but still short). For
bosons with an attractive self-interaction, we have shown similarly
that PBHs of mass 
 $\sim 1\, {\rm g}$ can form only in sufficiently massive inflaton clusters of
mass $M_h\sim 10^{14}\, {\rm g}$ or larger. The attractive self-interaction
facilitates the collapse of the quantum core (inflaton star). On the other hand,
we have suggested that,
in smaller inflaton clusters, for a sufficiently strong attractive
self-interaction,
the inflaton star could become unstable and collapse towards a dense inflaton 
star or explode in a bosenova (for
$f\sim 10^{14}\, {\rm GeV}$ the quantum core of an inflaton cluster of mass
$M_h\sim 1\,
{\rm g}$ collapses towards a dense axion star of mass $M_c\sim 10^{-4}\, {\rm
g}$). The collapse of the core cannot lead to a PBH in that
case because the core is not massive enough. The possibility to form  dense
inflaton stars and bosenova  was not considered by Padilla {\it et al.}
\cite{padilla} and deserves a specific study \cite{preparation}.

Our analytical results have been obtained from a Gaussian ansatz. This Gaussian
ansatz provides a good approximation of the exact maximum mass of
nonrelativistic  boson stars with an attractive self-interaction
\cite{prd1,prd2}. It also provides a good approximation of the exact
maximum mass of  noninteracting relativistic 
boson stars (see the Remarks at the end of Secs. \ref{sec_rqni} and
\ref{sec_rqatt}). It is expected to provide relatively accurate results in more
general
cases. 

Fermion stars such as white dwarfs and neutron stars also
possess a
maximum
mass due to special or general relativity. The study of DM halos made
of fermions (like sterile neutrinos) is also of interest and can be
investigated with methods similar to those exposed in this paper (see, e.g.,
\cite{prd1,mcmh,mcmhbh,wignerPH,modeldmF} for the development of the analogy
between
bosonic and fermionic DM).

\appendix

\section{Gaussian ansatz to evaluate the relativistic correction to the
quantum kinetic energy}
\label{sec_g}

For spherically symmetric density profiles, the relativistic correction to
the quantum kinetic energy of self-gravitating BECs  can be written as [see Eq.
(\ref{rq1})]
\begin{eqnarray}
E_{\rm R}=\frac{\hbar^2}{m^2c^2}\int_0^{+\infty} \left
(\frac{d\sqrt{\rho}}{dr}\right )^2\Phi(r) 4\pi r^2\, dr.
\label{g1}
\end{eqnarray}
Making a
Gaussian ansatz
for the density profile 
\begin{eqnarray}
\rho(r)=\frac{M}{R^3}\frac{1}{\pi^{3/2}}e^{-r^2/R^2},
\label{g2}
\end{eqnarray}
we get
\begin{eqnarray}
E_{\rm R}=\frac{\hbar^2M}{m^2c^2R^7}\frac{4}{\sqrt{\pi}}\int_0^{+\infty}
e^{-r^2/R^2}\Phi(r) r^4\, dr.
\label{g3}
\end{eqnarray}
We will evaluate $\Phi(r)$ in two different manners.

\subsection{First approach}

Following \cite{croon,choi} we make the approximation
\begin{eqnarray}
\Phi(r)=-\frac{GM(r)}{r},
\label{g4}
\end{eqnarray}
where
\begin{eqnarray}
M(r)=\int_0^r \rho(r') 4\pi {r'}^2\, dr'
\label{g5}
\end{eqnarray}
is the mass contained within the sphere of radius $r$.
In that case,
Eq. (\ref{g3}) becomes
\begin{eqnarray}
E_{\rm R}=-\frac{\hbar^2GM}{m^2c^2R^7}\frac{4}{\sqrt{\pi}}\int_0^{+\infty}
e^{-r^2/R^2}M(r) r^3\, dr.
\label{g6}
\end{eqnarray}
With the Gaussian ansatz from Eq. (\ref{g2}) we have
\begin{eqnarray}
M(r)=\frac{4M}{\sqrt{\pi}}{\cal M}\left (\frac{r}{R}\right ),
\label{g7}
\end{eqnarray}
where
\begin{eqnarray}
{\cal M}(x)=\int_0^x e^{-y^2}y^2\, dy=\frac{\sqrt{\pi}}{4}{\rm
erf}(x)-\frac{1}{2}xe^{-x^2}.
\label{g8}
\end{eqnarray}
Substituting Eqs. (\ref{g7}) and (\ref{g8}) into Eq.
(\ref{g6}) we obtain
\begin{eqnarray}
E_{\rm R}=-\frac{16}{\pi} \frac{G\hbar^2M^2}{m^2c^2R^3}\int_0^{+\infty} e^{-x^2}
{\cal M}(x) x^3\, dx.
\label{g9}
\end{eqnarray}
This leads to the expression from  Eq. (\ref{rq3}) with the coefficient
\begin{equation}
\chi_{\rm G}=\frac{16}{\pi}\int_0^{+\infty} e^{-x^2} {\cal M}(x) x^3\,
dx=\frac{7}{4\sqrt{2\pi}}\simeq 0.698.
\label{g10}
\end{equation}

\subsection{Second approach}

For spherically symmetric systems, the gravitational potential can be determined
(without approximation at that stage) from Newton's law
\begin{eqnarray}
\frac{d\Phi}{dr}(r)=\frac{GM(r)}{r^2}.
\label{g11}
\end{eqnarray}
Using Eq. (\ref{g7}) and integrating Eq. (\ref{g11}) with the condition
$\Phi(r)\sim -GM/r$ at infinity, we obtain 
\begin{equation}
\Phi(r)=-\frac{4GM}{\sqrt{\pi}R}\int_{r/R}^{+\infty}\frac{{\cal M}(x)}{x^2}\,
dx=-\frac{GM}{r}{\rm erf}\left (\frac{r}{R}\right ).
\label{g12}
\end{equation}
In that case, Eq. (\ref{g3}) becomes
\begin{eqnarray}
E_{\rm R}=-\frac{\hbar^2GM^2}{m^2c^2R^3}\frac{4}{\sqrt{\pi}}\int_0^{+\infty}
e^{-x^2}{\rm erf}(x) x^3\, dx.
\label{g13}
\end{eqnarray}
This leads to the expression from  Eq. (\ref{rq3}) with the coefficient
\begin{equation}
\chi_{\rm G}=\frac{4}{\sqrt{\pi}}\int_0^{+\infty} e^{-x^2} {\rm erf}(x) x^3\,
dx=\frac{5}{2\sqrt{2\pi}}\simeq 0.997.
\label{g14}
\end{equation}

\section{Relativistic complex SF}
\label{sec_ra}

In this Appendix, we discuss the main properties of a  relativistic  complex
self-interacting SF,
establish its hydrodynamic representation, make the TF approximation, and
determine its equation of state $P(\epsilon)$ for an arbitrary self-interaction
potential $V(|\varphi|^2)$. For a $|\varphi|^4$ self-interaction, we justify
the equation of state introduced by Colpi {\it et al.} \cite{colpi}. We also
justify
the GPP
equations
(\ref{gpp1}) and (\ref{gpp2}) in the nonrelativistic limit $c\rightarrow
+\infty$.

\subsection{Klein-Gordon-Einstein equations}
\label{sec_csf}

We consider a relativistic  complex SF $\varphi(x^\mu)=\varphi(x,y,z,t)$ which
is a continuous
function of space and time. It can represent the wavefunction of a
relativistic BEC \cite{chavmatos,action}. The total action of the system,  which
is the sum of the
Einstein-Hilbert action of general relativity $+$ the action of the SF, can be
written as
\begin{equation}
S=\int \left (\frac{c^4}{16\pi G}R+ \mathcal{L}\right) \sqrt{-g}\, d^4x,
\label{csf1}
\end{equation}
where $R$ is the Ricci scalar curvature, $\mathcal{L}=\mathcal{L}(\varphi,
\varphi^*,\partial_\mu\varphi,\partial_\mu\varphi^*)$
is the Lagrangian density of the SF,  and
$g={\rm det}(g_{\mu\nu})$ is the
determinant of
the metric tensor.  We consider a
canonical Lagrangian density of the form
\begin{eqnarray}
{\cal L}=\frac{1}{2}g^{\mu\nu}\partial_{\mu}\varphi^*\partial_{\nu}
\varphi-V_{\rm tot}(|\varphi|^2),
\label{csf2}
\end{eqnarray}
where the first term is the kinetic energy and the second term is minus the
potential energy. The potential energy can be decomposed into a rest-mass energy
term and a self-interaction energy term:
\begin{equation}
\label{csf3}
V_{\rm
tot}(|\varphi|^2)=\frac{m^2c^2}{2\hbar^2}|\varphi|^2+V(|\varphi|^2).
\end{equation}

The least action principle $\delta S=0$ with respect to variations
$\delta\varphi$
(or $\delta\varphi^*$),
which is equivalent to the
Euler-Lagrange equation
\begin{eqnarray}
\label{lh2rel}
D_{\mu}\left\lbrack \frac{\partial {\cal
L}}{\partial(\partial_\mu\varphi)^*}\right\rbrack-\frac{\partial {\cal
L}}{\partial\varphi^*}=0,
\end{eqnarray}
yields the KG equation
\begin{equation}
\label{csf4}
\square\varphi+2\frac{dV_{\rm tot}}{d|\varphi|^2}\varphi=0,
\end{equation}
where
$\square=D_{\mu}\partial^{\mu}=\frac{1}{\sqrt{-g}}\partial_{\mu}(\sqrt{-g}\, g^{
\mu\nu} \partial_{\nu})$ is the d'Alembertian
operator. For
a free massless SF ($V_{\rm tot}=0$), the KG equation reduces to
$\square\varphi=0$. On the other hand, using the decomposition from Eq.
(\ref{csf3}) we can rewrite
the KG equation (\ref{csf4}) as
\begin{equation}
\label{csf4b}
\square\varphi+\frac{m^2c^2}{\hbar^2}\varphi+2\frac{dV}{d|\varphi|^2}\varphi=0.
\end{equation}

The least action principle $\delta
S=0$ 
with respect to variations $\delta g^{\mu\nu}$ yields the Einstein field
equations
\begin{equation}
R_{\mu\nu}-\frac{1}{2}g_{\mu\nu}R=\frac{8\pi G}{c^4}T_{\mu\nu},
\label{ak21}
\end{equation}
where $R_{\mu\nu}$ is the Ricci tensor and $T_{\mu\nu}$ is the
energy-momentum (stress) tensor given by
\begin{eqnarray}
\label{em1}
T_{\mu\nu}&=&\frac{2}{\sqrt{-g}}\frac{\delta S}{\delta
g^{\mu\nu}}=\frac{2}{\sqrt{-g}}\frac{\partial (\sqrt{-g}{\cal L})}{\partial
g^{\mu\nu}}\nonumber\\
&=&2\frac{\partial {\cal L}}{\partial g^{\mu\nu}}-g_{\mu\nu}{\cal L}.
\end{eqnarray}
For a complex SF, the energy-momentum tensor takes the form
\begin{eqnarray}
\label{em1bb}
T_{\mu}^{\nu}=\frac{\partial {\cal L}}{\partial
(\partial_\nu\varphi)}\partial_\mu\varphi+\frac{\partial {\cal L}}{\partial
(\partial_\nu\varphi^*)}\partial_\mu\varphi^*-g_{\mu}^{\nu}{\cal L}.
\end{eqnarray}
For the Lagrangian
(\ref{csf2}), we get
\begin{eqnarray}
\label{em2}
T_{\mu\nu}=\frac{1}{2}(\partial_{\mu}\varphi^*\partial_{\nu}\varphi+\partial_{
\nu}\varphi^*\partial_{\mu}\varphi)
-g_{\mu\nu}{\cal L}.
\end{eqnarray}
Equations (\ref{csf4})
and (\ref{ak21}) with Eq. (\ref{em2}) form the KGE equations.

The conservation of
the energy-momentum tensor,
which results from the invariance of the Lagrangian density under continuous
translations in space and time (Noether theorem), reads
\begin{eqnarray}
\label{lh2r}
D_{\nu}T^{\mu\nu}=0.
\end{eqnarray}
The conservation of
the energy-momentum tensor is automatically included in
the Einstein equations through the contracted Bianchi identities. The
energy-momentum four vector is $P^{\mu}=\int T^{\mu 0}\sqrt{-g}\, d^3x$. 
Its time component $P^0$ is the energy while $(P^1,P^2,P^3)$ are the
components of the impulse ${\bf P}$. Each component of $P^\mu$ is conserved
in time, i.e., it is a constant of the motion. Indeed,
\begin{eqnarray}
\label{em1b}
{\dot P}^{\mu}&=&\frac{d}{dt}\int T^{\mu 0}\sqrt{-g}\, d^3x=c\int \partial_0
(T^{\mu 0}\sqrt{-g})\, d^3x\nonumber\\
&=&-c\int
\partial_i (T^{\mu i}\sqrt{-g})\, d^3x=0,
\end{eqnarray}
where we have used Eq. (\ref{lh2r}) with
$D_\mu V^\mu=\frac{1}{\sqrt{-g}}\partial_\mu(\sqrt{-g}V^{\mu})$ to get the third
equality.

The current of charge of a complex SF is given by
\begin{eqnarray}
\label{j1}
J^{\mu}=\frac{m}{i\hbar}\left \lbrack\varphi\frac{\partial {\cal L}}{\partial
(\partial_\mu\varphi)}-\varphi^* \frac{\partial {\cal L}}{\partial
(\partial_\mu\varphi^*)}\right\rbrack.
\end{eqnarray}
For the Lagrangian (\ref{csf2}), we obtain
\begin{eqnarray}
\label{charge1}
J_{\mu}=-\frac{m}{2i\hbar}
(\varphi^*\partial_\mu\varphi-\varphi\partial_\mu\varphi^*).
\end{eqnarray}
Using the KG equation (\ref{csf4}), one can show
that 
\begin{eqnarray}
\label{charge2}
D_{\mu}J^{\mu}=0.
\end{eqnarray}
This equation expresses the local conservation of the charge. The total charge
of the SF is $Q=\frac{e}{mc}\int J^0\sqrt{-g}\, d^3x$. Proceeding as above,
we easily find that $\dot Q=0$. The charge $Q$ is proportional to the number
$N$
of bosons
provided that antibosons are counted 
negatively \cite{landaulifshitz}. Therefore, Eq. (\ref{charge2})  also
expresses the local conservation of the boson number ($Q=Ne$). This conservation
law
results via the Noether theorem from the global $U(1)$ symmetry of the
Lagrangian, i.e., from the invariance of the Lagrangian density
under a global phase transformation $\varphi\rightarrow \varphi e^{-i\theta}$
(rotation) of the
complex SF. Note that $J_{\mu}$
vanishes for a real SF so that the charge and the particle number are not
conserved in that case.

\subsection{Hydrodynamic representation}
\label{sec_db}

We can write the KG equation (\ref{csf4}) under the form of 
hydrodynamic equations by using the de Broglie
transformation \cite{broglie1927a,broglie1927b,broglie1927c}. To that
purpose, we write the SF as 
\begin{equation}
\varphi=\frac{\hbar}{m}\sqrt{\rho}e^{i m \theta/\hbar},
\label{db4}
\end{equation}
where $\rho$ is the pseudo rest-mass density\footnote{We stress that $\rho$ is
{\it not} the rest-mass density. It is only
in the
nonrelativistic regime $c\rightarrow +\infty$ that $\rho$ coincides with the
rest-mass density.} and $\theta=S_{\rm tot}/m$ ($\sim$ phase) is the action by
unit of
mass. They satisfy
\begin{eqnarray}
\rho=\frac{m^2}{\hbar^2}|\varphi|^2\quad {\rm
and}\quad
\theta=\frac{\hbar}{2mi}\ln \left (\frac{\varphi}{\varphi^*}\right ).
\label{db2}
\end{eqnarray}
Substituting Eq. (\ref{db4}) into the Lagrangian density (\ref{csf2}), we obtain
\begin{eqnarray}
{\cal
L}=\frac{1}{2}g^{\mu\nu}\rho\partial_{\mu}\theta\partial_{\nu}
\theta+\frac{\hbar^2}{8m^2\rho}g^{\mu\nu}\partial_{\mu}\rho\partial_{\nu}
\rho-V_{\rm tot}(\rho)
\label{db5}
\end{eqnarray}
with
\begin{equation}
\label{db10}
V_{\rm
tot}(\rho)=\frac{1}{2}\rho c^2+V(\rho).
\end{equation}
The Euler-Lagrange equations for $\theta$ and $\rho$, resulting from the
least action principle, are 
\begin{equation}
\label{db6}
D_{\mu}\left \lbrack \frac{\partial
{\cal L}}{\partial(\partial_{\mu}\theta)}\right\rbrack-\frac{\partial
{\cal L}}{\partial\theta}=0,\quad D_{\mu}\left \lbrack \frac{\partial
{\cal L}}{\partial(\partial_{\mu}\rho)}\right\rbrack-\frac{\partial
{\cal L}}{\partial\rho}=0.
\end{equation}
They yield the equations of motion \cite{chavmatos,action}
\begin{eqnarray}
D_{\mu}\left ( \rho \partial^{\mu}\theta\right )=0,
\label{db8}
\end{eqnarray}
\begin{eqnarray}
\frac{1}{2}\partial_{\mu}\theta\partial^{\mu}\theta-\frac{\hbar^2}{2m^2}\frac{
\square\sqrt { \rho } } { \sqrt { \rho } } -  V'_{\rm
tot}(\rho)=0.
\label{db9}
\end{eqnarray}
The
same equations are obtained by substituting the de Broglie transformation from
Eq.
(\ref{db4}) into the
KG equation
(\ref{csf4}),
and separating the real and the imaginary parts.\footnote{The quantity
$v_{\mu}=\partial_{\mu}\theta$ could be interpreted as a pseudo quadrivelocity
but it does not satisfy the identity $v_\mu v^\mu=c^2$ \cite{chavmatos,action}.}
Equation (\ref{db8}) can be
interpreted as a continuity equation and Eq.
(\ref{db9}) can be interpreted as a  quantum relativistic Hamilton-Jacobi (or
Bernoulli)
equation with a relativistic covariant quantum potential
\begin{eqnarray}
Q_{\rm dB}=\frac{\hbar^2}{2m}\frac{\square\sqrt{\rho}}{\sqrt{\rho}}.
\label{db11}
\end{eqnarray}

The energy-momentum tensor is given, in
the hydrodynamic representation, by 
\begin{eqnarray}
\label{tmunuh}
T_{\mu}^{\nu}=\frac{\partial {\cal L}}{\partial
(\partial_\nu\theta)}\partial_\mu\theta+\frac{\partial {\cal L}}{\partial
(\partial_\nu\rho)}\partial_\mu\rho-g_{\mu}^{\nu}{\cal L}.
\end{eqnarray}
For the Lagrangian (\ref{db5}) we obtain
\begin{eqnarray}
T_{\mu\nu}=\rho\partial_{\mu}\theta\partial_{\nu}\theta+\frac{\hbar^2}{4m^2\rho}
\partial_{\mu}\rho\partial_{\nu}\rho-g_{\mu\nu}{\cal L}.
\label{em3gen}
\end{eqnarray}
This result can also be obtained from  Eq. (\ref{em2}) by using Eq.
(\ref{db4}).

The current of charge of a complex SF is given, in
the hydrodynamic representation, by
\begin{eqnarray}
J^{\mu}=-\frac{\partial {\cal L}}{\partial(\partial_{\mu}\theta)}.
\end{eqnarray}
For the Lagrangian (\ref{db5}), we obtain
\begin{eqnarray}
J_{\mu}=-\rho\partial_{\mu}\theta.
\label{charge4}
\end{eqnarray}
This result can also be obtained from  Eq. (\ref{charge1}) by using Eq.
(\ref{db4}). We then see that the continuity equation
(\ref{db8}) is
equivalent to Eq. (\ref{charge2}). It
expresses the local
conservation of the charge $Q$ of the SF (or the local conservation of the boson
number
$N$): $Q=Ne=-\frac{e}{mc}\int \rho\partial^0\theta \sqrt{-g}  \, d^3x$.

\subsection{TF approximation}
\label{sec_rtf}

In the classical limit or in the TF approximation ($\hbar\rightarrow 0$), the
Lagrangian from Eq.
(\ref{db5}) reduces to
\begin{eqnarray}
{\cal
L}=\frac{1}{2}g^{\mu\nu}\rho\partial_{\mu}\theta\partial_{\nu}
\theta-V_{\rm tot}(\rho).
\label{rtf1}
\end{eqnarray}
The Euler-Lagrange equations (\ref{db6}) 
yield the equations of motion
\begin{eqnarray}
D_{\mu}\left ( \rho \partial^{\mu}\theta\right )=0,
\label{rtf4}
\end{eqnarray}
\begin{eqnarray}
\frac{1}{2}\partial_{\mu}\theta\partial^{\mu}\theta-V'_{\rm
tot}(\rho)=0.
\label{rtf5}
\end{eqnarray}
The same equations are obtained by making the TF approximation in Eq.
(\ref{db9}), i.e., by neglecting the quantum potential. Equation (\ref{rtf4})
can be interpreted as a continuity equation and Eq.
(\ref{rtf5}) can be interpreted as a classical relativistic Hamilton-Jacobi (or
Bernoulli)
equation. We note that the continuity equation is not affected by the TF
approximation.

Assuming
$V_{\rm tot}'>0$, and using Eq. (\ref{rtf5}), we introduce the fluid
quadrivelocity
\begin{eqnarray}
u_{\mu}=-\frac{\partial_{\mu}\theta}{\sqrt{2V_{\rm
tot}'(\rho)}}c,
\label{rtf5b}
\end{eqnarray}
which satisfies the identity $u_{\mu}u^{\mu}=c^2$. The
energy-momentum
tensor is given by Eq. (\ref{tmunuh}). For the Lagrangian
(\ref{rtf1}) we obtain
\begin{eqnarray}
T_{\mu\nu}=\rho\partial_{\mu}\theta\partial_{\nu}
\theta-g_{\mu\nu}{\cal L}.
\label{em3}
\end{eqnarray}
This expression can also be obtained from Eq. (\ref{em3gen}) by making
the TF approximation. Using Eq.
(\ref{rtf5b}), we get
\begin{eqnarray}
T_{\mu\nu}=2\rho V'_{\rm tot}(\rho) \frac{u_{\mu}u_{\nu}}{c^2}-g_{\mu\nu}{\cal
L}.
\label{em3b}
\end{eqnarray}
The   energy-momentum tensor (\ref{em3b}) can be written under the perfect fluid
form
\begin{eqnarray}
T_{\mu\nu}=(\epsilon+P)\frac{u_{\mu}u_{\nu}}{c^2}-P g_{\mu\nu},
\label{em4}
\end{eqnarray}
where $\epsilon$ is the energy density and $P$ is the pressure, provided that we
make the identifications
\begin{eqnarray}
P={\cal L},\qquad \epsilon+P=2\rho V'_{\rm tot}(\rho).
\label{em5}
\end{eqnarray}
Therefore, the Lagrangian plays the role of the pressure of the fluid. 
Combining  Eq. (\ref{rtf1}) with the Bernoulli equation (\ref{rtf5}), we get
\begin{eqnarray}
{\cal L}=\rho V'_{\rm tot}(\rho)-V_{\rm tot}(\rho).
\label{em6}
\end{eqnarray}
Therefore, according to Eqs. (\ref{em5}) and (\ref{em6}), the energy density and
the pressure derived from the Lagrangian (\ref{rtf1}) 
are given by \cite{abrilphas,action}
\begin{eqnarray}
\epsilon=\rho V'_{\rm tot}(\rho)+V_{\rm
tot}(\rho)=\rho c^2+\rho V'(\rho)+V(\rho),
\label{rtf6}
\end{eqnarray}
\begin{eqnarray}
P=\rho V'_{\rm tot}(\rho)-V_{\rm
tot}(\rho)=\rho V'(\rho)-V(\rho),
\label{rtf7}
\end{eqnarray}
where we have used Eq. (\ref{db10}) to get the second
equalities.\footnote{Equations (\ref{rtf6}) and (\ref{rtf7}) 
can also be derived for
a cosmological homogeneous SF  from its
hydrodynamic representation or from the virial theorem in the fast oscillation
regime \cite{abrilphas}.} Eliminating
$\rho$ between Eqs. (\ref{rtf6}) and (\ref{rtf7}), we obtain the equation of
state $P(\epsilon)$. On the other hand, Eq. (\ref{rtf7}) can be integrated into
\cite{action}
\begin{eqnarray}
V(\rho)=\rho\int\frac{P(\rho)}{\rho^2}\, d\rho.
\label{etoilebis}
\end{eqnarray}
Equation (\ref{rtf7}) determines $P(\rho)$ as a function of $V(\rho)$ while Eq.
(\ref{etoilebis}) determines $V(\rho)$ as a function of $P(\rho)$.

\subsection{Rest-mass density}

In the TF approximation, using Eqs. (\ref{charge4}) and
(\ref{rtf5b}), we can write the
current as
\begin{eqnarray}
J_{\mu}=\rho\sqrt{\frac{2}{c^2}V'_{\rm tot}(\rho)}\, u_{\mu}.
\label{charge8}
\end{eqnarray}
The rest-mass density $\rho_{m}=nm$, which is related to the charge
density $\rho_e=ne$ by $\rho_{m}=(m/e)\rho_e$, is such that
\begin{eqnarray}
J_{\mu}=\rho_{m}  u_{\mu}.
\label{charge6w}
\end{eqnarray}
The continuity equation
(\ref{charge2}) can then be written as
\begin{eqnarray}
D_{\mu}(\rho_{m} u^{\mu})=0.
\label{charge7}
\end{eqnarray}
Comparing Eq. (\ref{charge8}) with Eq. (\ref{charge6w}), we
find that the
rest-mass density of the SF is related to the pseudo-rest mass density $\rho$
by
\begin{eqnarray}
\rho_{m}=\rho\sqrt{\frac{2}{c^2}V'_{\rm
tot}(\rho)}=\rho\sqrt{1+\frac{2}{c^2}V'(\rho)}.
\label{charge10c}
\end{eqnarray}
In general,
$\rho_{m}\neq \rho$ except (i) for a noninteracting SF ($V=0$), (ii) when $V$ is
constant, corresponding to the $\Lambda$FDM model (see
Appendix E of \cite{logosf}), (iii) and in the
nonrelativistic limit $c\rightarrow +\infty$.

{\it Remark:} More general results valid beyond the TF approximation are given
in \cite{action,csfpoly}.

\subsection{GPE equations}
\label{sec_grgp}

In order to recover the GPP equations in the
nonrelativistic limit $c\rightarrow +\infty$, we make the Klein
transformation (see, e.g., \cite{chavmatos})
\begin{eqnarray}
\varphi({\bf r},t)=\frac{\hbar}{m}e^{-i m c^2
t/\hbar}\psi({\bf r},t),
\label{kgp12}
\end{eqnarray}
where $\psi$ is the pseudo wave function. The pseudo rest-mass density [see
Eq. (\ref{db2})] is related to the pseudo wave function by
\begin{eqnarray}
\rho=|\psi|^2.
\label{nrl}
\end{eqnarray}
Mathematically, we can
always make the change of variables from Eq. (\ref{kgp12}). 
However, we emphasize that it is only in the nonrelativistic limit $c\rightarrow
+\infty$ that $\psi$ has the interpretation of a wave function and that
$\rho=|\psi|^2$ has the interpretation of a mass density. 

Substituting Eq. (\ref{kgp12}) into the KG equation (\ref{csf4}), we obtain
after simplification the general relativistic GP equation \cite{chavmatos,mabc}
\begin{eqnarray}
i\hbar
c\,\partial^0\psi-\frac{\hbar^2}{2m}\square\psi+\frac{1}{2}mc^2(g^{00}
-1)\psi\nonumber\\
+i\frac{\hbar c^2}{2}\square
t\, \psi-m\frac{dV}{d|\psi|^2}\psi=0,
\label{grgp1}
\end{eqnarray}
We note that $\square t$ can be written as $\square
t=-\frac{1}{c}g^{\mu\nu}\Gamma_{\mu\nu}^{0}$, where $\Gamma_{\mu\nu}^{\sigma}$
are the Christoffel symbols \cite{chavmatos}. We can similarly express the
energy-momentum tensor (\ref{em2}) of the SF which appears in the Einstein
equations (\ref{ak21}) in terms of the pseudo wavefunction $\psi$. This leads
to the Gross-Pitaevskii-Einstein (GPE) equations.

\subsection{Weak gravity limit}
\label{sec_nrl}

In the weak gravity limit of general relativity
$\Phi/c^2\ll 1$, using the simplest form of the conformal Newtonian gauge, the
line
element is given by
\begin{equation}
ds^2=c^2\left(1+2\frac{\Phi}{c^2}\right)dt^2-\left(1-2\frac{\Phi}{c^2}
\right)\delta_{ij}dx^idx^j,
\label{conf1}
\end{equation}
where $\Phi({\bf r},t)$ is the Newtonian
potential.\footnote{As in Refs \cite{abrilph,playa,chavmatos} 
we have neglected anisotropic
stresses and assumed that the lapse function $\Psi$ is equal to the Newtonian
potential $\Phi$. See
\cite{hn1,hn2} for a more general treatment.} In that
limit, the
Lagrangian of the SF is
\begin{eqnarray}
{\cal L}=\frac{1}{2c^2}\left (1-\frac{2\Phi}{c^2}\right )\left
|\frac{\partial\varphi}{\partial t}\right |^2-\frac{1}{2}\left
(1+\frac{2\Phi}{c^2}\right )|\nabla\varphi|^2\nonumber\\
-\frac{m^2c^2}{2\hbar^2}|\varphi|^2
-V(|\varphi|^2)
\label{lago}
\end{eqnarray}
and the KGE equations reduce to \cite{abrilph,playa,chavmatos} 
\begin{eqnarray}
\frac{1}{c^2}\frac{\partial^2\varphi}{\partial
t^2}-\left(1+\frac{4\Phi}{c^2}\right)\Delta\varphi
-\frac{4}{c^4}\frac{\partial\Phi}{
\partial t}\frac{\partial\varphi}{\partial
t}\nonumber\\
+\left (1+\frac{2\Phi}{c^2}\right) \frac{m^2
c^2}{\hbar^2}\varphi
+2\left(1+2\frac{\Phi}{c^2}\right)\frac{dV}{d|\varphi|^2}\varphi=0,\nonumber\\
\label{kge4}
\end{eqnarray}
\begin{eqnarray}
\frac{\Delta\Phi}{4\pi G}=\frac{T_0^0}{c^2}=\frac{1}{2c^4}\left
(1-\frac{2\Phi}{c^2}\right
)\left
|\frac{\partial\varphi}{\partial t}\right |^2 \nonumber\\
+\frac{1}{2c^2}\left (1+\frac{2\Phi}{c^2}\right
)|\nabla\varphi|^2+\frac{m^2}{2\hbar^2}|\varphi|^2
+\frac{1}{c^2}V(|\varphi|^2).
\label{kge11b}
\end{eqnarray}

Making the Klein transformation from Eq.
(\ref{kgp12}) in Eqs. (\ref{lago})-(\ref{kge11b}), we obtain
\cite{abrilph,playa,chavmatos}  
\begin{eqnarray}
{\cal L}&=&\frac{\hbar^2}{2m^2c^2}\left
(1-\frac{2\Phi}{c^2}\right )\left
|\frac{\partial\psi}{\partial t}\right |^2\nonumber\\
&+&\frac{i\hbar}{2m}\left
(1-\frac{2\Phi}{c^2}\right
)\left (\psi^*\frac{\partial\psi}{\partial t}-\psi\frac{\partial\psi^*}{\partial
t}\right )\nonumber\\
&-&\frac{\hbar^2}{2m^2}\left (1+\frac{2\Phi}{c^2}\right
)|\nabla\psi|^2-\Phi|\psi|^2-V(|\psi|^2),\nonumber\\
\end{eqnarray}
\begin{eqnarray}
i\hbar\frac{\partial\psi}{\partial t}-\frac{\hbar^2}{2m
c^2}\frac{\partial^2\psi}{\partial t^2}
+\frac{\hbar^2}{2 m}\left
(1+\frac{4\Phi}{c^2}\right )\Delta\psi-m\Phi \psi\nonumber\\
-\left
(1+\frac{2\Phi}{c^2}\right )m \frac{dV}{d|\psi|^2}\psi+\frac{2\hbar^2}{m
c^4}\frac{\partial\Phi}{\partial t}\left
(\frac{\partial \psi}{\partial t}-\frac{i m c^2}{\hbar}\psi\right
)=0,\nonumber\\
\label{kge13}
\end{eqnarray}
\begin{eqnarray}
\frac{\Delta\Phi}{4\pi G}=\frac{T_0^0}{c^2}=\left (1-\frac{\Phi}{c^2}\right
)|\psi|^2+\frac{1}{c^2}V(|\psi|^2)\nonumber\\
+\frac{\hbar^2}{2m^2c^4}\left
(1-\frac{2\Phi}{c^2}\right )\left
|\frac{\partial\psi}{\partial t}\right |^2
+\frac{\hbar^2}{2m^2c^2}\left (1+\frac{2\Phi}{c^2}\right
)|\nabla\psi|^2\nonumber\\
-\frac{\hbar}{mc^2}\left
(1-\frac{2\Phi}{c^2}\right ){\rm Im} \left
(\frac{\partial\psi}{\partial
t}\psi^*\right ).\qquad
\label{kgp14b}
\end{eqnarray}
We note the identity
\begin{eqnarray}
{\rm Im} \left
(\frac{\partial\psi}{\partial
t}\psi^*\right )=\frac{1}{2i}\left (\psi^*\frac{\partial\psi}{\partial
t}-\psi\frac{\partial\psi^*}{\partial
t}\right ).
\end{eqnarray}
Equations (\ref{kge13}) and (\ref{kgp14b}) form the  GPE equations in the weak
gravity
limit.  In the
nonrelativistic limit $c\rightarrow +\infty$, they reduce to the GPP
equations \cite{abrilph,playa,chavmatos}
\begin{eqnarray}
i\hbar\frac{\partial\psi}{\partial
t}=-\frac{\hbar^2}{2m}\Delta\psi+m\Phi\psi+m\frac{dV}{d|\psi|^2}\psi,
\label{nrl1}
\end{eqnarray}
\begin{eqnarray}
\Delta\Phi=4\pi G |\psi|^2.
\label{nrl2}
\end{eqnarray}
In that case, the Lagrangian of the SF is
\begin{eqnarray}
{\cal L}&=&\frac{i\hbar}{2m}\left (\psi^*\frac{\partial\psi}{\partial
t}-\psi\frac{\partial\psi^*}{\partial
t}\right )\nonumber\\
&-&\frac{\hbar^2}{2m^2}|\nabla\psi|^2-\Phi|\psi|^2-V(|\psi|^2).
\end{eqnarray}
The GPP equations (\ref{nrl1}) and (\ref{nrl2})
can also be rewritten as a single equation 
\begin{eqnarray}
\label{gty}
i\hbar \frac{\partial\psi}{\partial
t}=-\frac{\hbar^2}{2m}\Delta\psi+m\frac{dV}{d|\psi|^2}\psi\nonumber\\
-\psi\int \frac{Gm}{|{\bf r}-{\bf r}'|}|\psi|^2({\bf r}',t)\, d{\bf
r}'.
\end{eqnarray}

{\it Remark:} We note that, for a complex SF, the potential $V(|\psi|^2)$ which
occurs in the GP equation (\ref{nrl1}) is equal to the potential
$V(|\varphi|^2)$ which
occurs in the KG equation (\ref{csf4b}) up to the
change of function from Eq.
(\ref{kgp12}) leading to
\begin{eqnarray}
|\varphi|^2=\frac{\hbar^2}{m^2}|\psi|^2.
\label{kgp12b}
\end{eqnarray}
This equivalence is no more true for a real SF (see \cite{phi6} and Appendix
\ref{sec_rrsf}).

\subsection{Hydrodynamic representation of the GPE equations in the weak gravity
limit}
\label{sec_hyw}

We can write the GPE equations (\ref{kge13}) and (\ref{kgp14b}) in the form of
hydrodynamic equations by making the Madelung 
transformation \cite{madelung}
\begin{eqnarray}
\psi({\bf r},t)=\sqrt{\rho({\bf r},t)} e^{iS({\bf r},t)/\hbar},\label{kgp15}
\end{eqnarray}
\begin{eqnarray}
\rho=|\psi|^2, \qquad {\bf u}({\bf r},t)=\frac{\nabla S}{m},
\label{kgp16}
\end{eqnarray}
where $\rho$ is the pseudo rest-mass density, $S$ is the  pseudo action and
${\bf u}$
is the pseudo velocity field (they coincide with the mass density, the action
and the velocity field in the nonrelativistic limit $c\rightarrow +\infty$).

Substituting Eqs. (\ref{kgp15}) and (\ref{kgp16}) into the 
GPE equations (\ref{kge13}) and (\ref{kgp14b}), and 
separating the real and the imaginary
parts, we obtain the system of hydrodynamic equations
\cite{abrilph,playa,chavmatos}
\begin{eqnarray}
\frac{\partial\rho}{\partial t}+\nabla\cdot (\rho
{\bf u})=\frac{1}{mc^2}\frac{\partial }{\partial t}\left (\rho \frac{\partial
S}{\partial t}\right )\nonumber\\
+
\frac{4\rho}{mc^4}\frac{\partial\Phi}{\partial t}\left (mc^2-\frac{\partial
S}{\partial t}\right )-\frac{4\Phi}{c^2}\nabla\cdot (\rho {\bf
u}),
\label{kge15}
\end{eqnarray}
\begin{eqnarray}
\frac{\partial S}{\partial t}+\frac{(\nabla S)^2}{2 m}=-\frac{\hbar^2}{2
m
c^2}\frac{\frac{\partial^2\sqrt{\rho}}{\partial
t^2}}{\sqrt{\rho}}
+\left (1+\frac{4\Phi}{c^2}\right )\frac{\hbar^2}{2
m}\frac{\Delta\sqrt{\rho}}{\sqrt{\rho}}\nonumber\\
-\frac{2\Phi}{m c^2}(\nabla S)^2
-m\Phi
-\left (1+\frac{2\Phi}{c^2}\right ) m h(\rho)\nonumber\\
+\frac{1}{2 mc^2}\left (\frac{\partial
S}{\partial t}\right )^2
+\frac{\partial\Phi}{\partial
t}\frac{\hbar^2}{m c^4 \rho}\frac{\partial\rho}{\partial t},\qquad
\label{kge18}
\end{eqnarray}
\begin{eqnarray}
\frac{\partial {\bf u}}{\partial t}+({\bf u}\cdot
\nabla){\bf u}=-\frac{\hbar^2}{2m^2c^2}\nabla \left
(\frac{\frac{\partial^2\sqrt{\rho}}{\partial t^2}}{\sqrt{\rho}}\right
)\nonumber\\
+\frac{\hbar^2}{2m^2}\nabla\left\lbrack \left (1+\frac{4\Phi}{c^2}\right
)\frac{\Delta\sqrt{\rho}}{\sqrt{\rho}}\right\rbrack-\nabla\Phi-\frac{1}{\rho
}\nabla P\nonumber\\
-\frac{2}{c^2}\nabla (h\Phi)
-\frac{2}{c^2}\nabla
(\Phi
{\bf u}^2)
+\frac{1}{2m^2c^2}\nabla \left\lbrack\left (\frac{\partial S}{\partial
t}\right )^2\right \rbrack\nonumber\\
+\frac{\hbar^2}{m^2
c^4}\nabla \left (\frac{\partial\Phi}{\partial
t}\frac{1}{\rho}\frac{\partial\rho}{\partial t}\right ),
\label{kge16}
\end{eqnarray}
\begin{eqnarray}
\frac{\Delta\Phi}{4\pi G}=\frac{\hbar^2}{2m^2c^4}\left
(1-\frac{2\Phi}{c^2}\right )\left\lbrack
\frac{1}{4\rho}\left (\frac{\partial\rho}{\partial t}\right
)^2+\frac{\rho}{\hbar^2}\left (\frac{\partial S}{\partial t}\right
)^2\right\rbrack\nonumber\\
+\frac{\hbar^2}{2m^2c^2}\left (1+\frac{2\Phi}{c^2}\right )\left\lbrack
\frac{1}{4\rho}(\nabla\rho)^2+\frac{\rho}{\hbar^2}(\nabla
S)^2\right\rbrack\nonumber\\
-\left (1-\frac{2\Phi}{c^2}\right )\frac{\rho}{mc^2}\frac{\partial S}{\partial
t}+\left (1-\frac{\Phi}{c^2}\right
)\rho+\frac{1}{c^2}V(\rho),\nonumber\\
\label{kge17}
\end{eqnarray}
where $h(\rho)$ is the pseudo enthalpy defined by 
\begin{eqnarray}
h(\rho)=V'(\rho),
\label{hvp}
\end{eqnarray}
and $P(\rho)$ is the pseudo
pressure defined by the relation $h'(\rho)=P'(\rho)/\rho$, which  can be
integrated
into
 $P(\rho)=\rho h(\rho)-\int h(\rho)\, d\rho$, yielding
\begin{eqnarray}
P(\rho)=\rho V'(\rho)-V(\rho)=\rho^2 \left\lbrack
\frac{V(\rho)}{\rho}\right\rbrack'.
\label{kgp20g}
\end{eqnarray}
Eq. (\ref{kgp20g})
determine the equation of state $P(\rho)$ for a given self-interaction potential
$V(\rho)$. Inversely, for a given equation of
state, the self-interaction potential is given by
\begin{eqnarray}
V(\rho)=\rho\int \frac{P(\rho)}{\rho^2}\, d\rho.
\label{ge9bqw}
\end{eqnarray}
The
pseudo squared speed of sound is $c_s^2=P'(\rho)=\rho V''(\rho)$. The
hydrodynamic equations  (\ref{kge15})-(\ref{kge17}) have a clear physical
interpretation. Equation (\ref{kge15}), corresponding to the imaginary
part of the GPE equations, is the continuity equation expressing the
conservation of the charge of the SF $Q=-(e/m^2c^2)\int\rho(\partial S/\partial
t-mc^2)(1-4\Phi/c^2)\, d{\bf r}$. Equation
(\ref{kge18}),
corresponding to the real part of the GPE equations,
is the Hamilton-Jacobi (or Bernoulli) equation. Equation
(\ref{kge16}), obtained
by
taking the gradient of  Eq. (\ref{kge18}), is the momentum equation. Equation
(\ref{kge17}) is the Einstein equation. We stress
that the hydrodynamic equations  (\ref{kge15})-(\ref{kge17}) are equivalent to
the GPE equations (\ref{kge13})-(\ref{kgp14b}) which are themselves equivalent
to the 
KGE equations (\ref{kge4}) and (\ref{kge11b}).\footnote{We note that the
Bernoulli equation  (\ref{kge18}) is a second degree equation in $E=-\partial
S/\partial t$ which can be solved easily. We can then substitute the solution
into Eqs. (\ref{kge15}), (\ref{kge16}) and (\ref{kge17}) to get a closed
reduced system of equations (see Ref. \cite{chavmatos}).} The corresponding
Lagrangian is
\begin{eqnarray}
{\cal L}&=&\frac{\hbar^2}{2m^2c^2}\left
(1-\frac{2\Phi}{c^2}\right )\left\lbrack
\frac{1}{4\rho}\left (\frac{\partial\rho}{\partial t}\right
)^2+\frac{\rho}{\hbar^2}\left (\frac{\partial S}{\partial t}\right
)^2\right\rbrack\nonumber\\
&-&\frac{\hbar^2}{2m^2}\left (1+\frac{2\Phi}{c^2}\right )\left\lbrack
\frac{1}{4\rho}(\nabla\rho)^2+\frac{\rho}{\hbar^2}(\nabla
S)^2\right\rbrack\nonumber\\
&-&\left (1-\frac{2\Phi}{c^2}\right )\frac{\rho}{m}\frac{\partial S}{\partial
t}-\Phi\rho-V(\rho)
\end{eqnarray}

In the nonrelativistic limit
$c\rightarrow +\infty$, the Lagrangian reduces to
\begin{equation}
{\cal L}=-\frac{\rho}{m}\frac{\partial S}{\partial
t}-\frac{\rho}{2m^2}(\nabla
S)^2-\frac{\hbar^2}{8m^2}
\frac{(\nabla\rho)^2}{\rho}-\Phi\rho-V(\rho),
\end{equation}
and we obtain the quantum
Euler-Poisson equations
\cite{abrilph,playa,chavmatos} 
\begin{eqnarray}
\frac{\partial\rho}{\partial t}+\nabla\cdot (\rho
{\bf u})=0,
\label{nr3}
\end{eqnarray}
\begin{eqnarray}
\frac{\partial S}{\partial t}+\frac{(\nabla S)^2}{2 m}=\frac{\hbar^2}{2
m }\frac{\Delta\sqrt{\rho}}{\sqrt{\rho}}
-m\Phi-mh(\rho),
\label{nr6}
\end{eqnarray}
\begin{equation}
\frac{\partial {\bf u}}{\partial t}+({\bf u}\cdot
\nabla){\bf u}=
\frac{\hbar^2}{2m^2}\nabla\left( \frac{\Delta\sqrt{\rho}}{\sqrt{
\rho}} \right )-
\nabla\Phi-\frac{1}{\rho}\nabla P,
\label{nr4}
\end{equation}
\begin{eqnarray}
\Delta\Phi=4\pi G\rho,
\label{nr5}
\end{eqnarray}
where $h(\rho)$ and $P(\rho)$ are given by Eqs. (\ref{hvp}) and (\ref{kgp20g})
as before.

{\it Remark:} In the TF approximation, we can neglect the terms containing
$\hbar$. In that case, we have seen that the energy density $\epsilon$ and the
pressure $P$ are given by Eqs. (\ref{rtf6}) and (\ref{rtf7}). We note that Eq.
(\ref{rtf7}) coincides with
Eq. (\ref{kgp20g}). 
By contrast, $\epsilon\neq T_0^0$ even if we use the
Bernoulli equation to eliminate $\partial S/\partial t$.

\subsection{$|\varphi|^4$ potential}
\label{sec_phi4b}

We now consider a $|\varphi|^4$ (quartic) potential of the form
\begin{eqnarray}
V(|\varphi|^2)=\frac{\lambda}{4\hbar c}|\varphi|^4,
\label{pq1}
\end{eqnarray}
where $\lambda$ is the dimensionless self-interaction constant. For 
nonrelativistic BECs, the potential that occurs in the GP equation (\ref{gpp1})
is usually written as in Eq. (\ref{gpp2b}). Substituting Eq.
(\ref{kgp12b}) into Eq. (\ref{gpp2b}), we get
\begin{eqnarray}
V(|\varphi|^2)=\frac{2\pi a_s m}{\hbar^2}|\varphi|^4.
\label{pq2}
\end{eqnarray}
Comparing Eqs. (\ref{pq1}) and (\ref{pq2}), we obtain \cite{bectcoll}
\begin{eqnarray}
\frac{\lambda}{8\pi}=\frac{a_s m c}{\hbar}=\frac{a_s}{\lambda_C},
\label{pq3}
\end{eqnarray}
where $\lambda_C=\hbar/mc$  is the Compton wavelength of the bosons. For the
$|\varphi|^4$ model, the KG and GP equations take the form
\begin{equation}
\label{csf4phi4}
\square\varphi+\frac{m^2c^2}{\hbar^2}\varphi+\frac{8\pi
a_s m}{\hbar^2}|\varphi|^2\varphi=0,
\end{equation}
\begin{eqnarray}
i\hbar\frac{\partial\psi}{\partial
t}=-\frac{\hbar^2}{2m}\Delta\psi+m\Phi\psi+\frac{4\pi
a_s\hbar^2}{m^2}|\psi|^2\psi=0.
\label{mg1phi4}
\end{eqnarray}

In terms of the pseudo rest-mass density [see Eqs. (\ref{db4}) and (\ref{db2})],
the potential (\ref{pq2}) can be written as
\begin{eqnarray}
V(\rho)=\frac{2\pi a_s \hbar^2}{m^3}\rho^2.
\label{pq4}
\end{eqnarray}
Substituting Eq. (\ref{pq4}) into Eqs. (\ref{rtf6}) and (\ref{rtf7}) we obtain
\begin{eqnarray}
\epsilon=\rho c^2\left (1+\frac{6\pi a_s\hbar^2}{m^3c^2}\rho\right ),
\label{rtf6w}
\end{eqnarray}
\begin{eqnarray}
P=\frac{2\pi a_s\hbar^2}{m^3}\rho^2.
\label{rtf7w}
\end{eqnarray}
Eliminating $\rho$ between these two equations, we get 
\begin{eqnarray}
P=\frac{m^3c^4}{72\pi a_s\hbar^2}\left (\sqrt{1+\frac{24\pi
a_s\hbar^2}{m^3c^4}\epsilon}\mp 1\right )^2.
\label{rtf7qw}
\end{eqnarray}
This relativistic equation of state $P(\epsilon)$ was first obtained by Colpi
{\it et al.} \cite{colpi} in the context of boson stars. It was studied in
detail by Chavanis
and Harko \cite{chavharko} in connection to general relativistic BEC stars and
by Li {\it et al.} \cite{shapiro} and Su\'arez and Chavanis \cite{abrilphas} in
a BECDM cosmology (see also \cite{mlbec,action,csfpoly,ir1,ir2}
for related studies). This
equation of state reduces to that of an $n=1$ polytrope [see Eq. 
(\ref{rtf7w}) with $\epsilon\sim\rho c^2$] at low densities $\rho\ll
m^3c^2/|a_s|\hbar^2$
(nonrelativistic limit) and to the linear law $P\sim \epsilon/3$ similar to
the equation of state of the radiation at high
densities (ultrarelativistic limit).

{\it Remark:} Sometimes, a $|\varphi|^4$ self-interaction potential is written
as 
\begin{eqnarray}
V(|\varphi|^2)=\frac{m^2}{2\hbar^4}\lambda_s |\varphi|^4,
\label{pq5}
\end{eqnarray}
where $\lambda_s$ is the dimensional self-interaction constant. Comparing Eq.
(\ref{pq5})
with Eqs. (\ref{pq1}) and  (\ref{pq2}) we get
\begin{eqnarray}
\lambda_s=\frac{4\pi a_s \hbar^2}{m}=\frac{\lambda \hbar^3}{2m^2c}.
\label{pq6}
\end{eqnarray}

\section{Relativistic real SF}
\label{sec_rrsf}

In this Appendix, we discuss the main properties of a  relativistic real SF,
consider the instantonic potential of axions, take the nonrelativistic
limit, and justify the GPP equations (\ref{gpp1})-(\ref{gpp2b}) with 
$a_s<0$.

\subsection{Klein-Gordon-Einstein equations}
\label{sec_rkge}

For a real SF described by a canonical Lagrangian
\begin{eqnarray}
{\cal L}=\frac{1}{2}g^{\mu\nu}\partial_{\mu}\varphi\partial_{\nu}
\varphi-\frac{m^2c^2}{2\hbar^2}\varphi^2-V(\varphi),
\label{csf2j}
\end{eqnarray}
the KGE equations read
\begin{equation}
\label{rcsf4}
\square\varphi+\frac{m^2c^2}{\hbar^2}\varphi+\frac{dV}{d\varphi}=0,
\end{equation}
\begin{equation}
R_{\mu\nu}-\frac{1}{2}g_{\mu\nu}R=\frac{8\pi G}{c^4}T_{\mu\nu},
\label{rak21}
\end{equation}
with the energy-momentum tensor
\begin{equation}
\label{rem2}
T_{\mu\nu}=\partial_{\mu}\varphi\partial_{\nu}\varphi-g_{\mu\nu}{\cal L}.
\end{equation}

\subsection{The nonrelativistic limit}
\label{sec_nrlq}

In the nonrelativistic limit
$c\rightarrow +\infty$ where the SF displays rapid oscillations, the KGE
equations can be simplified by averaging over the oscillations. To that purpose,
we write 
\begin{equation}
\label{rkge15}
\varphi({\bf r},t)=\frac{1}{\sqrt{2}}\frac{\hbar}{m}\left\lbrack \psi({\bf
r},t)e^{-imc^2t/\hbar}+\psi^*({\bf r},t)e^{imc^2t/\hbar}\right\rbrack,
\end{equation}
where the complex wave function $\psi({\bf r},t)$ is a
slowly varying function of time (the fast oscillations $e^{imc^2t/\hbar}$ of the
SF have been factored out). This transformation (which is the counterpart
of the Klein transformation for a real SF) allows us to separate the fast
oscillations of the SF with pulsation $\omega=mc^2/\hbar$
caused by its rest
mass from the slow evolution of $\psi({\bf r},t)$. Using the simplest form 
of the Newtonian gauge [see Eq. (\ref{conf1})], substituting Eq. (\ref{rkge15})
into the KGE equations (\ref{rcsf4})-(\ref{rem2}) and averaging over the
oscillations we obtain the GPE equations (\ref{kge13}) and
(\ref{kgp14b}) with an effective potential $V_{\rm eff}(|\psi|^2)$ [see Secs. II
and III of Ref. \cite{phi6} and Appendix A of Ref. \cite{tunnel}
for the details of the derivation]. In the nonrelativistic limit
$c\rightarrow +\infty$, we obtain the GPP equations (\ref{nrl1}) and
(\ref{nrl2}) with $V_{\rm eff}(|\psi|^2)$ instead of $V(|\psi|^2)$. For a
real
SF, the effective potential $V_{\rm eff}(|\psi|^2)$ that
occurs in the GP equations (\ref{kge13}) and (\ref{nrl1})  is obtained from
the potential
$V(\varphi)$ that occurs in the KG equation (\ref{rcsf4}) by
first substituting Eq. (\ref{rkge15}) into  $V(\varphi)$, then averaging over
the oscillations. It is
different from the potential that one would obtain by directly replacing
$\varphi^2$ by
$(\hbar/m)^2|\psi|^2$ in the potential $V(\varphi)$.

{\it Remark:} General relativistic boson stars decribed by a real SF have a
naked singularity at the origin and are always
unstable \cite{js,jsarx}.  It is possible to
construct regular solutions which are periodic in time 
provided that  $M<M_{\rm max}^{\rm GR}=0.606\, M_P^2/m$ \cite{ssreal,alcubierre}
 but, on a
long timescale (which can nevertheless exceed the age of the universe), these
``oscillatons'' are
unstable and disperse to infinity or form a
black hole. Their instability is basically due to the fact that the
charge (boson number) is not conserved for a real SF. However, in the
nonrelativistic limit,
particle number conservation  is approximately restored and Newtonian boson
stars (like axion stars) can be stable (see Sec. \ref{sec_exact}).

\subsection{The instantonic potential of axions}
\label{sec_inst}

Axions are
hypothetical
pseudo-Nambu-Goldstone bosons of the
Peccei-Quinn phase transition associated with a $U(1)$ symmetry that
solves the strong charge parity (CP) problem of quantum chromodynamics (QCD).
The axion is a spin-$0$ particle with a very
small
mass  $m=10^{-4}\,
{\rm eV}/c^2$ and an extremely weak self-interaction (with a decay constant
$f=5.82\times 10^{10}\, {\rm GeV}$) arising
from
nonperturbative effects in QCD.
 Axions have
huge occupation numbers so they can be described by a classical 
relativistic quantum field theory with a real SF $\varphi({\bf r},t)$
whose
evolution is governed by the Klein-Gordon-Einstein (KGE) equations. The
instantonic
potential of axions is \cite{pq,wittenV,vv}
\begin{equation}
\label{inst1}
V(\varphi)=\frac{m^2cf^2}{\hbar^3}\left\lbrack 1-\cos\left
(\frac{\hbar^{1/2}c^{1/2}\varphi}{f}\right
)\right\rbrack-\frac{m^2c^2}{2\hbar^2}\varphi^2,
\end{equation}
where $m$ is the mass of the axion and $f$ is the axion decay
constant. For this potential, the KG equation (\ref{rcsf4}) takes the form
\begin{equation}
\Box\varphi+\frac{m^2c^{3/2}f}{\hbar^{5/2}}\sin\left
(\frac{\hbar^{1/2}c^{1/2}\varphi}{f}\right
)=0.
\label{kge5}
\end{equation}
This is the general relativistic sine-Gordon equation. Considering
the
dilute regime $\varphi\ll f/\sqrt{\hbar c}$ (which is valid in
particular in the nonrelativistic limit  $c\rightarrow
+\infty$)\footnote{According to Eq. (\ref{inst7}), the axion
decay
constant $f$ scales as $c^{3/2}$.} and expanding
the cosine term of Eq. (\ref{inst1}) in Taylor series, we obtain at leading
order
the
$\varphi^4$ potential
\begin{equation}
\label{inst3}
V(\varphi)=-\frac{m^2c^3}{24 f^2\hbar}\varphi^4.
\end{equation}
In that case, the KG
equation (\ref{rcsf4})
takes the form 
\begin{equation}
\Box\varphi+\frac{m^2c^2}{\hbar^2}\varphi-\frac{m^2c^3}{6f^2\hbar}\varphi^3=0.
\label{kge7}
\end{equation}

In the nonrelativistic limit, the effective potential  $V(|\psi|^2)$ appearing
in the GP
equation
(\ref{nrl1}) is given by \cite{ebycollapse,phi6}
\begin{equation}
\label{inst6}
V(|\psi|^2)=\frac{m^2cf^2}{\hbar^3}\left\lbrack
1-\frac{\hbar^3 c}{2f^2m^2}|\psi|^2-J_0\left
(\sqrt{\frac{2\hbar^{3}c|\psi|^2}{f^2m^2}}\right
)\right\rbrack,
\end{equation}
where $J_0$ is the Bessel function of zeroth order.\footnote{For convenience,
we write $V(|\psi|^2)$ instead of $V_{\rm eff}(|\psi|^2)$ now that we have
explained the meaning of the effective potential in Appendix \ref{sec_nrlq}.} If
we keep only the first
term in the expansion of Eq. (\ref{inst6}), we obtain
the
$|\psi|^4$ potential
\begin{equation}
\label{inst7}
V(|\psi|^2)=-\frac{\hbar^3
c^3}{16f^2m^2}|\psi|^4.
\end{equation}
This approximation is valid for dilute axion stars satisfying $|\psi|^2\ll
f^2m^2/\hbar^3c$. 
We note that $V(|\psi|^2)\equiv
\overline{V(\varphi)}$ is different from the expression that one would have
naively obtained by directly substituting $\varphi^2=(\hbar/m)^2|\psi|^2$
into Eq.
(\ref{inst1}). The difference is  already apparent in the first term of the
expansion of the
potential which involves a coefficient $-1/16$ [see Eq. (\ref{inst7})] instead
of $-1/24$ [see Eq. (\ref{inst3})]. They differ by a factor $2/3$. This is
because $\varphi$ is a real SF. Therefore, substituting $\varphi$ (exact) from
Eq. (\ref{rkge15}) into $V(\varphi)$, {\it then} averaging over the
oscillations,
is different from substituting $\varphi^2=(\hbar/m)^2|\psi|^2$
(already
averaged
over the oscillations) into $V(\varphi)$.

In general, a quartic potential is written as
\begin{equation}
V(\varphi)=\frac{\lambda}{4\hbar c}\varphi^4,
\label{kge8}
\end{equation}
where $\lambda$ is the dimensionless self-interaction constant. Comparing Eqs.
(\ref{inst3}) and (\ref{kge8}), we find that
\begin{equation}
\lambda=-\frac{m^2c^4}{6f^2}.
\label{kge9}
\end{equation}
On the other hand, comparing Eq. (\ref{inst7}) with Eq. (\ref{gpp2b}), we
obtain
\begin{equation}
\label{inst8}
a_s=-\frac{\hbar c^3m}{32\pi f^2}.
\end{equation}
Equations (\ref{kge9}) and (\ref{inst8}) then yield
\begin{eqnarray}
\frac{\lambda}{8\pi}=\frac{2a_s m c}{3\hbar}.
\label{kge31}
\end{eqnarray}
The
relation between $\lambda$
and $a_s$ is different for a real SF and for a complex SF (see Appendix
\ref{sec_ra}).
They differ by a factor $2/3$ for the reason
indicated previously.

We note that the self-interaction constant  $\lambda$ or the scattering length
$a_s$ is negative, so that the $\varphi^4$
self-interaction term for axions is {\it attractive}. This attraction is
responsible for the collapse of
dilute axion stars above the maximum mass $M_{\rm max}^{\rm NR}$ from Eq.
(\ref{dm16}) obtained in Refs. 
\cite{prd1,bectcoll}. The
next order
$\varphi^6$ term in the expansion of the potential (\ref{inst1})
has been considered in Refs. \cite{ebycollapse,phi6} and
turns out to
be repulsive. This
repulsion, that occurs at high densities and which has a relativistic origin,
may stop the collapse of dilute axion
stars and lead to the formation of dense axion stars \cite{braaten} (see,
however, footnote 18 concerning their possible instability with respect to
relativistic decay).

\section{Simple model of extended elementary particle}
\label{sec_smeep}

In the main text, we have considered a self-gravitating BEC made of $N$ bosons 
of individual mass $m$. In the nonrelativistic limit, the mass-radius relation
of the BEC is given by Eq. (\ref{pa1}). Using Eq. (\ref{lambda}), this relation
can be rewritten as
\begin{eqnarray}
M=\frac{a\frac{\hbar^2}{Gm^2R}}{1-b^2\frac{\lambda\hbar^3}{8\pi Gm^4cR^2}}.
\label{mo1}
\end{eqnarray}

We now introduce a  simple model of extended elementary particle.\footnote{This
model can be related to the models of extended particles listed in footnote
13.} We consider a quantum particle of mass $m$ and we assume that it is
confined by the gravitational potential created by its own wave function.
More specifically, we assume that the wave function $\psi({\bf r},t)$ governed
by the GP
equation  (\ref{gpp1}) determines the density profile $\rho({\bf r},t)$
of the particle through the relation $\rho=|\psi|^2$. The corresponding density
$\rho({\bf r},t)$ creates, via the Poisson equation (\ref{gpp2}), a
gravitational potential $\Phi({\bf r},t)$ which enters into the GP equation
(\ref{gpp1}). Finally, we identify the mass $M=\int\rho\, d{\bf r}$ produced by
$\rho$ with the mass $m$ of the particle. This model is similar to the model
introduced by Diosi \cite{diosi} who proposed that the spreading of the
wavepacket of a free particle is prevented by the gravitational potential
created by its own wave function $\psi$.  This interpretation gives to the
Schr\"odinger-Poisson
equation the status of a fundamental equation of physics (i.e. more than the
Schr\"odinger equation alone). For the sake of generality, we complete this
model by taking into account a possible self-interaction of the particle and
consider the GPP equations (\ref{gpp1}) and (\ref{gpp2}) with an arbitrary
value of $a_s$ (or $\lambda$). 

In this model, the mass-radius relation of the particle is obtained by setting
$M=m$ in Eq. (\ref{mo1}) yielding 
\begin{eqnarray}
R^2-a\frac{\hbar^2}{Gm^3}R-b^2\frac{\lambda\hbar^3}{8\pi Gm^4c}=0.
\label{mo2}
\end{eqnarray}
The solution of this second degree equation is
\begin{eqnarray}
R=\frac{a\hbar^2}{2Gm^3}\pm \sqrt{\frac{a^2\hbar^4}{4G^2m^6}+\frac{b^2\lambda\hbar^3}{8\pi Gm^4c}}.
\label{mo3}
\end{eqnarray}
We must select the sign $+$ when $\lambda\ge 0$ while the two
signs are allowed when $\lambda<0$.

\subsection{No self-interaction}
\label{sec_nosi}

For a noninteracting particle ($\lambda=0$),  we obtain the mass-radius relation
\begin{eqnarray}
R=a\frac{\hbar^2}{Gm^3}.
\label{mo4}
\end{eqnarray}
The exact prefactor is $9.95$. This returns the
result obtained in \cite{diosi,km,ak}.\footnote{These
authors did not give the value of the prefactor.}

{\it Remark:} If we apply this model to the electron of mass $m=m_e=9.11\times
10^{-28}\, {\rm g}$ we 
obtain a radius 
\begin{eqnarray}
R\sim \frac{\hbar^2}{Gm_e^3}\sim \left (\frac{M_P}{m_e}\right )^2\frac{\hbar}{m_e c}=2.20\times 10^{32}\, {\rm m}.
\label{mo5}
\end{eqnarray}
Since $M_P/m_e\sim 10^{20} \gg 1$, this radius  is much larger than the
Compton wavelength of the electron
$\lambda_e=\hbar/(m_e c)=3.86\times 10^{-13}\, {\rm m}$ which provides an
estimate of 
its typical size (see Appendix \ref{sec_mee} for more details). Therefore, this
model cannot
describe the electron. This is because self-gravity is negligible at the scale
of the electron (see Appendix \ref{sec_pm}).

\subsection{Repulsive self-interaction}

\begin{figure}[!h]
\begin{center}
\includegraphics[clip,scale=0.3]{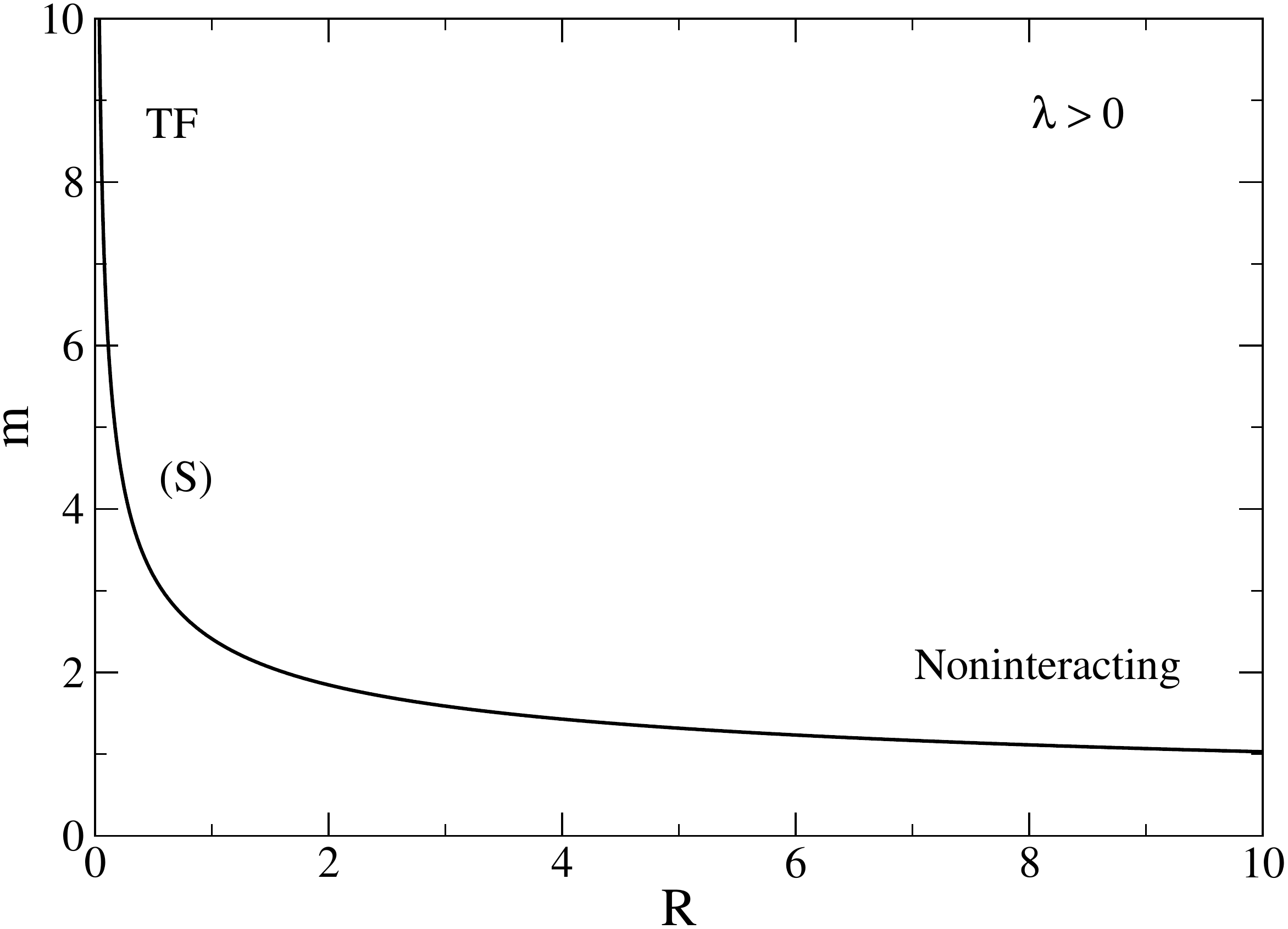}
\caption{Mass-radius relation of an extended elementary
particle with $\lambda>0$. The mass is normalized by
$m_a=(8\pi/\lambda)^{1/2}M_P$ and the radius by
$R_a=(\lambda/8\pi)^{3/2}l_P$ (this amounts to taking
$\hbar=c=G=\lambda/8\pi=1$ in the dimensional equations). We have taken
$a=9.946$ and $b=\pi$.}
\label{particleGNRpos}
\end{center}
\end{figure}

For a repulsive self-interaction ($\lambda>0$), the mass-radius relation is 
given by Eq. (\ref{mo3}) with the sign $+$. The radius $R$ of the particle
decreases with its  mass $m$ (see Fig. \ref{particleGNRpos}). For
$m\rightarrow 0$ (noninteracting limit) we recover Eq.  (\ref{mo4}). For
$m\rightarrow +\infty$ (TF limit)  we find that 
\begin{eqnarray}
R\sim b\sqrt{\frac{\lambda}{8\pi}}\left (\frac{\hbar^3}{Gm^4 c}\right )^{1/2}.
\label{mo7}
\end{eqnarray}
The exact prefactor is $\pi$. Therefore, the radius $R$ behaves like $m^{-3}$
when $m\ll m_a=(8\pi\hbar c/\lambda G)^{1/2}$ and like $m^{-2}$ when $m\gg
m_a=(8\pi\hbar c/\lambda G)^{1/2}$.

Actually, in the TF approximation, we can solve the problem exactly. Using Eq.
(\ref{lambda}), we can rewrite Eq. (\ref{dm7tf}) as
\begin{eqnarray}
\label{qw1}
\Delta\rho+\frac{8\pi G m^4 c}{\lambda \hbar^3}\rho=0.
\end{eqnarray}
The density profile of the extended particle, which is determined by Eq.
(\ref{qw1}), is given by [see Eq. (\ref{mg18})]
\begin{eqnarray}
\label{qw2}
\rho(r)=\frac{\rho_0 R}{\pi r}\sin \left (\frac{\pi r}{R}\right ).
\end{eqnarray}
Its radius is given by [see Eq. (\ref{dm15})]
\begin{eqnarray}
R\sim \pi\sqrt{\frac{\lambda}{8\pi}}\left (\frac{\hbar^3}{Gm^4 c}\right )^{1/2}
\label{mo7gt}
\end{eqnarray}
and its central density is given by [see Eq. (\ref{rhoM})] 
\begin{eqnarray}
\label{qw3}
\rho_0=\frac{\pi m}{4 R^3}.
\end{eqnarray}
On the other hand, according to Eq. (\ref{Etot}), the total energy
(gravitational $+$ internal) of the particle is
\begin{eqnarray}
\label{qw4}
E_{\rm tot}=-\frac{Gm^2}{2R}.
\end{eqnarray}
Finally, its pulsation is of the order of the inverse dynamical time [see Eq.
(\ref{tdyn})] 
\begin{eqnarray}
t_D\sim \frac{1}{\sqrt{G\rho_0}}\sim \left (\frac{R^3}{Gm}\right
)^{1/2}.
\end{eqnarray}

{\it Remark:} If we apply this model to the electron of mass 
$m_e=9.11\times 10^{-28}\, {\rm g}$
and radius $r_e=e^2/(m_e c^2)=2.82\times 10^{-15}\, {\rm m}$ (see Appendix
\ref{sec_mee}), we find that the self-interaction constant must be equal to
\begin{eqnarray}
\frac{\lambda}{8\pi}=\frac{\alpha^2}{\pi^2}\left (\frac{m_e}{M_P}\right
)^2=9.45\times 10^{-51},
\end{eqnarray}
where $\alpha$ is the fine-structure constant (\ref{alpha}).
However, the TF approximation is valid when $\lambda\gg (M_P/m_e)^2=5.71\times
10^{44}$ 
so we see that this condition is not satisfied for a particle of mass $m_e\ll
M_P$. Therefore, this model cannot describe the electron. This is because
self-gravity is negligible at the scale of the electron (see
Appendix \ref{sec_pm}).

\subsection{Attractive self-interaction}
\label{sec_gh}

\begin{figure}[!h]
\begin{center}
\includegraphics[clip,scale=0.3]{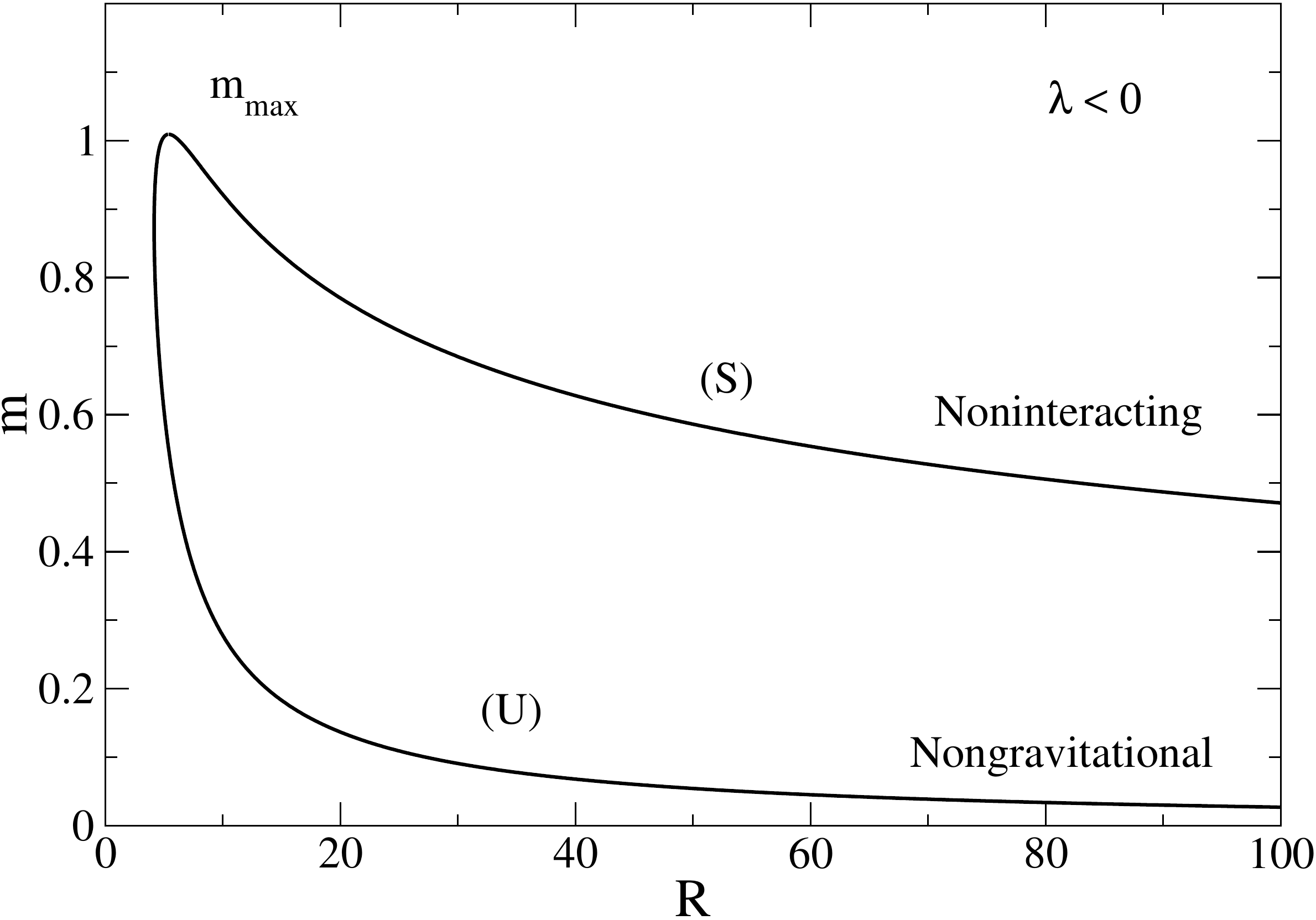}
\caption{Mass-radius relation of an extended elementary
particle with $\lambda<0$. The mass is normalized by
$m_a=(8\pi/|\lambda|)^{1/2}M_P$ and the radius by
$R_a=(|\lambda|/8\pi)^{3/2}l_P$ (this amounts to taking
$\hbar=c=G=|\lambda|/8\pi=1$ in the dimensional equations). We have taken
$a=11.1$ and $b=5.5$.}
\label{particleGNRneg}
\end{center}
\end{figure}

For an attractive self-interaction ($\lambda<0$), the two signs are allowed 
in the mass-radius relation from Eq. (\ref{mo3}). This determines two branches
of solutions. These two branches merge at the maximum mass (see Fig.
\ref{particleGNRneg})
\begin{eqnarray}
m_{\rm max}=\frac{a}{2b}\sqrt{\frac{8\pi}{|\lambda|}} \left (\frac{\hbar c}{G}\right )^{1/2},\label{mo9}
\end{eqnarray}
corresponding to the radius 
\begin{eqnarray}
R_*= \frac{4b^3}{a^2}\left (\frac{|\lambda|}{8\pi}\right )^{3/2} \left (\frac{G\hbar}{c^3}\right )^{1/2}.
\label{mo13}
\end{eqnarray}
The exact prefactors are $1.012$ and $5.5$. The branch associated with the sign
$+$ 
corresponds to stable solutions. For $R\rightarrow
+\infty$ and
$m\rightarrow 0$ (noninteracting limit) we recover Eq. (\ref{mo4}). The branch
associated with the sign $-$ corresponds to unstable solutions. For
$R\rightarrow 0$ and
$m\rightarrow 0$ (nongravitational limit) we
obtain
\begin{eqnarray}
R\sim\frac{b^2}{a}\frac{|\lambda|}{8\pi}\frac{\hbar}{mc},
\label{mo11}
\end{eqnarray}
where $\hbar/(mc)$ is the Compton wavelength of the particle. The exact
prefactor is $3.64$.

For $\lambda\sim 1$, the 
maximum mass is of the order of the Planck mass $M_P=(\hbar
c/G)^{1/2}=2.18\times 10^{-5}\, {\rm g}$ and the corresponding radius is of the
order of the Planck length $l_P=(G\hbar /c^3)^{1/2}=1.62\times 10^{-35}\, {\rm
m}$, which is also the semi-Schwarzschild radius $GM_P/c^2$ corresponding to the
Planck mass. It is interesting to see that the Schwarzschild radius enters into
the problem although we have used a nonrelativistic approach. This is due to the
definition of $\lambda$ in Eq. (\ref{lambda}). Actually, our nonrelativistic
approach is valid for $|\lambda|\gg 1$.
Therefore, the maximum mass is much smaller than the Planck mass and its radius
is much larger than the Planck length (or Schwarzschild radius). This is
relevant to describe an elementary particle. However, in that case, we
have to take into account electrostatic forces (see Appendix
\ref{sec_pm}), except if the particle is uncharged.

{\it Remark:} Eliminating $|\lambda|$ between Eqs. (\ref{mo9}) and (\ref{mo13}) we find that
\begin{eqnarray}
R_*=\frac{a}{2}\frac{\hbar^2}{Gm_{\rm max}^3}.
\label{mo11b}
\end{eqnarray}
This is the same scaling as in Eq. (\ref{mo4}). Therefore, as in
Appendix \ref{sec_nosi}, this
model cannot describe the electron because it
would yield a too large radius. Actually,
substituting $m=m_e$ and $R=r_e=e^2/(m_e c^2)$ (see Appendix
\ref{sec_mee}) in Eq. (\ref{mo2}) and introducing the fine structure constant
(\ref{alpha}) we get 
\begin{eqnarray}
\frac{\lambda}{8\pi}=\frac{1}{b^2}\alpha^2\left (\frac{m_e}{M_P}\right )^2-\frac{a}{b^2}\alpha.
\label{mo11c}
\end{eqnarray}
Since $m_e\ll M_P$, this formula reduces to
\begin{eqnarray}
\frac{\lambda}{8\pi}=-\frac{a}{b^2}\alpha,
\label{mo11d}
\end{eqnarray}
corresponding to the nongravitational limit ($m\ll m_{\rm max}$ and $R\ll R_*$).
Interestingly, in this limit, the mass-radius relation from Eq. (\ref{mo11}) is
consistent with the mass-radius relation of the electron [see Eq.
(\ref{mee2}) with $R=r_e$] provided that $\lambda$ is
given by Eq. (\ref{mo11d}), which is a pure number
$\lambda=-0.275\alpha=-2.00\times 10^{-3}$. Unfortunately, this equilibrium
state is unstable (see Sec. \ref{sec_att}).

\subsection{Tunnel effect}
\label{sec_tunnel}

For an attractive self-interaction ($\lambda<0$), since the stable equilibrium
state with $m<m_{\rm max}$ is only metastable,
the particle can overcome the barrier of potential by tunnel effect and collapse
(possibly becoming a black hole).  It has therefore a finite lifetime.
The lifetime of a self-gravitating BEC with an attractive self-interaction has
been calculated in \cite{tunnel} by using the instanton theory. It is found that
$t_{\rm life}\sim e^{N b(M)}$, where $N$ is the number of particles in the BEC
and $b(M)$ is a function related to the barrier of potential which tends to zero
when $M\rightarrow M_{\rm max}$. Since the number of bosons in a boson star is
huge, of the order of $N\sim 10^{50}-10^{100}$, the probability of a boson star
to collapse by tunnel effect  is completely 
negligible, being of order $e^{-N}$ (except when $M=M_{\rm max}$
where it scales as $N^{-1/5}$). However, in the case of an elementary particle,
we have $N=1$ so the tunneling probability is determined by the function
$b(m)$. 

When $m\rightarrow m_{\rm max}$, we can obtain the lifetime $t_{\rm
life}\sim 1/\Gamma$ of the particle by taking $N=1$ in Eq. (86) of
\cite{tunnel}. This yields 
\begin{eqnarray}
\Gamma\sim 12\left (\frac{8}{\pi^2}\right )^{1/4}\left\lbrack 2\left (1-\frac{m}{m_{\rm max}}\right )\right\rbrack^{7/8}(\alpha\sigma)^{1/4}\nonumber\\
\times e^{-\frac{24}{5}\sqrt{2}\left\lbrack 2\left (1-\frac{m}{m_{\rm max}}\right )\right\rbrack^{5/4}\sqrt{\alpha\sigma}}t_D^{-1}
\label{mo14}
\end{eqnarray}
where
\begin{eqnarray}
t_D=\left (\frac{\alpha}{\nu}\right )^{1/2}\frac{1}{\sqrt{G\rho_0}}=\frac{6\pi\zeta}{\nu}\left (\frac{\alpha}{\nu}\right )^{1/2}\frac{|a_s|\hbar}{Gm^2}
\label{mo15}
\end{eqnarray}
is the dynamical time. Within the Gaussian ansatz, the values 
of the coefficients are $\alpha_{\rm G}=3/2$, $\sigma_{\rm G}=3/4$, $\zeta_{\rm
G}=1/(2\pi)^{3/2}$, and  $\nu_{\rm G}=1/\sqrt{2\pi}$
\cite{prd1,tunnel}.  Using Eq. (\ref{lambda}) we can rewrite the dynamical time
as
\begin{eqnarray}
t_D\sim \frac{|\lambda|\hbar^2}{Gm^3c}.
\label{mo16}
\end{eqnarray}
For $m\sim m_{\rm max}$ and $R\sim R_*$ with [see Eqs. (\ref{mo9}) and
(\ref{mo13})]
\begin{eqnarray}
m_{\rm max}\sim  \frac{M_P}{\sqrt{|\lambda|}}\quad {\rm and}\quad R_*\sim |\lambda|^{3/2}\, l_P,
\label{mo17}
\end{eqnarray}
we obtain
\begin{eqnarray}
t_D\sim |\lambda|^{5/2}t_P,
\label{mo18}
\end{eqnarray}
where $t_P=(\hbar G/c^5)^{1/2}=5.39\times 10^{-44}\, {\rm s}$ is the Planck
time.

In order to have a long 
lifetime we need $|\lambda|\gg 1$ hence $m\ll M_P$ (this condition is actually
required by the validity of the nonrelativistic approximation). Therefore, in
this model, elementary particles with a mass $m\sim m_{\rm max}\ll M_P$ have a
long lifetime. However, when $m\ll M_P$, it is generally necessary to take into
account electrostatic interactions (see Appendix
\ref{sec_pm}), except if the particles
are neutral.

Alternatively, for a particle of mass $m\sim m_{\rm max}\sim M_P$,
corresponding 
to $|\lambda|\sim 1$, electrostatic interactions may be neglected (at least
marginally) but the lifetime of the particle is of the order of the
Planck time $t_P$. In that case, we have an elementary particle of mass $M_P$
and radius $l_P$ (of the order of the Schwarzschild radius). This ``Planck
particle'' (planckion) may be destabilized by tunnel effect and collapse
quasi-immediately towards a Planck
black hole of mass $M_P$ on a timescale $\sim t_P$. This scenario should be
confirmed by a general relativistic calculation.

\subsection{Relativistic effects}
\label{sec_releff}

In the main text, we have 
determined general equations giving the maximum mass $M_{\rm max}$ of a
relativistic self-gravitating gas of bosons of individual mass $m$ at $T=0$.
Taking $M=m$ in these equations, we obtain the maximum mass $m_{\rm max}$ of an
elementary particle described by the KGE equations (\ref{csf4b}) and
(\ref{ak21}) in which the gravitational field is produced by the wavefunction
$\varphi$ of the particle itself.

For a noninteracting particle, using Eq. (\ref{er1wy}),  we get
\begin{eqnarray}
\label{mo23}
m_{\rm max}=0.796\, M_P.
\end{eqnarray}
The maximum mass of the elementary particle is of the order of the Planck mass. 
This result may be connected to the result of Rosen \cite{rosenkg}.

For a particle with a repulsive self-interaction in the TF approximation
($\lambda\gg 1$), using Eq. (\ref{mtf}), we get
\begin{eqnarray}
\label{mo24}
m_{\rm max}=0.394\,\lambda^{1/6}\, M_P.
\end{eqnarray}
The maximum mass of the elementary particle is larger than the Planck mass.

For a particle with an attractive self-interaction in the nonrelativistic limit
($|\lambda|\gg 1$), using Eq. (\ref{dm16l}), we get
\begin{eqnarray}
\label{mo25}
m_{\rm max}=5.07\, \frac{M_P}{\sqrt{|\lambda|}}.
\end{eqnarray}
The maximum mass of the elementary particle is smaller than the Planck mass (see
Appendix \ref{sec_tunnel}).

\begin{figure}[!h]
\begin{center}
\includegraphics[clip,scale=0.3]{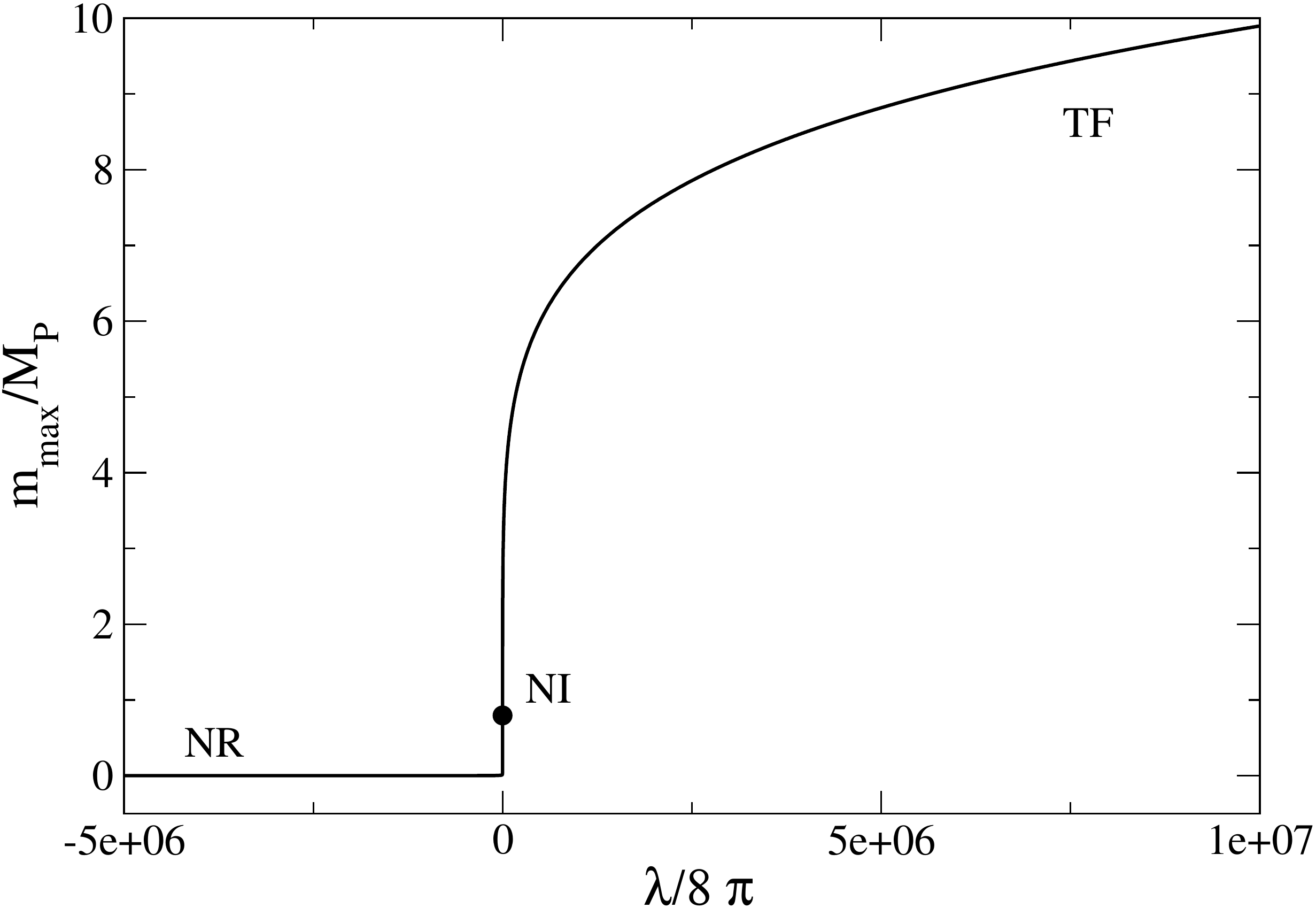}
\caption{Maximum mass $m_{\rm max}$ of a relativistic extended elementary
particle as a function of the self-interaction parameter $\lambda$.}
\label{lambdaMmax}
\end{center}
\end{figure}

More generally, for a particle with an arbitrary repulsive self-interaction, 
using the interpolation formula (\ref{e1jj}), we obtain the relation $m_{\rm
max}(\lambda)$ under the inverse form
\begin{eqnarray}
\label{mo26}
\frac{\lambda}{8\pi}=\frac{2.50\, \left (\frac{m_{\rm max}}{M_P}\right )^4-1}{0.235\, \left (\frac{M_P}{m_{\rm max}}\right )^2}.
\end{eqnarray}
On the other hand, for a particle with an arbitrary attractive self-interaction,
using the
interpolation formula (\ref{e1}), we obtain the relation $m_{\rm max}(\lambda)$
under the inverse form
\begin{eqnarray}
\label{mo27}
\frac{\lambda}{8\pi}=\frac{1-0.401\, \left (\frac{M_P}{m_{\rm max}}\right )^4}{0.391\, \left (\frac{M_P}{m_{\rm max}}\right )^2}.
\end{eqnarray}
Equation (\ref{mo26}) is a third degree equation for $(m_{\rm max}/M_P)^2$ 
while Eq. (\ref{mo27}) is a second degree equation for $(M_P/m_{\rm
max})^2$. The function $m_{\rm max}(\lambda)$ is plotted in
Fig. \ref{lambdaMmax}.

\begin{figure}[!h]
\begin{center}
\includegraphics[clip,scale=0.3]{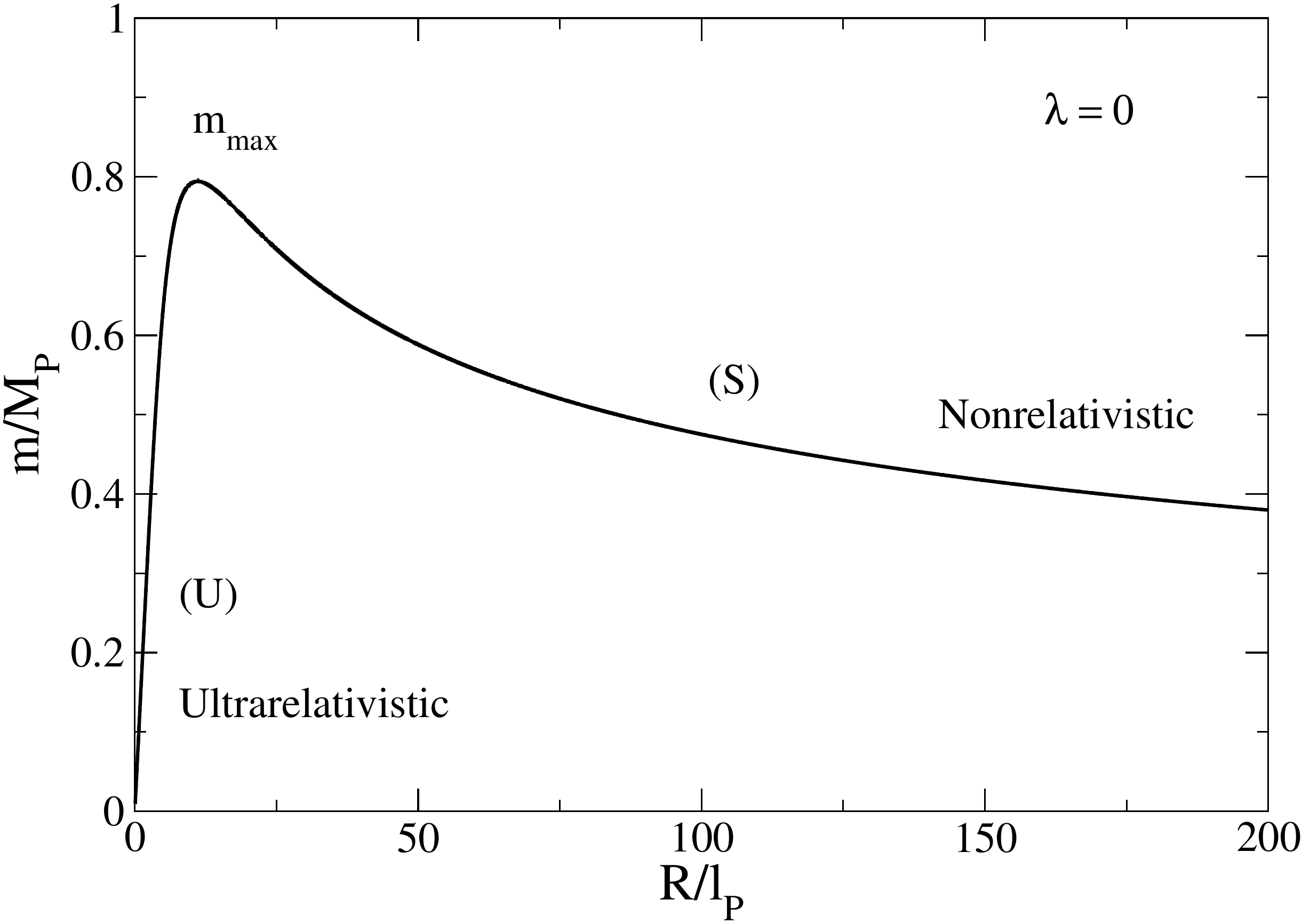}
\caption{Mass-radius
relation of a relativistic extended elementary
particle without self-interaction ($\lambda=0$).}
\label{particleGNRnegRELAT}
\end{center}
\end{figure}

{\it Remark:} For an attractive self-interaction, the relativistic mass-radius 
relation of the particle can be obtained from Eq. (\ref{mo3}) by making the
substitution (see Sec. \ref{sec_rq})
\begin{eqnarray}
\label{mo27b}
\lambda\rightarrow \lambda\left (1-\kappa\frac{8\pi Gm^2}{\lambda c \hbar}\right ),
\end{eqnarray}
yielding
\begin{eqnarray}
R=\frac{a\hbar^2}{2Gm^3}\pm\sqrt{\frac{a^2\hbar^4}{4G^2m^6}+\frac{b^2\lambda\hbar^3}{8\pi Gm^4c}\left (1-\kappa\frac{8\pi Gm^2}{\lambda c \hbar}\right )}.\nonumber\\
\label{mo3gv}
\end{eqnarray}
For $\lambda<0$ the mass-radius relation has a shape similar to that of
Fig. \ref{particleGNRneg}. For a noninteracting ($\lambda=0$) relativistic
particle, the mass-radius relation reduces to
\begin{eqnarray}
R=\frac{a\hbar^2}{2Gm^3}\pm\sqrt{\frac{a^2\hbar^4}{4G^2m^6}-\frac{\kappa b^2\hbar^2}{m^2c^2}}.
\label{mo3gvb}
\end{eqnarray}
It is plotted in Fig. \ref{particleGNRnegRELAT}. From this relation, we recover
the maximum mass (\ref{mo23}) and the
 nonrelativistic mass-radius relation (\ref{mo4}) for $m\ll m_{\rm max}$ and
$R\gg R_*$ (or
$c\rightarrow +\infty$). In the ultrarelativistic limit $m\ll m_{\rm max}$ and
$R\ll R_*$, we have $m\sim ac^2R/\kappa b^2G$ but these configurations are
unstable.

\subsection{Planck mass}
\label{sec_pm}

Let us introduce the ratio between the electrostatic force and the
gravitational force $F={e^2}/{Gm^2}$,
where $e$ is the elementary charge (the charge of the electron) 
and $m$ is the mass of the particle under consideration. Electrostatic and
gravitational forces are comparable ($F\sim 1$) when $m\sim e/\sqrt{G}$.
Introducing the fine-structure constant (\ref{alpha}) 
and treating $\alpha$ as a 
dimensionless number of order unity, the condition $F\sim 1$ can be rewritten as
$m\sim M_P$, where $M_P=(\hbar c/G)^{1/2}=2.18\times 10^{-5}\, {\rm g}$ is the
Planck mass. When $m\ll M_P$ (i.e. $e^2\gg Gm^2$), the electrostatic forces
overcome the
gravitational forces. For example, for the electron of mass $m_e=9.11\times
10^{-28}\, {\rm g}$, we have $F\sim 10^{40}$ (Weyl number). Therefore,
self-gravity is
completely negligible at the scale of the electron. When $m\gg M_P$ (i.e.
$Gm^2\gg e^2$), the
gravitational forces overcome the electrostatic forces. When $m\sim M_P$ (i.e.
$e^2\sim Gm^2$), corresponding to the Planck scale or the grand unification
energy scale, the gravitational and the electrostatic forces are
comparable. Therefore, models of extended elementary charged particles that
include
gravitational and electrostatic forces have masses of the order of the Planck
mass $M_P$ (up to a factor $\alpha$) \cite{ak}.

\section{Charged bosons}
\label{sec_elec}

In this Appendix, we consider 
the case of BECs made of charged bosons. We assume that the bosons have a mass
$m$ and carry a charge $e$. We stress that $e$ is not necessarily equal to the
elementary charge of the electron.

\subsection{Without self-gravity}
\label{sec_elecn}

We first ignore the gravitational interaction between bosons and consider the electrostatic GPP equations
\begin{eqnarray}
\label{elec1}
i\hbar \frac{\partial\psi}{\partial
t}=-\frac{\hbar^2}{2m}\Delta\psi+\frac{4\pi a_s\hbar^2}{m^2}|\psi|^2\psi+m\Phi\psi,
\end{eqnarray}
\begin{equation}
\label{elec2}
\Delta\Phi=-\frac{4\pi e^2}{m^2} |\psi|^2.
\end{equation}
Since the electrostatic interaction $e^2/r$ between two charges is similar to 
the gravitational interaction $-Gm^2/r$ between two masses (except for the
change of sign), the results of Sec. \ref{sec_pa} remain valid provided that we
make the substitution $Gm^2\rightarrow -e^2$ in the equations. Since the
electrostatic force between two charges is repulsive, and since the quantum
potential also has a repulsive nature, equilibrium states can exist only if the
self-interaction between bosons is attractive. Therefore, in the following, we
assume $a_s<0$. 

\begin{figure}[!h]
\begin{center}
\includegraphics[clip,scale=0.3]{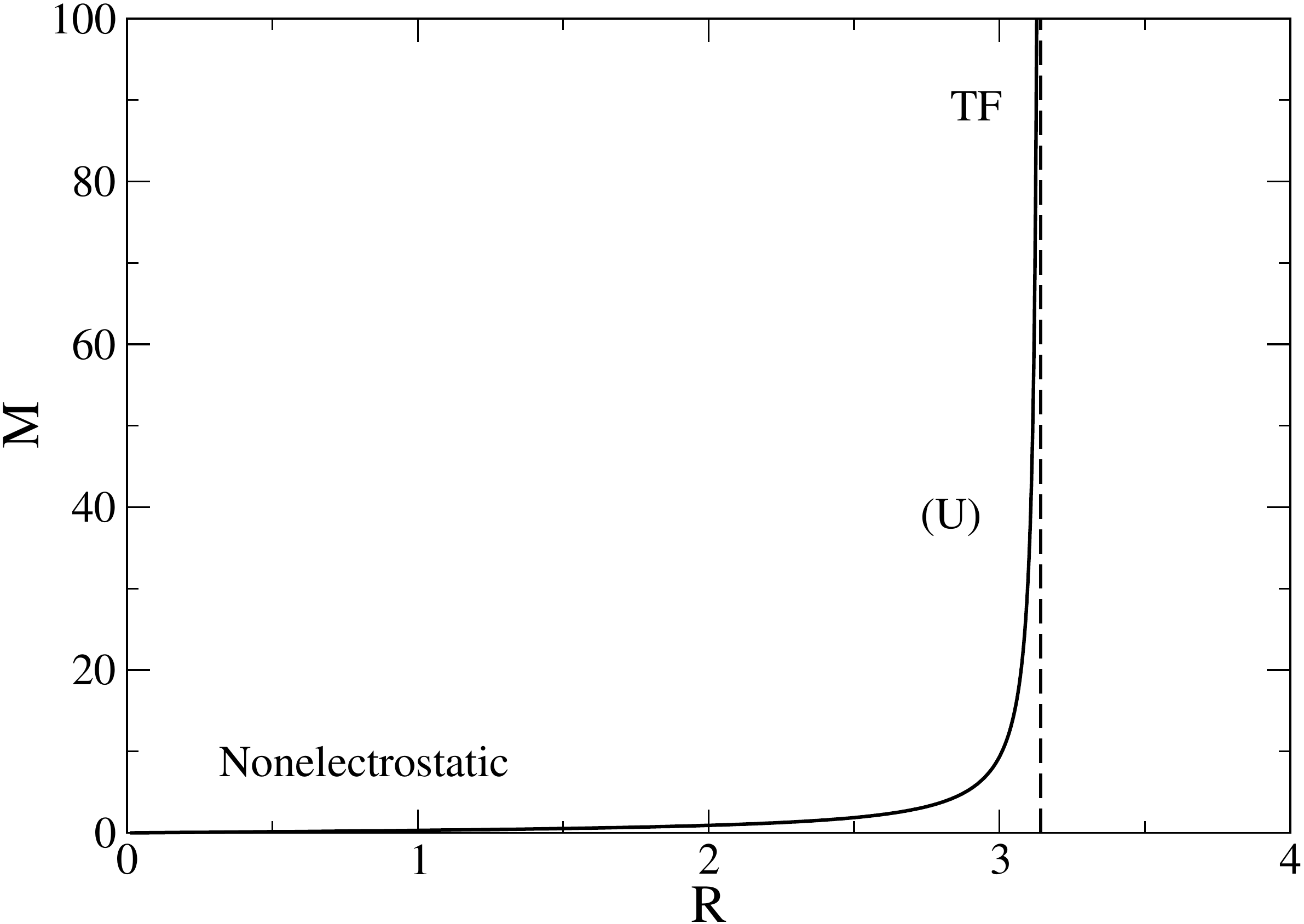}
\caption{Mass-radius relation of charged BECs with $a_s<0$ (for illustration
we have used Eq. (\ref{elec3}) with $a=2.71$ and $b=\pi$).
The mass is
normalized by
$M_a=(m\hbar^2/e^2|a_s|)^{1/2}$ and the radius by
$R_a=(|a_s|\hbar^2/e^2m)^{1/2}$.}
\label{RMcharge}
\end{center}
\end{figure}

In that case, the mass-radius relation from Eq. (\ref{pa1}) becomes
\begin{eqnarray}
M=\frac{a\frac{\hbar^2}{e^2R}}{b^2\frac{|a_s|\hbar^2}{e^2mR^2}-1}.
\label{elec3}
\end{eqnarray}
The radius 
increases monotonically with the mass (see Fig. \ref{RMcharge}) and tends to the
asymptotic value
\begin{eqnarray}
R_{\rm max}=b\left (\frac{|a_s|\hbar^2}{e^2m}\right )^{1/2}
\label{elec4}
\end{eqnarray}
when $M\rightarrow +\infty$ (TF limit). This corresponds to
the result from Eq. (\ref{pa5}) with the above substitution.  For $M\rightarrow
0$ (nonelectrostatic limit), we have
\begin{eqnarray}
R\sim \frac{b^2}{a}\frac{|a_s|M}{m},
\label{elec5}
\end{eqnarray}
returning the result from Eq. (\ref{pa9}). From these asymptotic results, we 
obtain $b=\pi$ and $a/b^2=0.275$ yielding $a=2.71$.

The total energy of the BEC [see Eq. (\ref{mra11})] is 
\begin{eqnarray}
E_{\rm tot}(R)=\sigma\frac{\hbar^2M}{m^2R^2}+\nu
\frac{e^2M^2}{m^2R}-\zeta\frac{2\pi
|a_s|\hbar^2M^2}{m^3R^3}.
\label{elec6}
\end{eqnarray}
We see that the equilibrium state is always 
unstable (it is a maximum of energy at fixed mass). This probably precludes
physical applications of this model.

{\it Remark:} In the nonelectrostatic limit ($e=0$ and $a_s<0$), 
we can use the exact results from Sec. \ref{sec_att} valid for nongravitational
BECs with
an attractive self-interaction ($G=0$ and $a_s<0$). In the TF limit ($\hbar=0$),
we can use the exact results from Sec. \ref{sec_rep} provided that we make the
substitutions $Gm^2\rightarrow -e^2$ and $a_s\rightarrow -|a_s|$. The density
profile of the BEC is given by Eq. (\ref{mg18}), its radius by Eq. (\ref{elec4})
with $b=\pi$, its central density by Eq. (\ref{rhoM}), its energy by
$E_{\rm tot}=M^2e^2/2m^2R_{\rm max}$ and the instability time by
$t_{D}\sim (m^2R_{\rm max}^3/Me^2)^{1/2}$. Therefore, in
the TF limit, a gas of charged bosons with an attractive self-interaction
($-e^2<0$ and $a_s<0$) has the same structure as a boson star with a repulsive
self-interaction ($G>0$ and $a_s>0$).
However, a boson star with a repulsive self-interaction is stable while a gas
of charged bosons  with an attractive self-interaction is unstable. In this
connection we note that the total energy $E_{\rm tot}$ is negative in the
gravitational case and positive in the electrostatic case.

\subsection{With self-gravity}
\label{sec_elecw}

If we take into account the gravitational force and the electrostatic 
force between bosons, we just have to make the substitution $Gm^2\rightarrow
Gm^2-e^2$ or, equivalently, $G\rightarrow G(1-e^2/Gm^2)$ in the equations
of Sec. \ref{sec_pa}. The effect of the electrostatic repulsion is to decrease
the value of the gravitational constant. Introducing the dimensionless parameter
\begin{eqnarray}
\alpha=\frac{e^2}{\hbar c},
\label{alphaeff}
\end{eqnarray}
coinciding with the fine structure constant only when
$e$ is the charge of the electron, this transformation can be rewritten as
\begin{eqnarray}
G\rightarrow G\left (1-\alpha\frac{M_P^2}{m^2}\right ),
\label{elec7}
\end{eqnarray}
where $M_P=(\hbar c/G)^{1/2}$ is the Planck mass. The gravitational attraction
prevails over
the electric repulsion (allowing stable equilibrium states) provided that
$e^2<e_c^2\equiv Gm^2$ i.e. $\alpha<\alpha_{\rm c}\equiv m^2/M_P^2$. When this
condition is fulfilled, the mass-radius 
relation  (\ref{pa1}) becomes 
\begin{eqnarray}
M=\frac{a\frac{\hbar^2}{(Gm^2-e^2)R}}{1-b^2\frac{a_s\hbar^2}{m(Gm^2-e^2)R^2}}.
\label{elec8}
\end{eqnarray}
Figures \ref{M-R-chi-pos-part1} and \ref{M-R-chi-neg-part1} remain valid with
the new scales $M_a$ and $R_a$ obtained by using the transformation
(\ref{elec7}).
There is always an equilibrium state (for any mass $M$) when $a_s\ge 0$ while 
an equilibrium state exists only below the maximum mass [see Eq. (\ref{pa8})
with Eq. (\ref{elec7})]
\begin{eqnarray}
M_{\rm max}=\frac{a}{2b}\, \frac{\hbar}{\sqrt{\frac{|a_s|}{m}(Gm^2-e^2)}}
\label{elec9}
\end{eqnarray}
when $a_s<0$. When $e^2>e_c^2=Gm^2$ the equilibrium states are
unstable.\footnote{The results of Appendix \ref{sec_elecn} can be generalized by
making the substitution $e^2\rightarrow e^2-Gm^2$.}

For relativistic bosons with a repulsive self-interaction ($a_s\ge 0$), 
substituting $R=kGM/c^2$ into Eq. (\ref{elec8}), we obtain\footnote{In
principle, we should take into account the electrostatic correction to the
Schwarzschild radius. This correction will be considered in a future work.}
\begin{eqnarray}
M_{\rm max}=\left (\frac{a}{k}\right )^{1/2}\frac{\hbar c}{Gm}\left (\frac{Gm^2}{Gm^2-e^2}\right )^{1/2}\sqrt{1+\frac{b^2}{ak}\frac{a_s c^2}{Gm}}.\nonumber\\
\label{elec10}
\end{eqnarray}
This is the general relativistic maximum mass 
of a charged boson star arising from the fact that its radius
approaches the Schwarzschild radius (see
Sec. \ref{sec_is}). The electrostatic corrections are encapsulated in the factor
\begin{eqnarray}
\left (\frac{Gm^2}{Gm^2-e^2}\right )^{1/2}.
\label{elec11}
\end{eqnarray}
Therefore, using the interpolation formulae of Sec. \ref{sec_summ} and
including the electrostatic corrections we get 
\begin{equation}
M_{\rm max}=0.633\frac{\hbar c}{Gm}\left (\frac{Gm^2}{Gm^2-e^2}\right )^{1/2}\sqrt{1+0.235\frac{a_s c^2}{Gm}}
\label{elec12}
\end{equation}
and
\begin{equation}
R_{*}=6.03\frac{\hbar}{mc}\left (\frac{Gm^2}{Gm^2-e^2}\right )^{1/2}\sqrt{1+0.101\frac{a_s c^2}{Gm}}.
\label{elec13}
\end{equation}

For relativistic bosons with an attractive self-interaction ($a_s\le 0$), making
the transformation from Eq. (\ref{rq7}) into Eq. (\ref{elec8}), we obtain the
mass-radius relation\footnote{In principle, we should take into account the
electrostatic correction in the relativistic quantum potential. This correction
will be considered in a future work.}
\begin{eqnarray}
M=\frac{a\frac{\hbar^2}{(Gm^2-e^2)R}}{1+b^2\frac{|a_s-\kappa\frac{Gm}{c^2}|\hbar^2}{m(Gm^2-e^2)R^2}}.
\label{elec14}
\end{eqnarray}
This mass-radius relation displays a maximum mass. Using the interpolation
formulae of Sec. \ref{sec_summ} and including the electrostatic corrections we
get 
\begin{eqnarray}
\label{elec15}
M_{\rm max}=1.012\, \frac{\hbar}{\sqrt{Gm\left |a_s-2.56 \frac{Gm}{c^2}\right
|}}\left (\frac{Gm^2}{Gm^2-e^2}\right )^{1/2},\nonumber\\
\end{eqnarray}
and
\begin{eqnarray}
\label{elec16}
R_*=5.5\, \left (\frac{\left |a_s-1.20 \frac{Gm}{c^2}\right |
\hbar^2}{Gm^3}\right )^{1/2}\left (\frac{Gm^2}{Gm^2-e^2}\right )^{1/2}.\nonumber\\
\end{eqnarray}

In the noninteracting limit ($a_s=0$), using Eqs. (\ref{elec12}) and (\ref{elec13}) or Eqs. (\ref{elec15}) and (\ref{elec16}), we obtain
\begin{equation}
M_{\rm max}=0.633\frac{\hbar c}{Gm}\left (\frac{Gm^2}{Gm^2-e^2}\right )^{1/2},
\label{elec17}
\end{equation}
\begin{equation}
R_{*}=6.03\frac{\hbar}{mc}\left (\frac{Gm^2}{Gm^2-e^2}\right )^{1/2}.
\label{elec18}
\end{equation}
For $e\rightarrow e_c=Gm^2$, taking $G=\hbar=c=m=1$ for convenience (we adopt
the same convention below), we find that $M_{\rm
max}\sim 0.448\, (1-e)^{-1/2}$ and $R_{*}\sim 4.26\, (1-e)^{-1/2}$.

In the TF limit (when $a_s>0$ with $a_s\gg Gm/c^2$), using Eqs. (\ref{elec12}) and (\ref{elec13}), we obtain
\begin{equation}
M_{\rm max}=0.307\, \left (\frac{a_s \hbar^2c^4}{G^3m^3}\right )^{1/2}\left (\frac{Gm^2}{Gm^2-e^2}\right )^{1/2},
\label{elec19}
\end{equation}
\begin{equation}
R_{*}=1.92\, \left (\frac{a_s \hbar^2}{Gm^3}\right )^{1/2} \left (\frac{Gm^2}{Gm^2-e^2}\right )^{1/2}.
\label{elec20}
\end{equation}
For $e\rightarrow e_c=Gm^2$, we find that $M_{\rm
max}\sim 0.217\, (1-e)^{-1/2}$ 
and $R_{*}\sim 1.36\, (1-e)^{-1/2}$. 

In the nonrelativistic limit (when $a_s<0$ with $|a_s|\gg Gm/c^2$), using  Eqs. (\ref{elec15}) and (\ref{elec16}), we obtain
\begin{eqnarray}
\label{elec21}
M_{\rm max}=1.012\, \frac{\hbar}{\sqrt{Gm |a_s|}}\left (\frac{Gm^2}{Gm^2-e^2}\right )^{1/2},
\end{eqnarray}
\begin{eqnarray}
\label{elec22}
R_*=5.5\, \left (\frac{ |a_s |
\hbar^2}{Gm^3}\right )^{1/2}\left (\frac{Gm^2}{Gm^2-e^2}\right )^{1/2}.
\end{eqnarray}
For $e\rightarrow e_c=Gm^2$, we find that $M_{\rm
max}\sim 0.716\, (1-e)^{-1/2}$ and $R_{*}\sim 3.89\, (1-e)^{-1/2}$.

Charged boson stars in general relativity  have been studied numerically by
Jetzer and van der Bij  in Refs. \cite{jetzerelec1,jetzerelec2} by solving the
Klein-Gordon-Maxwell-Einstein (KGME) equations numerically. Their asymptotic
results for $e\rightarrow e_c=1$,  displaying the scalings $M_{\rm max}\propto
(1-e)^{-1/2}$ and $R_{*}\propto (1-e)^{-1/2}$, are recovered by our analytical
approach.\footnote{These scalings do not amount to naively making the
substitution $Gm^2\rightarrow Gm^2-e^2$ in the relativistic 
formulae of uncharged boson stars.} For the maximum mass $M_{\rm max}$, they
obtained the prefactors $0.44\times 2^{1/4}=0.523$ and $0.226\times
2^{1/4}=0.269$ in the noninteracting and TF limits respectively.  These
numerical results are reasonable close to our approximate analytical results
$0.448$ and
$0.217$.

\subsection{Model of extended particles with electrostatic interactions}
\label{sec_epec}

We can use the foregoing results to take into account
electrostatic interactions in the models of relativistic elementary particles
considered in Appendix \ref{sec_releff}. We focus on the case where the
gravitational forces are more
important than the electrostatic forces (i.e.
$m>e/\sqrt{G}=\sqrt{\alpha}M_P$). 

\subsubsection{Nonrelativistic mass-radius relation}

We can easily generalize the nonrelativistic mass-radius relation $m(R)$ from
Eq. (\ref{mo3}) to a charged particle by making the substitution from Eq.
(\ref{elec7}). The general formula is a bit cumbersome but it takes a simple
form in appropriate limits. In the noninteracting case, we get [see Eq.
(\ref{mo4}) with
(\ref{elec7})]
\begin{eqnarray}
R=a\frac{\hbar^2}{Gm^3\left (1-\alpha \frac{M_P^2}{m^2}\right )}.
\label{mo4elec}
\end{eqnarray}
In the TF limit when $\lambda>0$, we get [see Eq. (\ref{mo7}) with
(\ref{elec7})]
\begin{eqnarray}
R\sim b\sqrt{\frac{\lambda}{8\pi}}\left \lbrack \frac{\hbar^3}{Gm^4\left
(1-\alpha \frac{M_P^2}{m^2}\right ) c}\right \rbrack^{1/2}.
\label{mo7elec}
\end{eqnarray}
The radius diverges when
$m\rightarrow \sqrt{\alpha}M_P$. We recover the nonelectrostatic case when
$m\gg \sqrt{\alpha}M_P$.

\subsubsection{Relativistic mass-radius relation when $\lambda\le 0$}

For an attractive self-interaction $\lambda\le 0$, we can easily generalize
the relativistic mass-radius relation $m(R)$ from
Eq. (\ref{mo3gv}) to a charged particle by making the substitution from Eq.
(\ref{elec7}) except in the term which is proportional to $\kappa$ since we
have assumed that it is not affected by electrostatic corrections (see footnote
48). The general formula is a bit cumbersome but for a
noninteracting ($\lambda=0$) particle, it reduces to
\begin{eqnarray}
R&=&\frac{a\hbar^2}{2G\left (1-\alpha \frac{M_P^2}{m^2}\right
)m^3}\nonumber\\
&\pm&\sqrt{\frac{a^2\hbar^4}{4G^2\left (1-\alpha
\frac{M_P^2}{m^2}\right )^2m^6}-\frac{\kappa b^2\hbar^2}{m^2c^2\left (1-\alpha
\frac{M_P^2}{m^2}\right )}}.\nonumber\\
\label{mo3gvbelec}
\end{eqnarray}

\subsubsection{Maximum mass}

We now determine how electrostatic corrections affect the maximum mass of the
particle calculated in Appendix \ref{sec_smeep}.

For a repulsive self-interaction, using Eqs. (\ref{lambda}) and
(\ref{alphaeff}) and making $M=m$ in Eq. (\ref{elec12}),
we find that the relation between the maximum particle mass $m_{\rm max}$ and
the dimensionless self-interaction constant $\lambda$ can be expressed under the
inverse form as
\begin{eqnarray}
\label{epec1}
\frac{\lambda}{8\pi}=\frac{2.50\frac{m_{\rm max}^2}{M_P^2}\left
(\frac{m_{\rm max}^2}{M_P^2}-\alpha\right )-1}{0.235 \frac{M_P^2}{m_{\rm
max}^2}}.
\end{eqnarray}
When $\alpha=0$ (i.e. $e=0$), we recover Eq. (\ref{mo26}).

For an attractive self-interaction, using Eqs. (\ref{lambda})
and (\ref{alphaeff}) and making $M=m$ in Eq.
(\ref{elec15}), 
we find that
\begin{eqnarray}
\label{epec2}
\frac{|\lambda|}{8\pi}=\frac{1.02}{\frac{m_{\rm
max}^2}{M_P^2}-\alpha}-2.56\frac{m_{ \rm
max}^2}{M_P^2}.
\end{eqnarray}
When $\alpha=0$, we recover Eq. (\ref{mo27}).

In the noninteracting case,  making $M=m$ in Eq. (\ref{elec17}), we obtain
\begin{eqnarray}
\label{epec3}
m_{\rm max}=M_P \left\lbrack \frac{\alpha}{2}+\sqrt{\frac{\alpha^2}{4}+0.401}\,
\right\rbrack^{1/2}. 
\end{eqnarray}
When $\alpha=0$, we recover Eq. (\ref{mo23}). When $\alpha\gg 1$, we get $m\sim
\sqrt{\alpha}\, M_P$.\footnote{Recall that $e$ is not necessarily the charge of
the electron so $\alpha$ can take large values.}

In the TF limit (when $\lambda>0$ and $\lambda\gg 1$),
making $M=m$ in Eq. (\ref{elec19}), we obtain
\begin{eqnarray}
\label{epec4}
\frac{\lambda}{8\pi}=10.6\, \frac{m_{\rm max}^4}{M_P^4}\left
(\frac{m_{\rm max}^2}{M_P^2}-\alpha\right ).
\end{eqnarray}
When $\alpha=0$, or when $\lambda\rightarrow +\infty$, we recover Eq.
(\ref{mo24}).

In the nonrelativistic limit (when $\lambda<0$ with $|\lambda|\gg 1$), making
$M=m$ in Eq.
(\ref{elec21}), we obtain
\begin{eqnarray}
\label{epec5}
m_{\rm max}=M_P \sqrt{1.02\,
\frac{8\pi}{|\lambda|}+\alpha}.
\end{eqnarray}
When $\alpha=0$, we recover Eq. (\ref{mo25}). When $|\lambda|\rightarrow
+\infty$, we find $m_{\rm
max}\rightarrow \sqrt{\alpha}M_P$.

\subsection{A simple model of extended electron}
\label{sec_elecsimple}

We can use the results of Appendix \ref{sec_elecn} to construct  a simple
model of extended electron.\footnote{This model can be related to the models of
extended particles listed in footnote 13. Unfortunately, the particle of our
model appears to be unstable.} We assume that the
wave function
$\psi({\bf r},t)$
governed by the  Schr\"odinger Eq. (\ref{elec1}) determines the density profile
$\rho({\bf r},t)$ of the electron through the relation $\rho=|\psi|^2$. The
corresponding density of charge $-\rho e/m$ creates, via the Poisson equation
(\ref{elec2}), an electric potential $(m/e)\Phi$  which enters into the
Schr\"odinger equation (\ref{elec1}). Finally, we identify the
mass $m$ appearing in the Schr\"odinger equation (\ref{elec1}) and the mass
$M=\int\rho\, d{\bf r}$ produced by $\rho$ with the mass $m_e$ of the electron,
following an argument similar to the one given by Diosi \cite{diosi} for  the
gravitational interaction (see Appendix \ref{sec_smeep}).\footnote{Vlasov
\cite{vlasov} argued that this
system of coupled Schr\"odinger-Poisson equations was imagined by Schr\"odinger
himself but Pitaevskii \cite{pita} contested that claim.} Of course,
since the
electrostatic
interaction and the quantum potential are both repulsive, this model does not
yield any equilibrium state (contrary to the gravitational model of Diosi
\cite{diosi} where the gravitational attraction compensates the repulsion
of the quantum potential). Therefore, we add an attractive $|\psi|^4$
self-interaction ($\lambda<0$) which opposes itself to the electrostatic
repulsion. This attractive term is similar in spirit to the Poincar\'e stress
\cite{poincare1905,poincareE} introduced in the
Abraham-Lorentz \cite{abraham,lorentz} electromagnetic model of the
electron to stabilize the particle (see Appendix \ref{sec_alm}). 
Using Eqs.
(\ref{elec3}) and
(\ref{lambda}), introducing the fine-structure constant (\ref{alpha}), and
taking $M=m_e$,  we obtain the electron mass-radius relation
\begin{eqnarray}
m_e^2+\frac{a e^2}{\alpha^2 R c^2}m_e-\frac{b^2|\lambda|e^4}{8\pi \alpha^3 c^4
R^2}=0,
\label{elec23}
\end{eqnarray}
where $a=2.71$ and $b=\pi$. The solution of this second degree equation is
\begin{eqnarray}
m_e c^2=\frac{e^2}{R}\left\lbrace
\sqrt{\frac{a^2}{4\alpha^4}+\frac{b^2|\lambda|}{8\pi\alpha^3}}-\frac{a}{
2\alpha^2}\right\rbrace.
\label{elec24}
\end{eqnarray}
It
can be written as
\begin{eqnarray}
m_e c^2=\chi(\lambda) \frac{e^2}{R}
\label{elec24a}
\end{eqnarray}
with the dimensionless constant
\begin{eqnarray}
\chi(\lambda)=\left\lbrace
\sqrt{\frac{a^2}{4\alpha^4}+\frac{b^2|\lambda|}{8\pi\alpha^3}}-\frac{a}{
2\alpha^2}\right\rbrace.
\label{elec24b}
\end{eqnarray}
Interestingly, this relation is similar to the relation appearing in the
Abraham-Lorentz \cite{abraham,lorentz} model
of the electron and in the Born-Infeld \cite{born1933,borninfeld} 
theory (see Appendix
\ref{sec_mee}).
Comparing Eq. (\ref{elec24a}) with Eq. (\ref{elec30b}) we get
\begin{eqnarray}
R=\chi(\lambda) r_e,
\label{elec24c}
\end{eqnarray}
where $r_e$ is the classical electron radius (\ref{mee2}).

\subsubsection{TF approximation}

In the TF approximation $|\lambda|/8\pi\gg a^2/(4b^2\alpha)$, the term
proportional to $m_e$ in Eq. (\ref{elec23}) can be neglected and Eq.
(\ref{elec24})
reduces to
\begin{eqnarray}
m_e c^2=\frac{b}{\alpha^{3/2}}\sqrt{\frac{|\lambda|}{8\pi}}\frac{e^2}{R}.
\label{elec25}
\end{eqnarray}
In that case, the equilibrium state results from the balance between the
repulsive electrostatic interaction and the attractive self-interaction.
Actually, in the TF approximation, we can solve the problem exactly. Making the
substitutions $Gm^2\rightarrow -e^2$ and $a_s\rightarrow -|a_s|$ in Eq.
(\ref{dm7tf}), using Eq. (\ref{lambda}) and introducing the fine-structure
constant (\ref{alpha}), we obtain\footnote{This equation is equivalent
to the Lane-Emden equation for a polytrope of index $n=1$ which describes the
balance between the gravitational attraction and the repulsive self-interaction.
In the present case, it describes the balance between the electrostatic
repulsion and the attractive self-interaction. However, in the gravitational
case (boson stars) the equilibrium is stable while in the electrostatic case
(electron) it is unstable.}
\begin{eqnarray}
\label{elec27}
\Delta\rho+\frac{8\pi m_e^2c^4\alpha^3}{|\lambda|e^4}\rho=0.
\end{eqnarray}
The density profile of the extended electron, which is determined by Eq.
(\ref{elec27}), is given by [see Eq. (\ref{mg18})]
\begin{eqnarray}
\label{elec27b}
\rho(r)=\frac{\rho_0 R}{\pi r}\sin\left (\frac{\pi r}{R}\right ).
\end{eqnarray}
Its radius is [see Eq. (\ref{dm15})] 
\begin{eqnarray}
\label{elec28}
R=\frac{\pi}{\alpha^{3/2}} \sqrt{\frac{|\lambda|}{8\pi}}\frac{e^2}{m_e c^2}
\end{eqnarray}
It can be written as Eq. (\ref{elec24c}) with
\begin{eqnarray}
\label{elec28b}
\chi_{\rm TF}(\lambda)=\frac{\pi}{\alpha^{3/2}} \sqrt{\frac{|\lambda|}{8\pi}}.
\end{eqnarray}
Its central
density is [see Eq. (\ref{rhoM})] 
\begin{eqnarray}
\label{elec29}
\rho_0=\frac{\pi m_e}{4R^3}.
\end{eqnarray}
The timescale of the instability is of order [see Eq.
(\ref{tdyn})]
\begin{eqnarray}
\label{elec29we}
t_D\sim \left (\frac{m^2}{\rho_0 e^2}\right )^{1/2}\sim \left
(\frac{m_eR^3}{e^2}\right )^{1/2}.
\end{eqnarray}
Finally, according to Eq. (\ref{Etot}), the total energy
(electrostatic $+$ internal) of the electron is\footnote{The energy of the
electron is positive implying that this configuration is unstable. By contrast,
the gravitational energy of a boson star is negative consistently  with the fact
that these objects are stable.}
\begin{eqnarray}
\label{elec30}
E_{\rm tot}=\frac{e^2}{2R}.
\end{eqnarray}
Combining Eq.
(\ref{elec30}) with Eq. (\ref{elec24a}) we find that
\begin{eqnarray}
\label{elec30a}
E_{\rm tot}=\frac{m_e c^2}{2\chi_{\rm TF}(\lambda)}.
\end{eqnarray}
If we impose that $R=r_e$, we get $\chi_{\rm TF}(\lambda)=1$. In that case,
the total energy of the electron is\footnote{We note that $E_{\rm tot}\neq m_e
c^2$. Alternatively, we could impose
$E_{\rm tot}=m_e c^2$ 
and deduce that the radius of the electron is given by $m_e c^2=e^2/(2R)$ [see
Eq. (\ref{elec30})] i.e. $R=e^2/(2m_e c^2)=r_e/2$. The relation $E=mc^2$
(sometimes with a prefactor $3/4$) has a long history in physics even before
Einstein's theory of relativity. It appeared at the end of the 19th century when
some researchers like Thomson noticed that the electromagnetic energy is
equivalent to mass. It was also used by Born and Infeld in their nonlinear
electrodynamics (see Appendix \ref{sec_mee}).} 
\begin{eqnarray}
\label{elec30b}
E_{\rm tot}=\frac{1}{2}m_e c^2,
\end{eqnarray}
where $m_e c^2$ is its rest-mass energy. On the other hand, its central
density is
\begin{eqnarray}
\label{elec29b}
\rho_0=\frac{\pi}{4}\rho_e,
\end{eqnarray}
where $\rho_e$ is the classical electron density
(\ref{mee4}). The timescale of the instability is of order
\begin{eqnarray}
t_D\sim t_e
\end{eqnarray}
where $t_e$ is the chronon (\ref{mee5}).

The condition $\chi_{\rm TF}(\lambda)=1$ implies that the self-interaction
constant is given by
\begin{eqnarray}
\label{elec32}
\frac{\lambda}{8\pi}=-\frac{\alpha^3}{\pi^2}=-3.94\times 10^{-8}.
\end{eqnarray}
This result shows 
that the TF approximation $|\lambda|/8\pi\gg a^2/(4b^2\alpha)=25.5$ is not
justified.
One has to take into account the contribution of the quantum potential and use
the more general
mass-radius relation from Eq. (\ref{elec24}).

\subsubsection{Nonelectrostatic limit}

In the nonelectrostatic limit $|\lambda|/8\pi\ll
a^2/(4b^2\alpha)=25.5$, the term
proportional to $m_e^2$ in Eq. (\ref{elec23}) can be neglected and Eq.
(\ref{elec24})
reduces to
\begin{eqnarray}
m_e c^2=\frac{b^2}{a}\frac{|\lambda|}{8\pi}\frac{e^2}{\alpha R}.
\end{eqnarray}
In that case, the equilibrium state results from the balance between the
repulsive quantum potential and the attractive self-interaction (see Appendix
\ref{sec_gh}). The total
energy is of order $E_{\rm tot}=0.393 m_ec^2/(\lambda/8\pi)^2$ [see Eq. (54) of
\cite{prd2} with Eq. (\ref{lambda})]. If we impose that
$R=r_e$ (i.e. $\chi_{\rm NE}(\lambda)=1$)  we get\footnote{This
yields $E_{\rm tot}=9.82\times 10^4\, m_e c^2$. Alternatively, we could impose
$E_{\rm tot}=m_ec^2$ and get $|\lambda|/8\pi=0.627$ and $R=313\,
r_e=8.82\times 10^{-13}\, {\rm m}$. In that case, the electron radius is of the
order of its Compton wavelength $\lambda_e=\hbar/(m_e c)=3.86\times 10^{-13}\,
{\rm m}$ (see Appendix \ref{sec_mee}), which is sensible since we are using a
quantum model.}
\begin{eqnarray}
\frac{|\lambda|}{8\pi}=\frac{a}{b^2}\alpha=2.00\times 10^{-3}.
\end{eqnarray}
This result is consistent with the condition of validity of the nonelectrostatic
limit. However, the equilibrium state is unstable.

\subsubsection{General case}

In the general case, if we impose that
$R=r_e$ (i.e. $\chi(\lambda)=1$) we get
\begin{eqnarray}
\label{elec32b}
\frac{|\lambda|}{8\pi}=\frac{a\alpha+\alpha^3}{b^2}.
\end{eqnarray}
Since $\alpha\ll 1$, we have in good approximation
\begin{eqnarray}
\label{elec32c}
\frac{|\lambda|}{8\pi}\simeq \frac{a}{b^2}\alpha=2.00\times 10^{-3},
\end{eqnarray}
corresponding to the nonelectrostatic limit that we have discussed previously.
This is also a regime of weak self-interaction since $|\lambda|/8\pi\ll
a^2/(4b^2\alpha)$.

In conclusion, our simple model of extended electron determines its density
profile (it can be obtained by solving the electrostatic
Gross-Pitaevskii-Poisson
equation
(\ref{elec1}) and
(\ref{elec2}) or, in good approximation, by solving the nonelectrostatic
Gross-Pitaevskii
equation alone) as well as its radius and central density. Eq. (\ref{elec24})
with $\chi(\lambda)=1$ is consistent with the relation $m_e c^2=e^2/r_e$ usually
introduced from qualitative considerations, but it is obtained here
from the solution of the electrostatic GPP equations. Unfortunately,  this
equilibrium state is unstable. The timescale of the
instability is [see Eq. (112) in \cite{prd1}) 
\begin{eqnarray}
\label{elec34}
t_u\sim \frac{m_e r_e^2}{\hbar}=6.87\times
10^{-26}\, {\rm s}.
\end{eqnarray}
In this model, the electron 
would have a very short lifetime. Therefore, although this model correctly
reproduces the relation between the mass and the radius of the electron, it
cannot account for its stability unless a particular mechanism to increase its
lifetime is found. Maybe, this model
could describe another elementary particle, different from the electron. In
that case, $e$ does not need to represent the elementary
charge and the lifetime of the particle may be enhanced.

\section{Models of extended electron}
\label{sec_mee}

In this Appendix we present basic equations applying to the electron and we
briefly recall the historical background.

\subsection{Basic equations}

The classical radius $r_e$ of the electron is defined  through the relation
\begin{eqnarray}
\label{mee1}
E=m_e c^2=\frac{e^2}{r_e}.
\end{eqnarray}
This equation expresses the equality (in order of magnitude) between the
rest-mass energy of the
electron and its electrostatic energy. This is a convenient manner to define the
``radius'' of the electron. This relation first appeared in the Abraham-Lorentz
\cite{abraham,lorentz} model of the extended
electron with electromagnetic mass and later in the Born-Infeld
\cite{born1933,borninfeld} theory
of
nonlinear electrodynamics. Recalling the value of the charge of the electron
$e=4.80\times 10^{-13}\, {\rm
g^{1/2}\, m^{3/2}\, s^{-1}}$ and its mass $m_e=9.11\times 10^{-28}\, {\rm
g}=0.511\, {\rm MeV/c^2}$,
we obtain
\begin{eqnarray}
\label{mee2}
r_e=\frac{e^2}{m_e c^2}=2.82\times 10^{-15}\, {\rm m}.
\end{eqnarray}
The Compton wavelength of the electron is
$\lambda_e=\hbar/(m_e c)=3.86\times 10^{-13}\, {\rm m}$. It is related to the
classical radius of the electron by
\begin{eqnarray}
\label{mee3}
\lambda_e=\frac{r_e}{\alpha}\simeq 137\, r_e,
\end{eqnarray}
where
\begin{eqnarray}
\label{alpha}
\alpha=\frac{e^2}{\hbar c}\simeq
\frac{1}{137}\simeq
7.30\times 10^{-3}
\end{eqnarray}
is Sommerfeld's fine-structure constant.\footnote{Since quantum effects enter
at a distance of the order $\lambda_e$ which is much larger than $r_e$, a
purely classical electromagnetic model of the electron is not relevant.} In
comparison, the Bohr (atomic) radius is $a_B=\hbar^2/(m_e
e^2)=r_e/\alpha^2=5.29\times 10^{-11}\, {\rm m}$. We have $r_e\ll \lambda_e\ll
a_B$. The
typical electron density is
\begin{eqnarray}
\label{mee4}
\rho_e=\frac{m_e}{r_e^3}=4.07\times 10^{16}\, {\rm g\, m^{-3}}.
\end{eqnarray}
The dynamical time associated with the electron is
\begin{eqnarray}
\label{mee5}
t_e=\left (\frac{m_e r_e^3}{e^2}\right
)^{1/2}=\frac{e^2}{m_e c^3}=\frac{r_e}{c}=3.32\times
10^{-24}\, {\rm
s}.
\end{eqnarray}
This is the time it takes for a light
wave to travel accross the
``size'' of an electron. This time first appeared in the Abraham-Lorentz
\cite{abraham,lorentz} theory of the extended electron when they tried to
calculate the recoil force on an accelerated charged particle caused by the
particle emitting electromagnetic radiation. This is also what Caldirola
\cite{chronon} called the
``chronon'', which is a sort of ``quantum of time''.

\subsection{Abraham-Lorentz model}
\label{sec_alm}

In the model of extended electron developed by Abraham \cite{abraham} and
Lorentz \cite{lorentz}, the
electron is considered as a spherical charge of radius $R$ (which
must be nonzero to avoid infinite energy accumulation) with a
charge $e$
uniformly distributed on its surface.  It was originally believed that the mass
of the electron
had a purely
electromagnetic nature. The electromagnetic energy of the electron
at rest (reducing to its electrostatic energy) is
\begin{eqnarray}
\label{al1}
E=\frac{1}{2}\, \frac{e^2}{R}.
\end{eqnarray}
On the other hand, the  electromagnetic impulse (Poynting vector) of the
electron in slow motion ($v\ll c$) is 
\begin{eqnarray}
\label{al2}
{\bf p}=\frac{2}{3}\, \frac{e^2}{Rc^2}{\bf v}.
\end{eqnarray}
It is proportional to the velocity like in the classical mechanical relation
${\bf p}=m{\bf
v}$. This suggests introducing the ``electromagnetic mass'' of the electron
\begin{eqnarray}
\label{al3}
m_e=\frac{2}{3}\, \frac{e^2}{Rc^2}.
\end{eqnarray}
This relation can also be written as
\begin{eqnarray}
\label{al4}
m_ec^2=\frac{2}{3}\, \frac{e^2}{R}.
\end{eqnarray}
Comparing Eq. (\ref{al4}) with Eq. (\ref{mee1}) we get
\begin{equation}
R=\frac{2}{3}r_e=1.88\times 10^{-15}\, {\rm m}.
\label{al5}
\end{equation}
This is the radius of the electron in the Abraham-Lorentz theory. In this sense,
the Abraham-Lorentz theory {\it justifies} the relation from
Eq. (\ref{mee1}) defining the classical radius of the
electron.

For an  arbitrary velocity, Lorentz \cite{lorentz1899} understood that
the charged sphere had to contract itself into an ellipsoid. Therefore, the
electrons undergo
length contraction in the line of motion. In his famous 1904
paper \cite{lorentz1904}, he
obtained
the relation 
\begin{eqnarray}
\label{al6}
{\bf p}=\frac{2}{3}\, \frac{e^2}{Rc^2}\frac{\bf v}{\sqrt{1-v^2/c^2}}.
\end{eqnarray}
In that case, the mass of the electron depends on the velocity as
\begin{eqnarray}
\label{al7}
m=\frac{m_e}{\sqrt{1-v^2/c^2}},
\end{eqnarray}
where the rest mass $m_e$ is given by Eq. (\ref{al3}).
This relation was discovered before Einstein's theory of relativity. It shows
that the
velocity cannot be larger than the speed of light. On the other hand, if we
combine Eqs. (\ref{al1}) and (\ref{al4}), we
get 
\begin{eqnarray}
\label{al8}
E=\frac{3}{4}m_ec^2
\end{eqnarray}
instead of $E=m_ec^2$. Abraham \cite{abraham1904,abraham} understood
that a purely charged sphere is unstable since the electric forces are all
repulsive. Poincar\'e
\cite{poincare1905,poincareE} solved
these problems by introducing additional forces of
nonelectromagnetic origin (Poincar\'e
stresses) that maintain the charges in the sphere. These stresses
prevent the electron from exploding and contribute to $1/4$ of the total energy
restoring the expected relation $E=m_ec^2$.
Poincar\'e stresses were also
thought as a dynamical explanation of Lorentz
length contraction. In this theory, the
mass of the electron has an
electromagnetic part and a part due to Poincar\'e stresses. It is impossible
that
all the mass is electromagnetic in origin. The result of Einstein's relativity
theory
\cite{einstein1905} then showed that the dependence of
mass on velocity is not characteristic of electromagnetic mass, but can
be derived very generally from the transformation law. After Einstein's theory
of
relativity, several authors 
pointed out that the electron's stability and the $4/3$ problem are two
different things \cite{rohrlich} However, binding forces (or
confining pressure) like the Poincar\'e stresses
are still necessary to prevent the electron from exploding due to Coulomb
repulsion.

{\it Remark:} The concept of  ``electromagnetic mass'' was
introduced by Thomson \cite{thomson1881} in 1881 (he discovered
the electron in 1897). He realized that the apparent mass of a charged body in
motion is larger
than the mass it would have if it were uncharged
(similar considerations were
already made by Stokes \cite{stokes} in hydrodynamics).
The electromagnetic mass was initially considered as a dynamical explanation of
the inertial mass of an object. This idea was further developed by Heaviside
\cite{heaviside1889}, Thomson \cite{thomson1893}, Searle \cite{searle1897},
Abraham \cite{abraham1903}, Lorentz \cite{lorentz1892,lorentz1904}, and was
incorporated in
the Abraham-Lorentz theory of the electron. Even before the work of Lorentz
\cite{lorentz1899}, Heaviside \cite{heaviside1889},
 Thompson \cite{thomson1893} and Searle \cite{searle1897} understood that the
mass of a charged body depends on
its velocity and becomes infinite when $v\rightarrow c$. They concluded that
no body can move at a speed greater than the speed of light. Heaviside
\cite{heaviside1889} and Thompson \cite{thomson1893} seem to be the first
to have isolated the factor
\begin{eqnarray}
\label{al9}
\frac{1}{\sqrt{1-v^2/c^2}}
\end{eqnarray}
in the energy (mass) of a charged body in motion.
Searle \cite{searle1897} obtained a different formula
\begin{eqnarray}
\label{al10}
E=E_0\left \lbrack\frac{c}{v}\ln\left (\frac{1+v/c}{1-v/c}\right
)-1\right\rbrack,
\end{eqnarray}
where $E_0=e^2/2R$ is the rest mass energy. For $v\ll c$, this formula becomes
\begin{eqnarray}
\label{al11}
E=E_0\left (1+\frac{2}{3}\frac{v^2}{c^2}+...\right ).
\end{eqnarray}
Using this result, and identifying the second term of the expansion
$\frac{2}{3}E_0\frac{v^2}{c^2}$ with the kinetic energy $\frac{1}{2}mv^2$, Wien
\cite{wien1900} obtained the relation
\begin{eqnarray}
\label{al12}
m=\frac{4}{3}\frac{E_0}{c^2}
\end{eqnarray} 
between the (electromagnetic) mass $m$ and the energy of the body at rest
$E_0$. It is apparently
the first time that
the famous formula $E=mc^2$ appeared in the literature (under the form
$m=\frac{4}{3}\mathfrak{E} A^2$)
with, however, the wrong prefactor $3/4$.

\subsection{Born-Infeld model}

The Maxwell equations of classical electrodynamics lead to the
description of an elementary charged particle like an electron as a singular
point in the
electromagnetic field. It was
initially thought that the mass of the electron was entirely due to its
electromagnetic self-energy. However, the self-energy of a point charge
(``point-electron'') in Maxwell's electrodynamics is infinite. Abraham
\cite{abraham} and Lorentz \cite{lorentz} respectively 
introduced the notion of rigid and contracting  electron with a finite size but
models of extended
electron need cohesive forces of nonelectromagnetic origin (Poincar\'e
stresses \cite{poincare1905,poincareE}) to stabilize the structure (see
Appendix \ref{sec_alm}). Therefore, modifications
of the Maxwell equations were suggested first by Mie \cite{mie} then
by Born and Infeld \cite{born1933,borninfeld}.
Using an analogy with special relativity, Born and Infeld developed a theory of
nonlinear electrodynamics which prevents the divergence of the electric field
produced by a point charge.  They introduced an electromagnetic Lagrangian
of the form
\begin{eqnarray}
\label{bi1}
{\cal L}_{\rm BI}=-b^2\sqrt{1-\frac{E^2-B^2}{b^2}},
\end{eqnarray}
which reduces to the Lagrangian of Maxwell electrodynamics
in the weak field limit.  This Lagrangian contains a
fundamental constant $b$ (``absolute field'') which plays the role of the speed
of light $c$ in special relativity. In the electrostatic case, the universal
constant $b$ is simply the upper limit of the field strengh ${\bf E}$. As a
result, the 
Born-Infeld equations which replace the Maxwell equations have a solution for
the point electron with the field everywhere finite and with a finite
self-energy.
Born and Infeld computed the electrostatic energy of the electron (which is now
finite) and
obtained the formula
\begin{eqnarray}
\label{bi2}
E=1.24\, \frac{4\pi e^2}{r_*},
\end{eqnarray}
where $r_*=(e/b)^{1/2}$. They proposed
to identify the electric energy with the mass of the
electron via the relation $E=m_e c^2$. This yields
\begin{eqnarray}
\label{bi3}
m_e c^2=1.24\, \frac{4\pi e^2}{r_*}.
\end{eqnarray}
Comparing Eq. (\ref{bi3}) with Eq. (\ref{mee1}) we get
\begin{equation}
r_*=1.24\times 4\pi\, r_e=4.39\times 10^{-14}\, {\rm m},
\label{bi46}
\end{equation}
which may be interpreted as the effective radius of the electron in the
Born-Infeld
theory. In this sense, the Born-Infeld theory {\it justifies} the relation from
Eq. (\ref{mee1}) defining the classical radius of the electron. In the 
Born-Infeld
electrodynamics, the electron has a finite radius because the electric field and
the electrostatic energy of a point charge are finite. Their theory of the
electron
can be considered
as a revival of the old
idea of the electromagnetic origin of mass; namely, that the
electron is a singularity in the electromagnetic field and
that its mass is purely electromagnetic.

\end{document}